\documentclass[prd,showpacs,floatfix,nofootinbib,twocolumn,amsmath,amssymb]{revtex4}
\usepackage{graphicx,graphics,color,dcolumn,booktabs,bm}
\usepackage{longtable,lscape}
\usepackage{txfonts}
\usepackage{overpic}
\usepackage{amssymb}
\usepackage{indentfirst}
\usepackage{epsfig}
\usepackage{feynmf}   
\usepackage{epstopdf}   
\usepackage{slashed}  
\usepackage{cases}
\usepackage{color}
\usepackage{multirow}
\usepackage{afterpage}

\graphicspath{{../property_of_higher_bottomonia_20170911/}}

\usepackage[colorlinks, citecolor=blue,anchorcolor=red,menucolor=red, linkcolor=red,filecolor=red,runcolor=red,urlcolor=blue,frenchlinks=red]{hyperref}

\begin{document}

\title{Higher bottomonium zoo}

\author{Jun-Zhang Wang$^{1,2}$}\email{wangjzh2012@lzu.edu.cn}
\author{Zhi-Feng Sun$^{1,2}$}\email{sunzhif09@lzu.edu.cn}
\author{Xiang Liu$^{1,2}$\footnote{Corresponding author}}\email{xiangliu@lzu.edu.cn}
\author{Takayuki Matsuki$^{3,4}$}\email{matsuki@tokyo-kasei.ac.jp}
\affiliation{
$^1$Research Center for Hadron and CSR Physics, Lanzhou University $\&$ Institute of Modern Physics of CAS, Lanzhou 730000, China\\
$^2$School of Physical Science and Technology, Lanzhou University,
Lanzhou 730000, China\\
$^3$Tokyo Kasei University, 1-18-1 Kaga, Itabashi, Tokyo 173-8602, Japan\\
$^4$Theoretical Research Division, Nishina Center, RIKEN, Wako, Saitama 351-0198, Japan
}

\begin{abstract}
In this work, we study higher bottomonia up to the $nL=8S$, $6P$, $5D$, $4F$, $3G$ multiplets using the modified Godfrey-Isgur (GI) model, which takes account of color screening effects. The calculated mass spectra of bottomonium states are in reasonable agreement with the present experimental data. Based on spectroscopy, partial widths of all allowed radiative transitions, annihilation decays, hadronic transitions, and open-bottom strong decays of each state are also evaluated by applying our numerical wave functions. Comparing our results with the former results, we point out difference among various models and derive new conclusions obtained in this paper. Notably, we find a significant difference between our model and the GI model when we study $D, F$, and $G$ and $n\ge 4$ states. Our theoretical results are valuable to search for more bottomonia in experiments, such as LHCb, and forthcoming Belle II.
\end{abstract}


\maketitle

\section{introduction}\label{sec1}
Since $J/\psi$ and $\Upsilon(1S)$ were observed in 1974 and 1977 \cite{Aubert:1974js,Augustin:1974xw,Herb:1977ek,Innes:1977ae}, respectively,
a heavy quarkonium has become an influential and attractive research field because its physical processes cover the whole energy range of QCD.
This energy range provides us an excellent place to study the properties of perturbative and non-perturbative QCD \cite{Brambilla:2010cs}. An exhaustive and deeper study of a heavy quarkonium helps people better understand the QCD characteristics. Since the bottomonium family is an important member of heavy quarkonia, people have made great effort to investigate a bottomonium experimentally and theoretically in the past years. Thirty years ago, a couple of higher excited bottomonia have been experimentally observed in succession, e.g., $\Upsilon(nS)$ with the radial quantum numbers $n$ from 2 to 6 and $P$-wave spin-triplet states $\chi_{bJ}(1P)$ and $\chi_{bJ}(2P)$  with $J=1,2,3$ \cite{Herb:1977ek,Innes:1977ae,Besson:1984bd,Lovelock:1985nb,Klopfenstein:1983nx,Pauss:1983pa,Han:1982zk,Eigen:1982zm}. With the upgrade and improvement of two $B$-Factories BaBar and Belle together with LHC, new breakthroughs in experiments have been achieved in recent years. First, the pseudoscalar partners $\eta_b(1S)$ and $\eta_b(2S)$ were identified by the BaBar Collaboration in 2008 and 2011, respectively \cite{Aubert:2008ba,Lees:2011mx}. Subsequently, the follow-up studies by the CLEO \cite{Bonvicini:2009hs,Dobbs:2012zn} and Belle Collaborations \cite{Mizuk:2012pb} confirmed the existence of these two bottomonia, where Belle did the most accurate measurements of the mass to date with values of $9402.4 \pm1.5\pm1.8$ MeV and $9999.0\pm3.5^{+2.8}_{-1.9}$ MeV, respectively \cite{Mizuk:2012pb}. The $\chi_{b1}(3P)$ state is a new state discovered by LHC in the chain decay of $\chi_{b1}(3P)\to\gamma\Upsilon(1S)/\gamma\Upsilon(2S)\to\gamma\mu^+\mu^-$ \cite{Aad:2011ih}. This state has been confirmed by the D$\emptyset$ Collaboration \cite{Abazov:2012gh}. BaBar discovered a possible signal of the $P$-wave spin-singlet $h_b(1P)$ state in 2011 \cite{Lees:2011zp}, and subsequently, Belle firstly and successfully confirmed the observation of the $h_b(1P)$ in the process of $\Upsilon(5S)$ $\rightarrow$$\pi^+\pi^-h_b(1P)$, where the first radial excited state $h_b(2P)$ has been also observed with mass value $10259.76 \pm 0.64^{+1.43}_{-1.03}$ MeV \cite{Adachi:2011ji}. Thus, both the ground and first radial excited states of the $P$-wave bottomonium have been fully established experimentally. However, this is not enough to make people satisfied because many of experimental information is still incomplete, including the total width and branching ratios of some significant decay channels, which still require both of experimental and theoretical efforts. With regard to the search for a $D$-wave bottomonium in the experiments, there also has some progresses by the CLEO and BaBar Collaborations \cite{Bonvicini:2004yj,delAmoSanchez:2010kz}, where spin-triplet $\Upsilon(1^3D_2)$ was observed with a 5.8 $\sigma$ significance while confirmation of $\Upsilon(1^3D_1)$ and $\Upsilon(1^3D_3)$ states is dubious on account of lower significance \cite{delAmoSanchez:2010kz}.

Such plentiful experimental achievements in the bottomonium family not only increase our motivation to search for more higher excited bottomonia in future experiments but also provide an excellent opportunity to test various theories and phenomenological models. To better understand the properties of a bottomonium, both the progress of experimental measurements and the calculation of theoretical and phenomenological methods are necessary. The lattice QCD is usually considered to be the most promising solution to the non-perturbative difficulty in low energy regions. In principle, the spectrum of the heavy quarkonium is directly obtained from the lattice QCD, but actually, it is quite troublesome because an additional large heavy quark mass $m_Q$ needs a tremendous computational effort than that of the light quarkonium \cite{Brambilla:2004wf}. In spite of this, the lattice QCD still has an advantage in the calculations of bottomonia \cite{Lewis:2012ir,Dowdall:2013jqa}. Before the lattice method becomes more comprehensive than the past, the mass spectra of hadrons are usually treated by the phenomenological model, namely, potential model, which takes into account the non-perturbative effects of QCD in the interaction potentials and can give satisfactory results consistent with the experiments.

In the past decades, a variety of potential models have been used to study the bottomonium system \cite{Godfrey:2015dia,Segovia:2016xqb,Li:2009nr,Deng:2016ktl,Ebert:2002pp,Ferretti:2013vua,Li:2015zda,  Vijande:2004he,Gupta:1986xt,Richardson:1978bt,Buchmuller:1980su,Martin:1980jx,Radford:2007vd,Motyka:1997di,Gonzalez:2003gx,Ding:1995he,Beyer:1992nd,
 Ding:1993uy,Wei-Zhao:2013sta,Gonzalez:2014nka},
 among which the most well-known model is the Godfrey-Isgur relativistic quark model. This model has been widely used in the study from light mesons to heavy mesons \cite{Godfrey:2015dia,Godfrey:1985xj}. Recently, Godfrey and Moats used the Godfrey-Isgur relativistic quark model for systematically investigating the properties of the $b\bar{b}$ system. This system contains a large number of higher excited states, and they have given the results of the production in $e^+e^-$ and $pp$ collisions so that experimentalists can look for and observe the most promising new bottomonia \cite{Godfrey:2015dia} in future. However, this is similar to the case of other meson families with abundant experimental information. That is, high-lying states are related to the higher mass values predicted by the GI model. In comparison with the measured results of $b\bar{b}$ states, one of the most salient particles of the model is the $\Upsilon(11020)$ or $\Upsilon(6S)$ whose {theoretical mass value is about 100 MeV larger than the experimental data}. In Ref. \cite{Song:2015nia}, the present authors proposed a modified GI model with color screening effects to investigate higher excited charm-strange mesons. Namely, they have taken into account the effect that the confinement potential is softened by the influence of light quarks induced from the vacuum in the long-range region \cite{Born:1989iv}, and have found that the addition of this effect can well describe the properties of higher charm-strange mesons. Hence, the recent study of bottomonia by the GI model \cite{Godfrey:2015dia} motivates us to explore what different conclusions can be drawn after considering screening effects in the GI model. With this motivation, we will carry out a most comprehensive study on the properties of the bottomonium family by the modified GI model with screening effects, primarily focusing on the properties of higher excited states.

 In this work, we calculate the mass spectra and decay behaviors of bottomonia using the modified GI model, including computations of radiative transition, hadronic transition, annihilation process, and OZI-allowed two-body strong decay, where the meson wave functions also come from the modified GI model. This model will be thoroughly introduced in the next section. After the $\chi^2$ fit with abundant experimental information, we can give a fairly good description for the mass spectra of bottomonia, where the masses of the higher excited states have been dramatically improved compared with the estimates by the GI model \cite{Godfrey:2015dia}. At the same time, we also predict the mass values of higher bottomonia, which are valuable for the experiments such as BaBar, Belle, and LHC to search for these missing particles. From the mass spectra of $b\bar{b}$ states, we discuss bottomonia by dividing them into those below and above open-bottom thresholds.

 For the states below the threshold, since strong decay channels are not open, radiative transitions and annihilation decays are usually dominant. Dipion hadronic transitions are, of course,  also major decay modes. Comparing calculated partial widths of these modes with those of the GI model, we find that the results between the two models are almost the same for the low-lying states but are different from them for the higher excited states such as $1F$ or $1G$ states. It is shown that the screening effect has little influence on the low-lying states but it is crucial for the higher states. For the states above the threshold, we calculate the strong decays using the $^3P_0$ model, where we adopt a meson wave function numerically obtained rather than the SHO wave function with a corresponding effective $\beta$ value. Because the states above the threshold have relatively large masses, an influence of the screening effect becomes more obvious for these states and our results that are largely different from the GI model should give a better prediction for the higher excited states.

 In addition, it is emphasized that the discussion and prediction on the higher bottomonia are the main points of our work derived from our calculations. We hope that the present investigation can not only reveal the inherent properties of the observed bottomonia but also provide valuable clues to the experimental search for more missing $b\bar{b}$ states in the future.

 This paper is organized as follows. In Sec. \ref{sec2}, we will give a brief introduction to the modified GI model with the screening effect and compare the results of the GI model and modified GI model. With our mass spectrum, the bottomonium spectroscopy will be analyzed in this section. Combining information of mass spectra and decay behaviors, we will study properties of the states bellow the $B\bar{B}$ threshold and those above the threshold in Secs. \ref{sec4} and \ref{sec5}, respectively, where plentiful predictions will be given. {{In Sec. \ref{sec:ccm}, we compare the results between the modified GI model and a coupled-channel quark model}}. We make a summary in Sec. \ref{sec6}. Finally, all the theoretical tools of various decay processes and physical quantities, like partial decay width, branching ratio, annihilation decay, radiative transition, hadronic transition, and total width, will be presented in Appendix \ref{sec3}.
 A complete list of interaction potentials of the modified GI model will be given in Appendix \ref{appendix:B}.

\section{spectrum}\label{sec2}
\subsection{Modified GI model with a screening effect}
In this work, we use the modified Godfrey-Isgur model with a screening potential \cite{Song:2015nia} to calculate bottomonium mass spectrum and wave function, which will be employed in the calculation of decay processes. The Godfrey-Isgur relativized quark model \cite{Godfrey:1985xj} is one of the most successful model in predicting mass spectra of mesons. Even though the GI model has achieved great success, for the higher orbital and radial excitations, the predicted masses are larger on the whole than the observed masses of newly discovered particles in recent decades. Coupled-channel effects usually play an important role for the higher excitations and a screening effect could partly substitute the coupled-channel effect \cite{Li:2009zu,Li:2009ad}. In 1985, Godfrey and Isgur proposed a relativized quark model motivated by QCD \cite{Godfrey:1985xj}. Compared with other models, relativistic corrections and universal one-gluon-exchange interaction as well as linear confinement potential are the most important features of this model. In the following, before presenting our modified GI model, we will introduce the GI model one by one. The Hamiltonian of a meson system is composed of kinetic energy and interaction between quark and anti-quark, and kinetic energy adopts a relativistic form as
\begin{equation}
H_0=\sqrt{p^2+m_1^2}+\sqrt{p^2+m_2^2},
\end{equation}
where $m_1$ and $m_2$ are constituent quark masses corresponding to quark and anti-quark, respectively. The interaction between quark and anti-quark includes short-range one-gluon-exchange potential $G(r)$ and long-range confinement $S(r)$, spin-orbit interaction, color hyperfine interaction (contact and tensor term). $G(r)$ and $S(r)$ have the following forms as
\begin{align}
G(r)&=-\frac{4\alpha_s(r)}{3r}
\end{align}
and
\begin{align}
S(r)&=br+c,
\end{align}
where $\alpha_s(r)$ is a running coupling constant. Relativistic contributions are divided into two parts. Firstly, the model makes a smearing transformation of $G(r)$ and $S(r)$. To express it shortly, we use a general function symbol $V(r)$ instead of $G(r)$ and $S(r)$. A special operation follows as
\begin{align}
\tilde{V}(r)=\int V(r')\rho(\mathbf{r}-\mathbf{r'})dr'^3
\end{align}
with
\begin{align}
\rho(\mathbf{r-r'})=\frac{\sigma^3}{\pi^{\frac{3}{2}}}e^{-\sigma^2}
(\mathbf{r-r'})^2,
\end{align}
where $\sigma$ is a smearing parameter.
Second, an important reflection of relativistic effects lies in the momentum dependence of interactions between quark and anti-quark. Therefore, a momentum-dependent factor is brought into the interactions. In this situations, a one-gluon-exchange potential $G(r)$ is modified as
\begin{align}
\tilde{G}(r)\rightarrow\left(1+\frac{p^2}{E_1E_2}\right)^{1/2}\tilde{G}(r)\left(1+\frac{p^2}{E_1E_2}\right)^{1/2}.
\end{align}
Tensor, contact, scalar spin-orbit, and vector spin-scalar terms should be changed as
\begin{align}
\frac{\tilde{V}(r)}{m_1m_2}\rightarrow\left(\frac{m_1m_2}{E_1E_2}\right)^{1/2+\varepsilon_i}
\frac{\tilde{V_i}(r)}{m_1m_2}\left(\frac{m_1m_2}{E_1E_2}\right)^{1/2+\varepsilon_i}, \label{eq:momf}
\end{align}
where $E_1$ and $E_2$ are the energies of quark and anti-quark and $\varepsilon_i$ corresponds to $i$-th type of interactions ($\varepsilon_c$, $\varepsilon_t$, $\varepsilon_{sov}$, and $\varepsilon_{sos}$). We readily notice that in the non-relativistic limit the factors become unity, and particular values of $\varepsilon_i$ can be obtained from Table \ref{MGIpara}.

After a brief review of the GI model, we will introduce a modified GI model with a screening potential. Our previous work \cite{Song:2015nia} presented the modified GI model with a color screening effect and revisited the properties of charm-strange mesons, subsequently, charm mesons have been also studied in the framework of the modified GI model \cite{Song:2015fha}, where higher excitations have been greatly improved. A screening effect can be achieved by modifying a confinement potential as
\begin{equation}
br\rightarrow \frac{b\left(1-e^{-\mu r}\right)}{\mu} \label{eq:br},
\end{equation}
where $\mu$ is a parameter which expresses the strength of a screening effect. As described in the preceding paragraphs, the modified confinement potential Eq. (\ref{eq:br}) also needs relativistic corrections. Details of techniques can be found in Ref. \cite{Song:2015nia}, {{ and a complete list of the interaction potentials in the modified GI model are listed in Appendix \ref{appendix:B}}}. It is worth mentioning that the modified GI model also employs simple harmonic oscillator (SHO) wave functions which can be considered as a complete basis set to expand exact meson wave functions. In the momentum space, an SHO wave function has the form
\begin{align}
\Psi_{nLM_L}(\mathbf{p})=R_{nL}(p, \beta)Y_{LM_L}(\Omega_p)
\end{align}
with
\begin{eqnarray}
&R_{nL}(p,\beta)=\frac{(-1)^{n-1}(-i)^L}{\beta^{3/2}}\sqrt{\frac{2(n-1)!}{\Gamma(n+L+1/2)}}\left(\frac{p}{\beta}\right)^L\\
&\times e^{\frac{-p^2}{2\beta^2}}L_{n-1}^{L+1/2}\left(\frac{p^2}{\beta^2}\right),
\nonumber
\end{eqnarray}
where $Y_{LM_L}(\Omega_p)$ is a spherical harmonic function, $L_{n-1}^{L+1/2}\left(\frac{p^2}{\beta^2}\right)$ is an associated Laguerre polynomial, and $\beta$ is a parameter of an oscillator radial wave function. In the next subsection, we calculate bottomonium mass spectra via the modified GI model, which help us well understand the $b\bar{b}$ family and are also used for the subsequent decay processes.

\begin{figure}[tbp]
\begin{center}
\scalebox{0.9}{\includegraphics{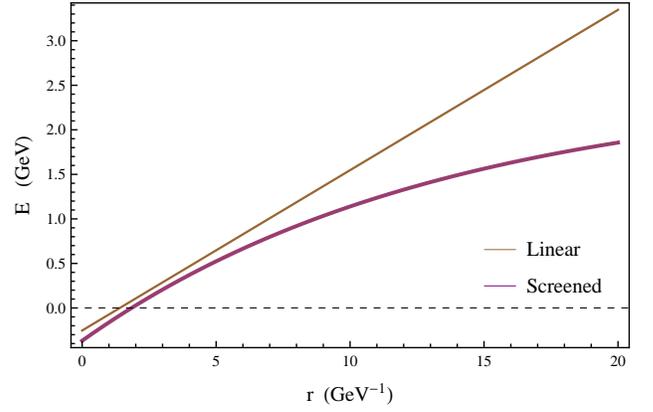}}
\caption{ Comparison of confinement potential curves of the modified GI model with screening effects and the GI model. Our screening effect parameter $\mu$ is 0.07426 GeV and other confinement parameters $b$ and $c$ are found in Table \ref{MGIpara}. \label{Fig2:confine}}
 \end{center}
\end{figure}

\subsection{Mass spectrum}
\label{sub2B}

\begin{table}[htbp]
\renewcommand\arraystretch{1.3}
\caption{Parameters of the modified GI relativistic quark model with screening effects in bottomonium system. \label{MGIpara}}
\begin{center}
{\tabcolsep0.20in
\begin{tabular}{cccc}
\toprule[1pt]\toprule[1pt]
Parameter &  Modified GI model &GI model \cite{Godfrey:1985xj} \\
 \midrule[1pt]
 $\epsilon_{sov}$&-0.04098&-0.035\\
$\epsilon_c$&-0.15384& -0.168\\
$\epsilon_t$&0.02597& 0.025\\
$\epsilon_{sos}$&0.05777&0.055\\
$ m_b$ &5.027 &4.977\\
 $b$&0.21355&0.18\\
 $c$& -0.36804&-0.253\\
$\mu$   & 0.07426&-\\
\bottomrule[1pt]\bottomrule[1pt]
\end{tabular}
}
\end{center}
\end{table}

\begin{table}[htbp]
\renewcommand\arraystretch{1.3}
\caption{
Experimentally observed mass values of bottomonia which are used to calculate the minimu $\chi^2$ and to obtain the model parameters listed in Table \ref{MGIpara}. Experimental values and errors in the fitting are listed in the last two columns, respectively.
The notation ${\rm n}$ in the ${\chi^2}/{\rm n}$ is the number of selected particles here ($=18$) and all mass values are in units of MeV.
\label{fitting}}
\begin{center}
{\tabcolsep0.022in
\begin{tabular}{ccccccc}
\toprule[1pt]\toprule[1pt]
States&$n^{2S+1}L_J$ & This work & GI \cite{Godfrey:1985xj}  & Experiment \cite{Olive:2016xmw} &Error  \\
 \midrule[1pt]
 $\eta_b(1S)$           &$1^1 {S}_0$&9398   &  9394  & ${9399\pm2.3}$ &5.0\\
 $\eta_b(2S)$           &$2^1 {S}_0$&9989   &  9975  & ${9999\pm3.5^{+2.8}_{-1.9}}$&5.0 \\
 $\Upsilon(1S)$         &$1^3 {S}_1$&9463   &  9459  & ${9460.3\pm0.26}$&5.0\\
 $\Upsilon(2S)$         &$2^3 {S}_1$&10017  &  10004 & ${10023.3\pm0.31}$&5.0\\
 $\Upsilon(3S)$         &$3^3 {S}_1$&10356  &  10354 & ${10355.2\pm0.5}$&5.0\\
 $\Upsilon(4S)$         &$4^3 {S}_1$&10612  &  10633 & ${10579.4\pm1.2}$&5.0\\
 $\Upsilon(10860)$      &$5^3 {S}_1$&10822  &  10875 & ${10881.8^{+1.0}_{-1.1}\pm 1.2}$&5.0\\
 $\Upsilon(11020)$      &$6^3 {S}_1$&11001  &  11092 & ${11003.0 \pm1.1^{+0.9}_{-1.0}}$&5.0\\
 $h_b(1P)$              &$1^1 {P}_1$& 9894  &  9881  & ${9899.3\pm0.8}$&5.0\\
 $h_b(2P)$              &$2^1 {P}_1$&10259  &  10250 & ${10259.8\pm0.5\pm1.1}$&5.0\\
 $\chi_{b0}(1P)$        &$1^3 {P}_0$&9858   & 9845   & ${9859.4\pm0.42\pm0.31}$&5.0\\
 $\chi_{b0}(2P)$        &$2^3 {P}_0$&10235  & 10225  & ${10232.5\pm0.4\pm0.5}$&5.0\\
 $\chi_{b1}(1P)$        &$1^3 {P}_1$&9889   & 9875   & ${9892.8\pm0.26\pm0.31}$&5.0\\
 $\chi_{b1}(2P)$        &$2^3 {P}_1$&10255  & 10246  & ${10255.5\pm0.22\pm0.50}$&5.0\\
 $\chi_{b1}(3P)$        &$3^3 {P}_1$&10527  & 10537  & ${10512.1\pm2.1\pm0.9}$&5.0\\
 $\chi_{b2}(1P)$        &$1^3 {P}_2$&9910   & 9896   & ${9912.2\pm0.26\pm0.31}$&5.0\\
 $\chi_{b2}(2P)$        &$2^3 {P}_2$&10269  & 10261  & ${10268.7\pm0.22\pm0.50}$&5.0\\
 $\Upsilon(1D)$         &$1^3 {D}_2$&10162  & 10147  & ${10163.7\pm1.4}$&5.0\\

 \midrule[1pt]
${\chi^2}/{\rm n}$ &    & 11.3&31.4  \\
\bottomrule[1pt]\bottomrule[1pt]
\end{tabular}
}
\end{center}
\end{table}

\begin{table}[htbp]
\renewcommand\arraystretch{1.25}
\caption{ The mass spectrum of the predicted bottomonia. All mass values are in units of MeV.\label{spectrum}}
\begin{center}
{\tabcolsep0.05in
\begin{tabular}{lcc|ccc}
\toprule[1pt]
\toprule[1pt]
    States&     This work & GI \cite{Godfrey:1985xj} &  States &This work  & GI \cite{Godfrey:1985xj} \\
\midrule[1pt]
 $\eta_b(3^1S_0)$& 10336  &10333 & $\Upsilon_1(4^3D_1)$& 10871  &10927   \\
 $\eta_b(4^1S_0)$& 10597  &10616 &$\Upsilon_2(4^3D_2)$& 10876   &10934 \\
 $\eta_b(5^1S_0)$& 10810  &10860 &$\Upsilon_3(4^3D_3)$& 10880   &10939  \\
 $\eta_b(6^1S_0)$& 10991  &11079 &$\eta_{b2}(5^1D_2)$&  11046     &11143   \\
 $\eta_b(7^1S_0)$& 11149  &11281      &$\Upsilon_1(5^3D_1)$& 11041  &11137\\
 $\Upsilon(7^3S_1)$& 11157  &11294     &$\Upsilon_2(5^3D_2)$& 11045   &11143\\
 $\eta_b(8^1S_0)$&  11289  & 11470     &$\Upsilon_3(5^3D_3)$& 11049   &11148\\
 $\Upsilon(8^3S_1)$& 11296  &11481      &$h_{b3}(1^1F_3)$  & 10366        &10354\\
 $h_b(3^1P_1)$    &  10530  &10540      &$\chi_{b2}(1^3F_2)$& 10362       &10350\\
 $\chi_{b0}(3^3P_0)$& 10513  &10521      &$\chi_{b3}(1^3F_3)$& 10366      &10354\\
 $\chi_{b2}(3^3P_2)$& 10539  &10549      &$\chi_{b4}(1^3F_4)$& 10369        &10358\\
 $h_b(4^1P_1)$    &  10751  &10790      &$h_{b3}(2^1F_3)$  &  10609       &10619\\
 $\chi_{b0}(4^3P_0)$& 10736  &10773      &$\chi_{b2}(2^3F_2)$& 10605     &10615\\
 $\chi_{b1}(4^3P_1)$& 10749  &10787      &$\chi_{b3}(2^3F_3)$& 10609     &10619\\
 $\chi_{b2}(4^3P_2)$& 10758  &10797      &$\chi_{b4}(2^3F_4)$& 10612      &10622\\
 $h_b(5^1P_1)$    &  10938  &11013      &$h_{b3}(3^1F_3)$   & 10812        &10853\\
 $\chi_{b0}(5^3P_0)$& 10926  &10998      &$\chi_{b2}(3^3F_2)$& 10809      &10849\\
 $\chi_{b1}(5^3P_1)$& 10936  &11010      &$\chi_{b3}(3^3F_3)$& 10812       &10853\\
 $\chi_{b2}(5^3P_2)$& 10944  &11020      &$\chi_{b4}(3^3F_4)$& 10815        &10856\\
 $h_b(6^1P_1)$    &  11101  & 11218     &$h_{b3}(4^1F_3)$  &   10988        &11066\\
 $\chi_{b0}(6^3P_0)$& 11090  &11204      &$\chi_{b2}(4^3F_2)$& 10985       &11063\\
 $\chi_{b1}(6^3P_1)$& 11099  &11215      &$\chi_{b3}(4^3F_3)$& 10988        &11066\\
 $\chi_{b2}(6^3P_2)$& 11106  &11224      &$\chi_{b4}(4^3F_4)$& 10990        &11069\\
 $\eta_{b2}(1^1D_2)$& 10163  &10148      &$\eta_{b4}(1^1G_4)$& 10534      &10530\\
 $\Upsilon_1(1^3D_1)$& 10153  &10137      &$\Upsilon_3(1^3G_3)$& 10533    &10528\\
 $\Upsilon_3(1^3D_3)$& 10170  &10155      &$\Upsilon_4(1^3G_4)$& 10535    &10530\\
 $\eta_{b2}(2^1D_2)$& 10450  &10450      &$\Upsilon_5(1^3G_5)$& 10536     &10532\\
 $\Upsilon_1(2^3D_1)$& 10442  &10441      &$\eta_{b4}(2^1G_4)$& 10747      &10770\\
 $\Upsilon_2(2^3D_2)$& 10450  &10449      &$\Upsilon_3(2^3G_3)$& 10745     &10769\\
 $\Upsilon_3(2^3D_3)$& 10456  &10455      &$\Upsilon_4(2^3G_4)$& 10747      &10770\\
 $\eta_{b2}(3^1D_2)$&  10681  &10706      &$\Upsilon_5(2^3G_5)$& 10748      &10772\\
 $\Upsilon_1(3^3D_1)$& 10675  &10698      &$\eta_{b4}(3^1G_4)$&  10929      &10988\\
 $\Upsilon_2(3^3D_2)$& 10681  &10705      &$\Upsilon_3(3^3G_3)$& 10928     &10986\\
 $\Upsilon_3(3^3D_3)$& 10686  &10711      &$\Upsilon_4(3^3G_4)$& 10929     &10988\\
 $\eta_{b2}(4^1D_2)$&  10876  &10934      &$\Upsilon_5(3^3G_5)$& 10931     &10989\\

\bottomrule[1pt]
\bottomrule[1pt]
\end{tabular}
}
\end{center}
\end{table}

\begin{figure*}[htbp]
\begin{center}
\scalebox{1.05}{\includegraphics{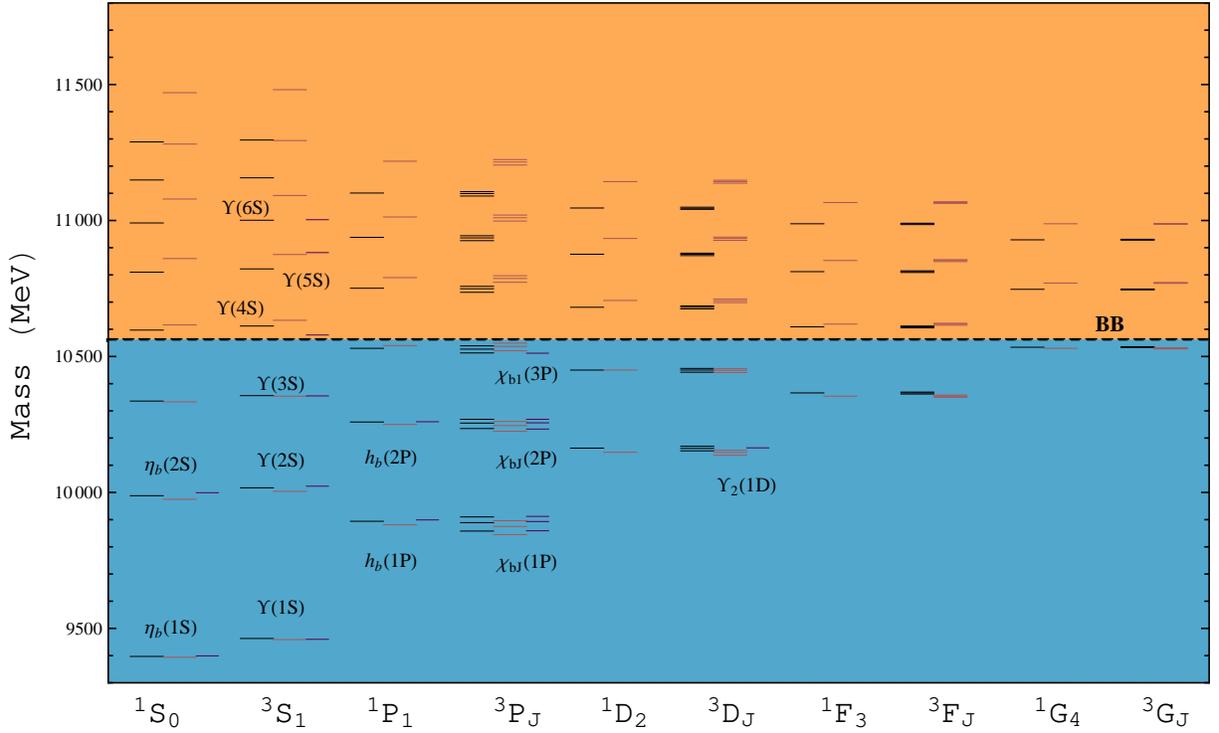}}
\caption{Mass spectrum of bottomonium. Here, bottomonia are classified by the quantum number $^{2S+1}L_J$, from left to right successively in each category , black, pink, and purple lines stand for our results from the modified GI model, the predictions of the GI model \cite{Godfrey:1985xj,Godfrey:2015dia}, and experimental values taken from PDG \cite{Olive:2016xmw}, respectively. The position of the open-bottom threshold is identified by the dashed line. In addition, the notation $J$ means all the total angular momentum numbers of triplet states such as $J=0,1,2$ for the $P$-wave states and $J=1,2,3$ for the $D$-wave states and so on. \label{Fig1:spectrum}  }
 \end{center}
\end{figure*}

\begin{figure*}[htbp]
\begin{center}
\scalebox{1.1}{\includegraphics{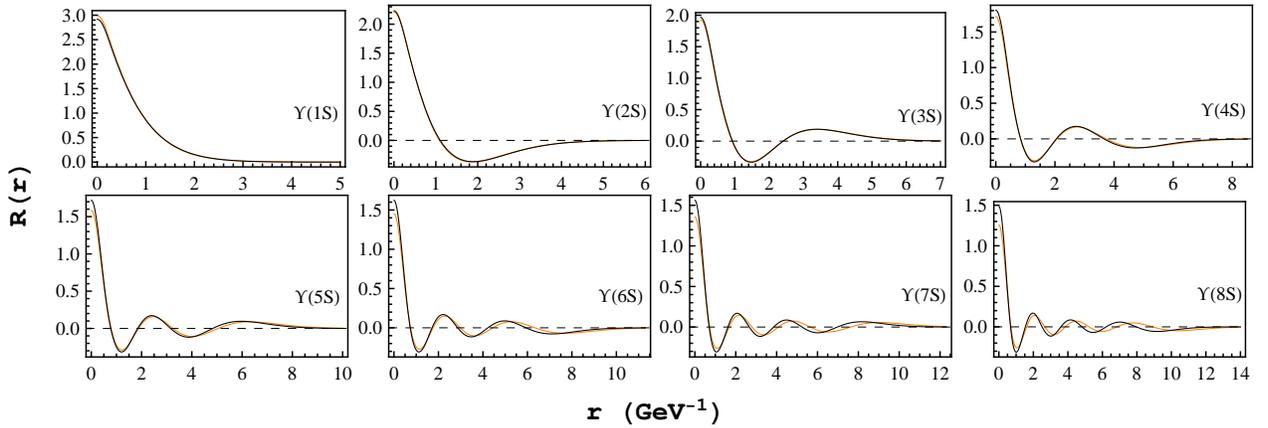}}
\caption{The radial wave functions of the $S$-wave $\Upsilon(nS)$ state in the position space, which includes radial quantum number from one to eight. The orange and black curves represent the results from our modified GI model and GI model, respectively   . \label{Fig:3wavefunction} }
 \end{center}
\end{figure*}

Due to the introduction of a new parameter $\mu$ in our modified GI model, we need to combine experimental information to scale all available model parameters. At the same time, the richness of bottomonia  in the experiment also provides us with an excellent opportunity to determine these model parameters, where the $\chi^2$ fitting method is adopted. The $\chi^2$ fitting method is to find the minimum $\chi^2$ value, thereby, to obtain a set of corresponding fitting parameters in which the theoretical predictions of phenomenological model and experimental results are most consistent on the whole. $\chi^2$ is defined as
 \begin{equation}
 \chi^2=\sum_i \left(\frac{\mathcal{V}_{Exp}(i)-\mathcal{V}_{Th}(i)}{\mathcal{V}_{Er}(i)}\right)^2,
 \end{equation}
 where the $\mathcal{V}_{Exp}(i)$, $\mathcal{V}_{Th}(i)$ and $\mathcal{V}_{Er}(i)$ are experimental value, theoretical value and error of $i$-th data, respectively.
We select eighteen bottomonia to fit our model parameters as shown in Table \ref{fitting} that have been established in the experiments. In this Table, all experimental masses are taken from PDG \cite{Olive:2016xmw} and a uniform value of $\mathcal{V}_{Er}(i)=5.0$ MeV is chosen as a fitting error for all the states, which is larger than their respective experimental uncertainty. The reason is that the experimental errors of these particles are relatively small and unevenly distributed. In other words, if the error of a particle is much smaller than the others, its mass will be too precise so that it is unfavorable to the overall fitting. The final fitting $\chi^2$ value is given as 11.3 in conjunction with a uniform error and all corresponding  fitting parameters of the modified GI model  are listed in Table \ref{MGIpara}, where parameters of the GI model are also given for comparison.

 In Table \ref{fitting}, the theoretical mass values of the  GI model are also listed and the corresponding $\chi^2$ value is calculated to be 31.4. Comparing $\chi^2$ values, one can easily see that the fitted modified GI model well improves the whole description of the bottomonium spectrum. Although the GI model was successful in investigating the bottomonium spectrum, comparing with the observed masses, there are two main shortcomings in the theoretical estimation of the GI model. The first one is that the theoretical mass values of low-lying states are universally 10-20 MeV smaller than the expeimental values. The second is that the theoretical predictions for higher excited states become larger than the experiments. This is because the screening effect is not taken into account, especially for the $\Upsilon(6S)$ state whose mass difference is close to 100 MeV. It is worth mentioning that the experimental mass value of $\Upsilon(10860)$ or $\Upsilon(5S)$ state is overestimated. That is, the recent BaBar experiment tends to give a 30 MeV higher than the initial experimental value \cite{Olive:2016xmw}. This means that the theoretical prediction of the modified GI model for this state will be inevitably small and this can also explain why the fitted $\chi^2$ value can not be much smaller. In fact, other studies of non-relativistic potential models that consider a screening effect \cite{Li:2009nr,Deng:2016ktl,Ding:1993uy} are also not very satisfactory for the description of this state, $\Upsilon(5S)$. Except for $\Upsilon(10860)$, our results for the bottomonium spectrum are pretty good that can be seen from Table \ref{fitting}. Our model not only solves the two above shortcomings of the GI model but also gives a fairly precise theoretical estimate, where the calculated mass values of many particles are almost the same as experimental results within the fitting errors. Based on our excellent description for the bottomonium spectrum with experimental information, we also predict the mass values for higher bottomonia from $S$-wave to $G$-wave that have not yet been observed in Table \ref{spectrum}.

 In order to facilitate readers to understand the bottomonia more intuitively, the mass spectra are also shown in Fig. \ref{Fig1:spectrum}, which could give an overall grasp of the spectroscopy and is convenient for the comparison among the results of two models and experiments. As seen from Fig. \ref{Fig1:spectrum}, after introducing the screening effect, the impact of a screening potential on the ground states and low-lying states is not so obvious. In contrast, the screening effect begins to be important in the region of higher excited states, especially for those with larger radial quantum numbers $n$. The greater the $n$ value the more prominent this effect is because of the more and more obvious mass differences between the GI model and modified GI model. This feature can also be reflected from the wave function of bottomonia and we take the $S$-wave $\Upsilon$ family as an example, whose radial wave functions are shown in Fig. \ref{Fig:3wavefunction}. Here the wave functions of the $\Upsilon(mS)$ with $m=1,\cdots, 4$ are almost undistinguishable but the higher radial excited states appear to be visibly different in the positions of nodes and radial distributions. Of course, to thoroughly understand the impact of the screening effect and the nature of the bottomonium family, we need to analyze their decay behaviors, which is also the main task of the later sections.


\begin{table}[bp]
\renewcommand\arraystretch{1.3}
\caption{The mass values of the bottom and bottom-strange mesons involved in the present calculation. Because of the tiny measured mass difference between $B^{\pm}$ and $B_0$, we use the experimental mass of $B_0$ as an input, i.e., $B(1^1S_0)$ meson. \label{bottommeson} }
\begin{center}
{\tabcolsep0.10in
\begin{tabular}{lcc}
\toprule[1pt]\toprule[1pt]
State & $n^{2S+1}L_J$ & Input mass (MeV) \\
\midrule[1pt]
$B$            & $1^1S_0$      & 5280 \cite{Olive:2016xmw} \\
$B^*$        & $1^3S_1$      & 5325  \cite{Olive:2016xmw}\\
$B(1^3P_0)$  & $1^3P_0$      & 5700   \\
$B(1P_1)$  & $\cos(\theta)|1^1P_1\rangle+\sin(\theta)|1^3P_1\rangle$      & 5717 \\
$B(1P_1^{\prime})$ & $-\sin(\theta)|1^1P_1\rangle+\cos(\theta)|1^3P_1\rangle$      & 5726 \cite{Olive:2016xmw} \\
$B(1^3P_2)$  & $1^3P_2$      & 5740 \cite{Olive:2016xmw}\\
$B_s$  & $1^1S_0$      & 5367 \cite{Olive:2016xmw}\\
$B_s^*$  & $1^3S_1$      & 5415 \cite{Olive:2016xmw}\\

\bottomrule[1pt]\bottomrule[1pt]
\end{tabular}
}
\end{center}
\end{table}

\begin{table}[tp]
\renewcommand\arraystretch{1.4}
\caption{ The calibration of the parameter $\gamma$ in the $^3P_0$ model. $\Gamma_{Exp.}$ and $\Gamma_{Th.}$ denote the experimental and theoretical widths, respectively. $\Gamma_{Error}$ represents fitting error. For $b\bar{b}$ system, we obtain $\gamma=7.09$ with $\chi^2/n=1.2229$ where $n=3$. All width results are in units of MeV. \label{gammavalue}}
\begin{center}
{\tabcolsep0.18in
\begin{tabular}{lccc}
\toprule[1pt]\toprule[1pt]
Bottomonium  & $\Gamma_{Exp.}$ \cite{Olive:2016xmw}  & $\Gamma_{Th.}$  & $\Gamma_{Error}$  \\
\midrule[1pt]
$\Upsilon(4S)$& $20.5\pm2.5$  &24.6747 &2.5\\
$\Upsilon(10860)$& $48.5^{+1.9+2.0}_{-1.8-2.8}$ &45.5839  & 3.33  \\
$\Upsilon(11020)$& $39.3^{+1.7+1.3}_{-1.6-2.4}$ &38.3302  & 2.88 \\
\midrule[1pt]
$\gamma$         & 7.09       \\
\bottomrule[1pt]\bottomrule[1pt]
\end{tabular}
}
\end{center}
\end{table}

\section{analysis of states below open-bottom thresholds}\label{sec4}

\begin{table*}[htbp]
\renewcommand\arraystretch{1.3}
\caption{Partial widths and branching ratios of annihilation decay, radiative transition, and hadronic transition and total widths for the $S$-wave $\eta_b$ below the
open-bottom thresholds. Experimental results are taken from the PDG \cite{Olive:2016xmw}. The theoretical results of the Godfrey-Isgur relativized quark model \cite{Godfrey:2015dia}
 and the nonrelativistic constituent quark model \cite{Segovia:2016xqb} are summarized in the rightmost columns. The width results are in units of keV.\label{decay1etab}}
\begin{center}
{\tabcolsep0.05in
\begin{tabular}{lllrclrclrclrc}
\toprule[1pt]\toprule[1pt]
&  & \multicolumn{2}{c}{This work} & & \multicolumn{2}{c}{Expt. \cite{Olive:2016xmw}} & & \multicolumn{2}{c}{GI \cite{Godfrey:2015dia}}  & & \multicolumn{2}{c}{Ref. \cite{Segovia:2016xqb}} \\
\cmidrule[1pt]{3-5}\cmidrule[1pt]{6-7}\cmidrule[1pt]{9-11}\cmidrule[1pt]{12-13}
State   & Channels & Width & $\mathcal{B}$(\%)   & &  Width & $\mathcal{B}$(\%)  & &   Width & $\mathcal{B}$(\%) & &   Width & $\mathcal{B}$(\%) \\
\midrule[1pt]
$\eta_b(1S)$   &  $gg$    & 17.9 MeV & $\sim$100   & & $\cdots$  & $\cdots$   & & 16.6 MeV  & $\sim$100    & & 20.18 MeV & $\sim$100 &\\
        & $\gamma\gamma$  &  1.05    & $5.87\times10^{-3}$  & & $\cdots$  & $\cdots$  & & 0.94      & $5.7\times10^{-3}$ & & 0.69 & $3.42\times10^{-3}$ \\
               &Total     & 17.9 MeV & 100 & & $10^{+5}_{-4}$ MeV  &$\cdots$    & & 16.6 MeV & 100    & & 20.18 MeV & 100 \\

$\eta_b(2S)$  &  $gg$    & 8.33 MeV & $\sim$100   & & $\cdots$  & $\cdots$   & & 7.2 MeV  & $\sim$100    & & 10.64 MeV &99.86 \\
        & $\gamma\gamma$  & 0.489    & $5.86\times10^{-3}$  & & $\cdots$  & $\cdots$  & & 0.41      & $5.7\times10^{-3}$ & & 0.36 & $3.38\times10^{-3}$ \\
              & $h_b(1^1P_1)\gamma$ & 2.467 & $2.96\times10^{-2}$ & & $\cdots$    & $\cdots$    & & 2.48   & $3.4\times10^{-2}$  & & 2.85 & $2.68\times10^{-2}$ \\
              & $\Upsilon(1^3S_1)\gamma$ & 0.0706 & $8.47\times10^{-4}$  & & $\cdots$   &  $\cdots$   & & 0.068   & $9.4\times10^{-4}$  & & 0.045 & $4.22\times10^{-4}$ \\
               & $\eta_b(1^1S_0)\pi \pi$ &10.3   & 0.124  & &  $\cdots$  &  $\cdots$   & & 12.4   & 0.17  & & 11.27 &0.1058 \\
               &Total      & 8.34 MeV & 100     & &  $<$24 MeV    &  $\cdots$   & & 7.2 MeV   & 100  & & 10.66 MeV  & 100 \\

$\eta_b(3S)$  &  $gg$    & 5.51 MeV & $\sim$100   & & $\cdots$  & $\cdots$   & & 4.9 MeV  & $\sim$100    & & 7.94 MeV &99.93 \\
        & $\gamma\gamma$  &  0.323    & $5.85\times10^{-3}$  & & $\cdots$  & $\cdots$  & & 0.29      & $5.9\times10^{-3}$ & & 0.27 & $3.4\times10^{-3}$ \\
              & $h_b(1^1P_1)\gamma$ &  1.12 & $2.03\times10^{-2}$  & & $\cdots$   &  $\cdots$   & & 1.3   & $2.6\times10^{-2}$  & &0.0084 &$1.06\times10^{-4}$ \\
              & $h_b(2^1P_1)\gamma$ & 2.88 & $5.22\times10^{-2}$  & & $\cdots$   &  $\cdots$   & & 2.96   & $6\times10^{-2}$  & & 2.60  & $3.27\times10^{-3}$ \\
              & $\Upsilon(1^3S_1)\gamma$ & 0.0732 & $1.33\times10^{-3}$ & & $\cdots$   &  $\cdots$   & & 0.074   & $1.5\times10^{-3}$  & & 0.0051 & $ 6.42\times10^{-4}$ \\
              & $\Upsilon(2^3S_1)\gamma$ & 0.0111 & $2.01\times10^{-4}$  & & $\cdots$   & $\cdots$    & & 0.0091   & $1.8\times10^{-4}$   & & 0.0092  & $1.16\times10^{-4}$ \\
              & $\eta_b(1^1S_0)\pi \pi$ &3.18  & $5.76\times10^{-2}$  & & $\cdots$   & $\cdots$    & & $1.70\pm0.22$   & $3.5\times10^{-2}$  & & 1.95  &$2.45\times10^{-3}$\\
              & $\eta_b(2^1S_0)\pi \pi$ &0.501  & $9.08\times10^{-3}$ & &  $\cdots$  &  $\cdots$   & & $1.16\pm0.10$   & $2.4\times10^{-2}$  & &0.34  &$4.28\times10^{-3}$ \\
               &Total      & 5.52 MeV & 100   & &  $\cdots$  & $\cdots$    & & 4.9 MeV   & 100  & & 7.95 MeV  & 100 \\

$\eta_b(4S)$  &  $gg$    & 4.03 MeV & 99   & & $\cdots$  & $\cdots$   & & 3.4 MeV  & $\sim$100    & & $\cdots$ &$\cdots$ \\
        & $\gamma\gamma$  &  0.237    & $5.82\times10^{-3}$  & & $\cdots$  & $\cdots$  & & 0.20   & $5.9\times10^{-3}$ & &$\cdots$ & $\cdots$ \\
              & $h_b(1^1P_1)\gamma$ & 0.688 & $1.69\times10^{-2}$  & & $\cdots$   & $\cdots$    & & $\cdots$ & $\cdots$  & & $\cdots$  &$\cdots$ \\
              & $h_b(2^1P_1)\gamma$ & 0.732 & $1.8\times10^{-2}$  & & $\cdots$   & $\cdots$    & & $\cdots$& $\cdots$ & &$\cdots$ &$\cdots$ \\
              & $h_b(3^1P_1)\gamma$ & 1.50 & $3.69\times10^{-2}$  & & $\cdots$   & $\cdots$    & & 1.24   & $3.6\times10^{-2}$  & &$\cdots$  & $\cdots$\\
              & $\Upsilon(1^3S_1)\gamma$ & 0.0594 & $1.46\times10^{-3}$  & &  $\cdots$  &  $\cdots$   & & $\cdots$  & $\cdots$  & &$\cdots$  &$\cdots$ \\
              & $\Upsilon(2^3S_1)\gamma$ & 0.0141 & $3.46\times10^{-4}$  & & $\cdots$   &  $\cdots$   & & $\cdots$ & $\cdots$  & &$\cdots$  &$\cdots$ \\
              & $\Upsilon(3^3S_1)\gamma$ & 0.00308 & $7.57\times10^{-5}$  & & $\cdots$   & $\cdots$    & & $\cdots$  & $\cdots$  & &$\cdots$  &$\cdots$ \\
              & $\eta_b(1^1S_0)\pi^+ \pi^-$ &24.6  & 0.604  & &  $\cdots$  &  $\cdots$   & & $2.03\pm0.29$  & $6\times10^{-2}$  & &$\cdots$  &$\cdots$ \\
              & $\eta_b(2^1S_0)\pi^+ \pi^-$ &0.183  & $4.5\times10^{-3}$   & & $\cdots$   & $\cdots$    & & $1.9\pm0.36$   & $5.6\times10^{-2}$  & &$\cdots$  &$\cdots$ \\
               &Total      & 4.07  MeV & 100    & &  $\cdots$  &  $\cdots$   & & 3.4 MeV   & 100  & &$\cdots$  &$\cdots$ \\

\bottomrule[1pt]\bottomrule[1pt]
\end{tabular}
}
\end{center}
\end{table*}

\begin{table*}[htbp]
\renewcommand\arraystretch{1.3}
\caption{ Partial widths and branching ratios of annihilation decay, radiative transition, and hadronic transition and total widths for the $S$-wave $\Upsilon$ below the open-bottom
thresholds. Experimental results are taken from the PDG \cite{Olive:2016xmw}. The theoretical results of the GI model \cite{Godfrey:2015dia}
 and the nonrelativistic constituent quark model \cite{Segovia:2016xqb} are summarized in the rightmost columns. The width results are in units of keV.\label{decay1upsilon}}
\begin{center}
{\tabcolsep0.04in
\begin{tabular}{llcrccccllcllc}
\toprule[1pt]\toprule[1pt]
&  & \multicolumn{2}{c}{This work} & & \multicolumn{2}{c}{Expt. \cite{Olive:2016xmw}} & & \multicolumn{2}{c}{GI \cite{Godfrey:2015dia}}  & & \multicolumn{2}{c}{Ref. \cite{Segovia:2016xqb}} \\
\cmidrule[1pt]{3-5}\cmidrule[1pt]{6-7}\cmidrule[1pt]{9-11}\cmidrule[1pt]{12-13}
State   & Channels & Width & $\mathcal{B}$(\%)   & &  Width & $\mathcal{B}$(\%)  & &   Width & $\mathcal{B}$(\%) & &   Width & $\mathcal{B}$(\%) \\
\midrule[1pt]
$\Upsilon(1S)$ & $\ell^+\ell^-$ &  1.65 &  2.89   & & $1.340\pm0.018$  & $2.38\pm0.11$    & & 1.44 & 2.71    & & 0.71  &1.31 & \\
               &  $ggg$          & 50.8 &  89   & & $44.13\pm1.09 \footnotemark[2]$  &  $81.7\pm0.7$   & & 47.6 & 89.6    & & 41.63 &77.06 \\
               & $\gamma gg$     & 1.32 &  2.31   & & $1.19\pm0.33 \footnotemark[2]$   &  $2.20\pm0.60$   & & 1.2 & 2.3    & & 0.79 & 1.46\\
         & $\gamma\gamma\gamma$ &  $1.94\times10^{-5}$ & $3.4\times10^{-5}$   & & $\cdots$   & $\cdots$    & & $1.7\times10^{-5}$ & $3.2\times10^{-5}$    & & $3.44\times10^{-6}$ & $6.37\times10^{-6}$ \\
        & $\eta_b(1^1S_0)\gamma$ & 0.00952 & $1.67\times10^{-2}$  & &$\cdots$   & $\cdots$    & & 0.010 & $1.9\times10^{-2}$    & & 0.00934   &$1.73\times10^{-2}$ \\
                    &Total      &  57.1  & 100   & & $54.02\pm1.25$  & $\cdots$    & & 53.1 & 100    & & 44.6 $\footnotemark[1]$ &$\cdots$  \\

$\Upsilon(2S)$ & $\ell^+\ell^-$ &  0.821 & 1.84    & & $0.612\pm0.011$  & $1.91\pm0.16$    & & 0.73 & 1.8    & & 0.37  &1.16  \\
               &  $ggg$          & 28.4 &  63.8   & & $18.80\pm1.59 \footnotemark[2]$  &  $58.8\pm1.2$   & & 26.3 & 65.4    & & 24.25 &75.83 \\
               & $\gamma gg$     & 0.739 &  1.66   & & $0.60\pm0.10 \footnotemark[2]$  & $1.87\pm0.28$    & & 0.68 & 1.7    & & 0.46 & 1.44\\
         & $\gamma\gamma\gamma$ &  $1.09\times10^{-5}$ & $2.45\times10^{-5}$   & &  $\cdots$  &  $\cdots$   & & $9.8\times10^{-6}$ & $2.4\times10^{-5}$    & & $2.00\times10^{-6}$ & $6.25\times10^{-6}$ \\
               & $\chi_{b0}(1^3P_0)\gamma$ & 0.907 & 2.04 && $1.22\pm0.16 \footnotemark[2]$  & $3.8\pm0.4$ && 0.91  & 2.3  &&1.09   &3.41       \\
               & $\chi_{b1}(1^3P_1)\gamma$ & 1.60 & 3.60 && $2.21\pm0.22 \footnotemark[2]$ & $6.9\pm0.4$  && 1.63  & 4.05  && 1.84  &5.75       \\
               & $\chi_{b2}(1^3P_2)\gamma$ & 1.86 & 4.18 && $2.29\pm0.22 \footnotemark[2]$ & $7.15\pm0.35$  && 1.88  & 4.67  && 2.08  &6.50       \\
               & $\eta_b(1^1S_0)\gamma$ & 0.0688 & 0.155 && $0.012\pm0.005 \footnotemark[2]$ & $(3.9\pm1.5)\times10^{-2}$ &&  0.081 &0.20   && 0.0565  & 0.18      \\
               & $\eta_b(2^1S_0)\gamma$ & $5.82\times10^{-4}$ & $1.31\times10^{-3}$  && $\cdots$ &$\cdots$  && $5.9\times10^{-4}$  & $1.5\times10^{-3}$  && $5.80\times10^{-4}$  & $1.81\times10^{-3}$      \\
               & $\Upsilon(1^3S_1)\pi \pi$ & 8.46 & 19 && $8.46\pm0.71 \footnotemark[2]$ & $26.45\pm0.48$ && 8.46   &21.0   && 8.57  &  26.80     \\
                    &Total      & 44.5  & 100   && $31.98\pm2.63$  & $\cdots$ && 40.2  & 100  && 39.5$\footnotemark[1]$  &  $\cdots$     \\

$\Upsilon(3S)$ & $\ell^+\ell^-$ &  0.569 & 1.59    & & $0.443\pm0.008$  & $2.18\pm0.21$    & & 0.53 & 1.8    & & 0.27  &1.33  \\
               &  $ggg$          & 21.0 &  58.7   & & $7.25\pm0.85 \footnotemark[2]$  & $35.7\pm2.6$    & & 19.8 & 67.9    & & 18.76 &92.32 \\
               & $\gamma gg$     & 0.547 &  1.53   & & $0.20\pm0.04 \footnotemark[2]$   & $0.97\pm0.18$    & & 0.52 & 1.8    & & 0.36 & 1.77\\
         & $\gamma\gamma\gamma$ &  $8.04\times10^{-6}$ & $2.25\times10^{-5}$   & &  $\cdots$   &   $\cdots$   & & $7.6\times10^{-6}$ & $2.6\times10^{-5}$    & & $1.55\times10^{-6}$ & $7.63\times10^{-6}$ \\
               & $\chi_{b0}(1^3P_0)\gamma$ & 0.0099 & $2.77\times10^{-2}$ && $0.055\pm0.010 \footnotemark[2]$  & $0.27\pm0.04$  && 0.01  & 0.03  && 0.15  &  0.74     \\
               & $\chi_{b1}(1^3P_1)\gamma$ &  0.0363 & 0.101 && $0.018\pm0.010 \footnotemark[2]$  & $(9\pm5)\times10^{-2}$ && 0.05  & 0.2  && 0.16  & 0.79      \\
               & $\chi_{b2}(1^3P_2)\gamma$ & 0.359 & 1.0 && $0.20\pm0.03 \footnotemark[2]$  & $0.99\pm0.13$ && 0.45  & 1.5  && 0.0827  &  0.41     \\
               & $\chi_{b0}(2^3P_0)\gamma$ & 1.06 & 2.96 && $1.20\pm0.16 \footnotemark[2]$  & $5.9\pm0.6$ && 1.03  & 3.54  && 1.21  & 5.96      \\
               & $\chi_{b1}(2^3P_1)\gamma$ & 1.96 & 5.47 && $2.56\pm0.34 \footnotemark[2]$  & $12.6\pm1.2$ && 1.91  & 6.56  &&  2.13 & 10.48      \\
               & $\chi_{b2}(2^3P_2)\gamma$ & 2.37 & 6.62 && $2.66\pm0.41 \footnotemark[2]$  & $13.1\pm1.6$ && 2.30  & 7.90  && 2.56  & 12.60      \\
               & $\eta_b(1^1S_0)\gamma$ & 0.0604 & 0.169 && $0.010\pm0.002 \footnotemark[2]$ & $(5.1\pm0.7)\times10^{-2}$ && 0.060  & 0.20  && 0.057  &  0.28     \\
               & $\eta_b(2^1S_0)\gamma$ & 0.0118 & $3.3\times10^{-2}$ && $<0.014 \footnotemark[2] $  & $<$0.062 (90\%C.L) && 0.19   &0.65   && 0.011  &$5.41\times10^{-2}$       \\
               & $\eta_b(3^1S_0)\gamma$ & $3.37\times10^{-4}$ & $9.4\times10^{-4}$  &&  $\cdots$ &  $\cdots$ && $2.5\times10^{-4}$  & $8.6\times10^{-4}$  && $6.58\times10^{-4}$    &  $3.24\times10^{-3}$      \\
               & $\Upsilon(1^3S_1)\pi \pi$ &6.17  & 17.2  && $1.34\pm0.13 \footnotemark[2]$ & $6.57\pm0.15$ &&1.34   & 4.60  && 1.77  & 8.71      \\
               & $\Upsilon(2^3S_1)\pi \pi$ &0.479  & 1.34 && $0.95\pm0.10 \footnotemark[2]$  & $4.67\pm0.23$  && 0.95  & 3.3  && 0.42  &  2.07     \\
               &Total      & 35.8  & 100  && $20.32\pm1.85$ & $\cdots$  && 29.1  & 100  && 28.5 $\footnotemark[1]$  &    $\cdots$   \\

\bottomrule[1pt]\bottomrule[1pt]
\end{tabular}
}
\end{center}
\footnotetext[1]{From the summation of partial widths of Ref. \cite{Segovia:2016xqb}.}.
\footnotetext[2]{From the calculation of combining experimental total widths and branching ratios of the PDG \cite{Olive:2016xmw}.}
\end{table*}

\begin{table*}[htbp]
\renewcommand\arraystretch{1.35}
\caption{ Partial widths and branching ratios of annihilation decay and radiative transition and total widths for the $1P$ bottomonium states.
 Experimental results are taken from the PDG \cite{Olive:2016xmw}. The theoretical results of the GI model \cite{Godfrey:2015dia}
 and the nonrelativistic constituent quark model \cite{Segovia:2016xqb} are summarized in the rightmost columns. The width results are in units of keV.\label{decay11P}}
\begin{center}
{\tabcolsep0.04in
\begin{tabular}{llcrcllcllcllc}
\toprule[1pt]\toprule[1pt]
&  & \multicolumn{2}{c}{This work} & & \multicolumn{2}{c}{Expt. \cite{Olive:2016xmw}} & & \multicolumn{2}{c}{GI \cite{Godfrey:2015dia}}  & & \multicolumn{2}{c}{Ref. \cite{Segovia:2016xqb}} \\
\cmidrule[1pt]{3-5}\cmidrule[1pt]{6-7}\cmidrule[1pt]{9-11}\cmidrule[1pt]{12-13}
State   & Channels & Width & $\mathcal{B}$(\%)   & &  Width & $\mathcal{B}$(\%)  & &   Width & $\mathcal{B}$(\%) & &   Width & $\mathcal{B}$(\%) \\
\midrule[1pt]
$h_b(1^1P_1)$ &  $ggg$          & 44.7 &  56.5   & &  $\cdots$  & $\cdots$    & & 37 & 51    & & 35.26 &44.68 &\\
              & $\eta_b(1^1S_0)\gamma$ & 34.4 & 43.5  && $\cdots$   &  $52^{+6}_{-5}$   && 35.7      & 49     &&  43.66  & 55.32     \\
              & $\chi_{b0}(1^3P_0)\gamma$ & $9.01\times10^{-4}$ & $1.14\times10^{-3}$  && $\cdots$    & $\cdots$     && $8.9\times10^{-4}$       & $1.2\times10^{-5}$     && $8.61\times10^{-4}$   & $1.09\times10^{-3}$     \\
              & $\chi_{b1}(1^3P_1)\gamma$ & $9.23\times10^{-6}$ &  $1.17\times10^{-5}$ && $\cdots$    &  $\cdots$    &&  $1.0\times10^{-5}$     & $1.4\times10^{-5}$     && $1.15\times10^{-5}$   & $1.46\times10^{-5}$     \\
               &Total      &  79.1 & 100    &&  $\cdots$   &  $\cdots$    && 73      &   100   && 78.92   &  100   \\

$\chi_{b0}(1^3P_0)$&  $\gamma \gamma$          & 0.199 &  $5.87\times10^{-3}$   & & $\cdots$   & $\cdots$     & & 0.15 & $5.8\times10^{-3}$    & & 0.12 &$5.91\times10^{-3}$ \\
                  &  $gg$          & 3.37 MeV &  99.4   & & $\cdots$   & $\cdots$     & & 2.6 MeV & $\sim$100   & & 2.00 MeV &98.61 \\
                   & $\Upsilon(1^3S_1)\gamma$ & 22.8 & 0.673  && $\cdots$    &  $1.76\pm0.35$   &&  23.8     &  0.92    && 28.07   &1.38      \\
               &Total      & 3.39 MeV & 100     && $\cdots$    &  $\cdots$    &&  2.6 MeV     &  100    && 2.03 MeV   & 100    \\

$\chi_{b1}(1^3P_1)$ &  $q\bar{q}g$          & 81.7 & 74.3   & & $\cdots$   & $\cdots$     & & 67 & 70    & & 71.53 &66.73 \\
                   & $\Upsilon(1^3S_1)\gamma$ & 28.3 & 25.7  &&  $\cdots$   & $33.9\pm2.2$    &&  29.5     &   31   && 35.66   & 33.27     \\
               &Total      &  110 & 100    && $\cdots$    & $\cdots$     &&  96     & 100     && 107.19    & 100    \\

$\chi_{b2}(1^3P_2)$ &  $\gamma \gamma$          & 0.0106 &  $5.41\times10^{-3}$    & &$\cdots$    & $\cdots$     & & $9.3\times10^{-3}$ & $5.2\times10^{-3}$    & & $3.08\times10^{-3}$ &$2.51\times10^{-3}$ \\
                  &  $gg$          & 165 & 84.2   & & $\cdots$   & $\cdots$     & & 147 & 81.7    & & 83.69 &68.13 \\
                    & $\Upsilon(1^3S_1)\gamma$ & 31.4 & 16.0  && $\cdots$    & $19.1\pm1.2$    &&  32.8     & 18.2     &&39.15    &31.87      \\
                    & $h_{b}(1^1P_1)\gamma$ & $9.37\times10^{-5}$ & $4.78\times10^{-5}$   && $\cdots$    & $\cdots$     && $9.6\times10^{-5}$      &  $5.3\times10^{-5}$    &&  $8.88\times10^{-5}$  & $7.23\times10^{-5}$     \\
              &Total      &  196 & 100   && $\cdots$    &  $\cdots$    &&  180     &  100    && 122.84   &  100    \\
\bottomrule[1pt]\bottomrule[1pt]
\end{tabular}
}
\end{center}
\end{table*}

\begin{table*}[htbp]
\renewcommand\arraystretch{1.25}
\caption{ Partial widths and branching ratios of annihilation decay, radiative transition, and hadronic transition and total widths for the $2P$ bottomonium states.
Experimental results are taken from the PDG \cite{Olive:2016xmw}. The theoretical results of the GI model \cite{Godfrey:2015dia}
 and the nonrelativistic constituent quark model \cite{Segovia:2016xqb} are summarized in the rightmost columns. The width results are in units of keV.\label{decay12P}}
\begin{center}
{\tabcolsep0.045in
\begin{tabular}{llcrcllcllcllc}
\toprule[1pt]\toprule[1pt]
&  & \multicolumn{2}{c}{This work} & & \multicolumn{2}{c}{Expt. \cite{Olive:2016xmw}} & & \multicolumn{2}{c}{GI \cite{Godfrey:2015dia}}  & & \multicolumn{2}{c}{Ref. \cite{Segovia:2016xqb}} \\
\cmidrule[1pt]{3-5}\cmidrule[1pt]{6-7}\cmidrule[1pt]{9-11}\cmidrule[1pt]{12-13}
State   & Channels & Width & $\mathcal{B}$(\%)   & &  Width & $\mathcal{B}$(\%)  & &   Width & $\mathcal{B}$(\%) & &   Width & $\mathcal{B}$(\%) \\
\midrule[1pt]
$h_b(2^1P_1)$ &  $ggg$          & 64.6 &  69.6   & &  $\cdots$  & $\cdots$    & & 54 & 64    & & 52.70 &57.82 &\\
              & $\eta_b(1^1S_0)\gamma$ & 10.8 & 11.6   && $\cdots$   &  $22\pm5$  && 13.0     & 15   &&  14.90  &  16.35    \\
              & $\eta_b(2^1S_0)\gamma$ & 15.0 & 16.2   &&  $\cdots$  & $48\pm13$   &&  14.1    & 17   &&  17.60  &  19.31    \\
              & $\eta_{b2}(1^1D_2)\gamma$ & 1.81 & 1.95  &&  $\cdots$  & $\cdots$   && 1.7     & 2.0   && 5.36   &  5.88    \\
              & $\chi_{b0}(1^3P_0)\gamma$ & 0.00556 & $5.99\times10^{-3}$  && $\cdots$   & $\cdots$   && $3.2\times10^{-4}$     & $3.8\times10^{-4}$   &&  0.0364  &  $3.99\times10^{-2}$    \\
              & $\chi_{b1}(1^3P_1)\gamma$ & 0.00130 & $1.40\times10^{-3}$  && $\cdots$   & $\cdots$   && 0.0011     & $1.3\times10^{-3}$   && 0.00128   &  $1.41\times10^{-3}$    \\
\end{tabular}
}
\end{center}
\end{table*}

\begin{table}[htbp]
\renewcommand\arraystretch{1.4}
\caption{ Partial widths and branching ratios of annihilation decay, radiative transition, and hadronic transition and total widths for $3^1P_1$ and $3^3P_0$ bottomonium states.
 The width results are in units of keV.\label{decay13Pa}}
\begin{center}
{\tabcolsep0.1in
\begin{tabular}{cllrc}
\toprule[1pt]\toprule[1pt]
State         &  Channels & Width & $\mathcal{B}$(\%)  \\
\midrule[1pt]
$h_b(3^1P_1)$ & $ggg$                  & 71.1 &  74.6    \\
              & $\eta_b(1^1S_0)\gamma$ & 3.91 &  4.10    \\
              & $\eta_b(2^1S_0)\gamma$ & 4.60 & 4.83     \\
              & $\eta_b(3^1S_0)\gamma$ & 9.94 & 10.4     \\
              & $\eta_{b2}(1^1D_2)\gamma$ & 0.0585 & $6.14\times10^{-2}$     \\
               & $\eta_{b2}(2^1D_2)\gamma$ & 1.44 &  1.51    \\
              & $\chi_{b0}(1^3P_0)\gamma$ & 0.00510 &  $5.35\times10^{-3}$    \\
\end{tabular}
}
\end{center}
\end{table}

\begin{table}[htbp]
\renewcommand\arraystretch{1.4}
\caption{ Partial widths and branching ratios of annihilation decay, radiative transition, and hadronic transition and total widths for $3^3P_1$ and $3^3P_2$ bottomonium states.
 The width results are in units of keV.\label{decay13Pb}}
\begin{center}
{\tabcolsep0.1in
\begin{tabular}{cllrc}
\toprule[1pt]\toprule[1pt]
State         &  Channels & Width & $\mathcal{B}$(\%)  \\
\midrule[1pt]
$\chi_{b1}(3^3P_1)$& $q\bar{q} g$ & 126 &   86.9   \\
                   & $\Upsilon(1^3S_1)\gamma$ & 1.62 &  1.12    \\
                   & $\Upsilon(2^3S_1)\gamma$ & 2.49 &  1.72   \\
                    & $\Upsilon(3^3S_1)\gamma$ & 8.36 &  5.77    \\
                   & $\Upsilon_1(1^3D_1)\gamma$ & 0.00418 & $2.88\times10^{-3}$  \\
                   & $\Upsilon_2(1^3D_2)\gamma$ & 0.0615 &  $4.24\times10^{-2}$    \\
                   & $\Upsilon_1(2^3D_1)\gamma$ & 0.425 &   0.293   \\
\end{tabular}
}
\end{center}
\end{table}

\begin{table*}[htbp]
\renewcommand\arraystretch{1.3}
\caption{ Partial widths and branching ratios of annihilation decay, and radiative transition and total widths for the $1D$ bottomonium states.
 Experimental results are taken from the PDG \cite{Olive:2016xmw}. The theoretical results of the GI model \cite{Godfrey:2015dia}
 and the nonrelativistic constituent quark model \cite{Segovia:2016xqb} are summarized in the rightmost columns. The width results are in units of keV.\label{decay11D}}
\begin{center}
{\tabcolsep0.035in
\begin{tabular}{llcrcllcllcllc}
\toprule[1pt]\toprule[1pt]
&  & \multicolumn{2}{c}{This work} & & \multicolumn{2}{c}{Expt. \cite{Olive:2016xmw}} & & \multicolumn{2}{c}{GI \cite{Godfrey:2015dia}}  & & \multicolumn{2}{c}{Ref. \cite{Segovia:2016xqb}} \\
\cmidrule[1pt]{3-5}\cmidrule[1pt]{6-7}\cmidrule[1pt]{9-11}\cmidrule[1pt]{12-13}
State   & Channels & Width & $\mathcal{B}$(\%)   & &  Width & $\mathcal{B}$(\%)  & &   Width & $\mathcal{B}$(\%) & &   Width & $\mathcal{B}$(\%) \\
\midrule[1pt]
$\eta_{b2}(1^1D_2)$ &  $gg$          & 0.657 &  2.6   & & $\cdots$   & $\cdots$     &&  1.8 &  6.6  &&  0.37 &2.07 \\
                    & $h_{b}(1^1P_1)\gamma$ & 24.3 & 96  &&  $\cdots$    &  $\cdots$      && 24.9     & 91.5       && 17.23     & 96.58    \\
                    & $\Upsilon_1(1^3D_1)\gamma$ & $2.56\times10^{-5}$ & $1.01\times10^{-4}$    &&  $\cdots$    &  $\cdots$      && $\cdots$     & $\cdots$       && $\cdots$     & $\cdots$    \\
                     & $\Upsilon_2(1^3D_2)\gamma$ & $4.28\times10^{-8}$ & $1.69\times10^{-7}$  && $\cdots$     &  $\cdots$      && $\cdots$     &  $\cdots$      &&  $\cdots$    & $\cdots$    \\
                    & $\eta_b(1^1S_0)\pi^+ \pi^-$ &0.241  & 0.953  && $\cdots$     &        && 0.35     &  1.3      &&  0.16    &  0.9   \\
               &Total      & 25.3 & 100 &&  $\cdots$    &  $\cdots$      && 27.2     & 100       &&  17.84    & 100    \\
\end{tabular}
}
\end{center}
\end{table*}

\begin{table}[htbp]
\renewcommand\arraystretch{1.4}
\caption{ Partial widths and branching ratios of annihilation decay, radiative transition, and hadronic transition and total widths for $2^1D_2$ and $2^3D_1$ bottomonium states.
 The width results are in units of keV.\label{decay12Da}}
\begin{center}
{\tabcolsep0.1in
\begin{tabular}{cllrc}
\toprule[1pt]\toprule[1pt]
State         &  Channels & Width & $\mathcal{B}$(\%)  \\
\midrule[1pt]
$\eta_{b2}(2^1D_2)$ &  $gg$          & 1.22 & 5.33       \\
                    & $h_{b}(1^1P_1)\gamma$ & 3.36 & 14.7     \\
                    & $h_{b}(2^1P_1)\gamma$ & 16.8 &  73.4    \\ & $h_{b3}(1^1F_3)\gamma$ & 1.35 &  5.90    \\
                    & $\Upsilon_1(1^3D_1)\gamma$ & 0.00102 &  $4.46\times10^{-3}$    \\
                     & $\Upsilon_2(1^3D_2)\gamma$ & $2.57\times10^{-4}$ & $1.12\times10^{-3}$     \\
                      & $\Upsilon_3(1^3D_3)\gamma$ & $4.69\times10^{-5}$ & $2.05\times10^{-4}$     \\
\end{tabular}
}
\end{center}
\end{table}

\begin{table}[htbp]
\renewcommand\arraystretch{1.4}
\caption{ Partial widths and branching ratios of annihilation decay, radiative transition, and hadronic transition and total widths for $2^3D_2$ and $2^3D_3$ bottomonium states.
 The width results are in units of keV.\label{decay12Db}}
\begin{center}
{\tabcolsep0.1in
\begin{tabular}{cllrc}
\toprule[1pt]\toprule[1pt]
State         &  Channels & Width & $\mathcal{B}$(\%)  \\
\midrule[1pt]
$\Upsilon_2(2^3D_2)$ &  $ggg$          & 1.65 &  7.05      \\
                     & $\chi_{b1}(1^3P_1)\gamma$ & 2.81 &  12.0   \\
                     & $\chi_{b2}(1^3P_2)\gamma$ & 0.489 & 2.09     \\
                     & $\chi_{b1}(2^3P_1)\gamma$ & 13.1 &  55.9    \\
                     & $\chi_{b2}(2^3P_2)\gamma$ & 3.96 &  16.9    \\
                     & $\chi_{b2}(1^3F_2)\gamma$ & 0.164 & 0.700     \\
                     & $\chi_{b3}(1^3F_3)\gamma$ & 1.21 & 5.17     \\
\end{tabular}
}
\end{center}
\end{table}

\begin{table}[htbp]
\renewcommand\arraystretch{1.4}
\caption{Partial widths and branching ratios of annihilation decay, radiative transition, and hadronic transition and total widths for the $1F$ bottomonium states.
 The widths are in units of keV.\label{decay11F}}
\begin{center}
{\tabcolsep0.1in
\begin{tabular}{cllrc}
\toprule[1pt]\toprule[1pt]
State         &  Channels & Width & $\mathcal{B}$(\%)  \\
\midrule[1pt]
$h_{b3}(1^1F_3)$  & $\eta_{b2}(1^1D_2)\gamma$ & 22.0 & $\sim$ 100     \\
                  & $\chi_{b2}(1^3F_2)\gamma$ & $1.95\times10^{-6}$ & $8.86\times10^{-6}$     \\
                  & $h_{b}(1^1P_1)\pi \pi$ & 0.0412&  0.187    \\
               &Total  & 22.0 & 100               \\

$\chi_{b2}(1^3F_2)$ & $gg$          & 0.834 &  3.53      \\
                    & $\Upsilon_1(1^3D_1)\gamma$ &  19.4 & 82.1     \\
                   & $\Upsilon_2(1^3D_2)\gamma$ & 3.26 &  13.8    \\
                    & $\Upsilon_3(1^3D_3)\gamma$ & 0.0852 & 0.360      \\
                      & $\chi_{b0}(1^3P_0) \pi \pi$ & 0.0351 & 0.149     \\
                      & $\chi_{b1}(1^3P_1) \pi \pi$ & 0.0192 &  $8.12\times10^{-2}$     \\
                       & $\chi_{b2}(1^3P_2) \pi \pi$ & 0.00191 & $8.08\times10^{-3}$     \\
               &Total      & 23.6 & 100        \\

$\chi_{b3}(1^3F_3)$   & $gg$          &0.0672 & 0.305     \\
                      & $\Upsilon_2(1^3D_2)\gamma$ & 19.7 & 89.4     \\
                      & $\Upsilon_3(1^3D_3)\gamma$ & 2.26 &  10.3    \\
                       & $\chi_{b1}(1^3P_1) \pi \pi$ & 0.00107 & $4.86\times10^{-3}$     \\
                       & $\chi_{b2}(1^3P_2) \pi \pi$ & 0.00311 & $1.41\times10^{-2}$    \\
               &Total      & 22.0 & 100               \\

$\chi_{b4}(1^3F_4)$ & $gg$          & 0.05 &  0.235      \\
                    & $\Upsilon_3(1^3D_3)\gamma$ & 21.2 &  99.5    \\
                    & $h_{b3}(1^1F_3)\gamma$ & $1.15\times10^{-6}$ & $5.40\times10^{-6}$     \\
                    & $\chi_{b2}(1^3P_2) \pi \pi$ & 0.0354 &  0.166   \\
               &Total      & 21.3 & 100        \\

\bottomrule[1pt]\bottomrule[1pt]
\end{tabular}
}
\end{center}
\end{table}

\begin{table}[htbp]
\renewcommand\arraystretch{1.4}
\caption{Partial widths and branching ratios of annihilation decay, radiative transition, and hadronic transition and total widths for the $1G$ bottomonium states.
 The widths are in units of keV.\label{decay11G}}
\begin{center}
{\tabcolsep0.1in
\begin{tabular}{cllrc}
\toprule[1pt]\toprule[1pt]
State         &  Channels & Width & $\mathcal{B}$(\%)  \\
\midrule[1pt]
$\eta_{b4}(1^1G_4)$ & $gg$          & $6.61\times10^{-4}$ & $3.13\times10^{-3}$       \\
                    & $h_{b3}(1^1F_3)\gamma$ & 21.1 & $\sim$ 100     \\
                    & $\Upsilon_3(1^3G_3)\gamma$ & $3.33\times10^{-8}$ &  $1.58\times10^{-7}$    \\
                    & $\eta_{b2}(1^1D_2)\pi \pi$ &0.00580  &   $2.75\times10^{-2}$   \\
               &Total      & 21.1  & 100               \\

$\Upsilon_3(1^3G_3)$ & $\chi_{b2}(1^3F_2)\gamma$ & 20.1 & 92.2      \\
                     & $\chi_{b3}(1^3F_3)\gamma$ & 1.67 & 7.66     \\
                     & $\chi_{b4}(1^3F_4)\gamma$ & 0.0256 & 0.117     \\
                     & $\Upsilon_1(1^3D_1)\pi \pi$ &0.00634  & $2.91\times10^{-2}$     \\
                     & $\Upsilon_2(1^3D_2)\pi \pi$ &0.00123  & $5.64\times10^{-3}$     \\
                     & $\Upsilon_3(1^3D_3)\pi \pi$ &$5.74\times10^{-5}$  & $2.63\times10^{-4}$     \\
               &Total      & 21.8 & 100        \\

$\Upsilon_4(1^3G_4)$ & $\chi_{b3}(1^3F_3)\gamma$ & 20.1 & 93.9     \\
                     & $\chi_{b4}(1^3F_4)\gamma$ & 1.30 & 6.07     \\
                     & $\eta_{b4}(1^1G_4)\gamma$ & $4.28\times10^{-8}$ & $2.0\times10^{-7}$     \\
                     & $\Upsilon_2(1^3D_2)\pi \pi$ &0.00530  & $2.48\times10^{-2}$     \\
                     & $\Upsilon_3(1^3D_3)\pi \pi$ &$7.50\times10^{-4}$  & $3.50\times10^{-3}$     \\
               &Total      & 21.4 & 100               \\

$\Upsilon_5(1^3G_5)$ & $\chi_{b4}(1^3F_4)\gamma$ & 21.1 &  $\sim$ 100    \\
                     & $\eta_{b4}(1^1G_4)\gamma$ & $3.42\times10^{-7}$ &  $1.62\times10^{-6}$    \\
                     & $\Upsilon_3(1^3D_3)\pi \pi$ &0.00476 & $2.26\times10^{-2}$      \\
               &Total      & 21.1 & 100        \\

\bottomrule[1pt]\bottomrule[1pt]
\end{tabular}
}
\end{center}
\end{table}

\subsection{The $\eta_b$ states}

The $\eta_b$ family with a quantum number $^1S_0$ is the partner of spin-triplet $\Upsilon$ states. The ground state $\eta_b(1S)$ and radial excited state
$\eta_b(2S)$ have been established in the experiment, and their measured average masses are $9399.0\pm2.3$ MeV and $9999\pm3.5^{+2.8}_{-1.9}$ MeV \cite{Olive:2016xmw}, respectively, which
are in pretty good agreement with our theoretical values in Table \ref{fitting}. There is a more interesting physical quantity, i.e., the hyperfine mass splitting between the spin-singlet and spin-triplet states $\Delta m(nS)=m[\Upsilon(nS)]-m[\eta_b(nS)]$, which reflects the spin-dependent interaction and
can be used to test various theoretical models. For the $1S$ state, our theoretical result of hyperfine mass splitting is 65 MeV within the upper
limit of error, which is well consistent with the experimental result of $62.3\pm3.2$ MeV \cite{Olive:2016xmw} and the lattice calculation of $60.3\pm7.7$ MeV \cite{Meinel:2010pv}.
The $\Delta m(2S)$ from our modified GI model is
estimated as 28 MeV which is also equally well consistent with the measured value of $24.3^{+4.0}_{-4.5}$ MeV \cite{Mizuk:2012pb} and the lattice computation of $23.5-28.0$ MeV
\cite{Meinel:2010pv}. It is worth mentioning that the predicted mass splittings from the GI model and our
modified GI model are exactly the same, although there are some differences in the respective masses.

Together with the successful description for the mass of $\eta_b$ system, we also explore their decay behaviors. We list partial widths and branching ratios
of electromagnetic decays, annihilation decay, and hadronic transitions for $\eta_b$ states below the open-bottom thresholds from $\eta_b(1S)$ up to $\eta_b(4S)$
in Table \ref{decay1etab}. Here, the OZI-allowed two-body strong decay channels do not yet open so that the annihilation into the two gluons is dominant, whose corresponding
branching ratio is almost $100\%$. From Table \ref{decay1etab}, we see that the decay widths from our modified GI model are not so different from those of the GI results \cite{Godfrey:2015dia}. This is because screening effects are weak for the wave function of low-lying states as discussed in Section. \ref{sec2}. The decay predictions of $\eta_b(1S)$ and $\eta_b(2S)$
are also consistent with those of nonrelativistic constituent quark model \cite{Segovia:2016xqb}.

For the two unobserved $\eta_b(3S)$ and $\eta_b(4S)$ state, we predict the masses of 10336 and 10597 MeV and total widths of 5.52 and 4.07 MeV, respectively. The corresponding
hyperfine mass splitting $\Delta m(3S)$ is 20 MeV while the result of nonrelativistic constituent quark model \cite{Segovia:2016xqb} is 19 MeV, and our estimate for $\Delta m(4S)$
is 15 MeV. These predictions require further validation by the future experiments. In addition to the dominant two gluon annihilation decay, some other possible main decay channels of
$\eta_b(3S)$ are $\eta_b(1S)\pi\pi$ and $h_b(2P)\gamma$, both of which have almost the same branching ratios. The decay mode $ h_b(1P)\gamma$ is also estimated to be
important in our calculation, but the corresponding prediction of nonrelativistic constituent quark model \cite{Segovia:2016xqb} is two orders of magnitude smaller than our result. Similarly, the hadronic transition to $\eta_b(1S)\pi\pi$ and E1 radiative transition to $h_b(3P)\gamma$ are predicted
to be important decay modes of $\eta_b(4S)$ with the branching ratios of $(9.1 \times 10^{-3})$ and $(3.7 \times 10^{-4})$, respectively.

\subsection{The $\Upsilon$ states}

From Fig. \ref{Fig1:spectrum}, we can see clearly that $\Upsilon(4S)$ state is slightly above the open-bottom threshold. Hence, the states
below the thresholds are only $\Upsilon(1S)$, $\Upsilon(2S)$, and $\Upsilon(3S)$, which were first discovered bottomonia together in the E288 Collaboration at Fermilab by studying produced muon pairs in a regime of invariant masses larger than 5 GeV \cite{Herb:1977ek,Innes:1977ae}. At present, these three particles do not bring much controversy due to the high accuracy of the experimental measurements for their masses, which can be usually matched by the calculation of most of the potential models and the
lattice QCD. The mass differences between our theoretical estimates and experimental center values for these three states are 3, 5 and 1 MeV, respectively, which also indicates the
reliability of our modified GI quark model in the mass spectrum.
In addition, the BaBar collaboration \cite{Lees:2011bv} measured the level difference $m[\Upsilon(3S)]-m[\Upsilon(2S)]$ with $331.50\pm0.02\pm0.13$ MeV.
Our corresponding theoretical result of 339 MeV is also in good agreement with that of BaBar.

Partial widths and branching ratios of electromagnetic decays, annihilation decay, and hadronic transitions and the total widths for $\Upsilon(1S)$, $\Upsilon(2S)$, and $\Upsilon(3S)$
are listed in Table \ref{decay1upsilon}. Compared with the $S$-wave spin-singlet $\eta_b$ family, the experimental information of the spin-triplet $\Upsilon$ states is clearly much more
abundant, which includes total widths and partial rates of most of the decay processes. Next, we firstly start from the analysis of the common decay properties of $\Upsilon(1S)$, $\Upsilon(2S)$, and
$\Upsilon(3S)$. From Table \ref{decay1upsilon}, the annihilation decay to three gluons is dominant since each branching ratio is 89\%, 63.8\% and 58.7\% from our model, respectively, and the contributions to the total width from other three annihilation decay modes $\ell^+\ell^-$, $\gamma gg$, and $\gamma\gamma\gamma$
are much smaller. Especially, the $\gamma\gamma\gamma$ decay is difficult to search for in the experiment due to the very small branching ratio,
the $10^{-5}\sim 10^{-6}$ order of magnitude, and leptonic annihilation decay and $\gamma gg$ decay modes have almost the same predicted partial widths. All the experimental widths and branching ratios
basically agree with our theoretical calculations although the errors in $\Upsilon(3S)$ state are large. Additionally,
it needs to be emphasized that the experimental partial widths marked as $b$ shown in Table \ref{decay1upsilon} are obtained by combining measured total widths and branching ratios of the
PDG \cite{Olive:2016xmw}. 

The M1 radiative transition $\Upsilon(1S)\rightarrow \eta_b(1S) \gamma$ is the unique electromagnetic decay of $\Upsilon(1S)$ state, which has no experimental information until now.
The calculations of three models, our work and Refs. \cite{Godfrey:2015dia,Segovia:2016xqb}  give a consistent estimate, 0.01 keV.
Numerous radiative decay modes are opened for $\Upsilon(2S)$ including the transition to $\chi_{bJ} (1P)$, all of which are measured by experiment. By comparison, we can see that the
experimental data of radiative transition of $\Upsilon(2S)$ can be well reproduced by our model and that of Refs. \cite{Godfrey:2015dia,Segovia:2016xqb} except for the decay channel $\eta_b(1S) \gamma$. It should be mentioned
that the calculations of branching ratios by the nonrelativistic constituent quark model \cite{Segovia:2016xqb} adopt the measured total widths, which is the reason why our theoretical results
are smaller but the estimates of partial widths are close to the vlaues of Ref. \cite{Segovia:2016xqb}.
At last, the hadronic transition $\Upsilon(2S)\rightarrow \Upsilon(1^3S_1)\pi \pi$ has been used to fix the unknown model parameter $C_1$ in the QCD multipole expansion approach, Eq.~(\ref{hadronic amplitude}) in Appendix \ref{sec3}.

There are some difficulties in the theoretical description of $\Upsilon(3S)$ as a whole. Our theoretical total width of $\Upsilon(3S)$
is 35.8 keV, which is larger than the PDG result of $20.32\pm 1.85$ keV. The excess mainly comes from the annihilation mode of $ggg$ and hadronic mode of
$\Upsilon(1S)\pi\pi$. Considering the uncertainty of the phenomenological models, we make a comparison in radiative transitions by using experimental partial widths
rather than directly measured branching ratios. Our predictions for $\chi_{bJ}(2P) \gamma$ are satisfactory and in spite of some deviations, the $\chi_{bJ}(1P) \gamma$ and
$\eta_b(1S) \gamma$ modes can be ensured in the order of magnitude.

\subsection{The $P$-wave $\chi_{bJ}$ and $h_b$ states}\label{Pwave}

The $P$-wave states $\chi_{bJ}(1P)$ and $\chi_{bJ}(2P)$ with $J=1,2,3$ were firstly discovered in the search for the radiative processes
$\Upsilon(2S)\rightarrow \chi_{bJ}(1P) \gamma$ \cite{Han:1982zk} and $\Upsilon(3S)\rightarrow \chi_{bJ}(2P) \gamma$ \cite{Eigen:1982zm} in 1982. The corresponding spin-singlet partners
$h_b(1P)$ and $h_b(2P)$ had been missing until 2012, which were firstly observed by the Belle collaboration at the same time \cite{Adachi:2011ji}. In the same year, 2012, the Large Hadron Collider brought good news about the confirmation of the first observed $3P$ bottomonium state  $\chi_{b1}(3P)$ in the chain decay of the radiative transition to $\Upsilon(1S)\gamma$ and $\Upsilon(2S)\gamma$
and to $\Upsilon(1S,2S) \rightarrow \mu^{+}\mu^{-}$. This experiment reported the measured mass of $10530 \pm5 \pm 9$ MeV \cite{Aad:2011ih}, whose central value is consistent with our theoretical prediction of 10527 MeV in Table \ref{fitting} although statistical and systematic uncertainties are relatively large. Several subsequent measurements from LHCb give a
mass value about 15 MeV less than that in Ref. \cite{Aad:2011ih}, which are incompatible with the predictions of most of the potential models. Therefore, further study on $\chi_{b1}(3P)$ in more experiments
is necessary. Furthermore, the predicted masses of other $3P$ states are listed in Table \ref{spectrum}. Table \ref{fitting} shows that our theoretical masses for all of $1P$ and $2P$ states are in remarkable agreement with the experiments with the error of less than 5 MeV, which again proves the validity of the modified GI model with a screening effect.

Partial widths and branching ratios of radiative transition, annihilation decay, and hadronic transition and the total widths for $1P$ to $3P$ bottomonium states are successively given in Tables~\ref{decay11P}-\ref{decay13Pb}. Because there are currently no measured total widths or partial widths available for the observed 1$P$ and $2P$ states, we will analyze the decay properties of $P$-wave bottomonium states directly from the branching ratios. For $1P$ states, there are actually not many open decay channels and the annihilation processes
of multi-gluon or hybrid $q\bar{q}g$ final state have dominant contributions to the total width. As for radiative transitions, the CLEO collaboration released the most accurate measurement on electromagnetic process of
$\chi_{bJ}(1P)\rightarrow \Upsilon(1S)\gamma$ with $J=1,2,3$, whose branching ratios are $1.76\pm0.30\pm0.78 \%$, $33.1\pm1.8\pm1.7 \%$, and
$18.6\pm1.1\pm0.9 \%$ \cite{Kornicer:2010cb}, respectively.

In the past few years, the Belle collaboration has studied the decay mode of $h_b(1P)\rightarrow \eta_b(1S)\gamma$, whose
branching ratio is $49.2\pm5.7^{+5.6}_{-3.3} \%$ in 2012 \cite{Mizuk:2012pb} and $56 \pm8 \pm4 \%$ in 2015 \cite{Tamponi:2015xzb}. Combining these results with the average ratios of PDG \cite{Olive:2016xmw}, we can see that the theoretical estimates shown in Table \ref{decay11P} are also generally supported by experimental data.

Different from the case of $1P$ states, the measurements of $E1$ radiative decays $\chi_{bJ}(2P)\rightarrow\Upsilon(1S,2S)\gamma$ with $J=1,2,3$  and
 $h_b(2P) \rightarrow \eta_b(1S,2S)\gamma$ are much larger than theoretical results as shown in Table \ref{decay12P}.
It may be due to an overestimation of annihilation decay calculation
or the experimental measurement error. After all, the uncertainties of some
experimental data are quite large.

In addition, it is worth noting that our results of critical hadronic decay channel of $\chi_{bJ}(2P)\rightarrow \chi_{bJ}(1P) \pi\pi$ with non-flip $J$
and $h_b (2P)\rightarrow h_b(1P) \pi\pi$ calculated by the QCD multi-pole expansion approach are roughly consistent with those of the GI model \cite{Godfrey:2015dia}. However, our results for the
process of $\chi_{bJ}(2P)\rightarrow \chi_{bJ^{\prime}}(1P) \pi\pi$ with $J \neq J^{\prime}$ are incompatible with the GI model, where our estimates are quite suppressed similar to the calculations of
nonrelativistic constituent quark model \cite{Segovia:2016xqb}. The analogous situation also exists in the $3P$ and $D$-wave bottomonium states, which
needs to be identified by the future experimental measurements.

Although the first $3P$ state has been experimentally observed, the experimental information of its decay behaviors is still lacking. Our predictions for various decay channels of $3^1P_1$,
$3^3P_0$ and $3^3P_1$, $3^3P_2$ states are listed in Tables \ref{decay13Pa} and \ref{decay13Pb}, respectively. {{By comparing our results with those of the GI model \cite{Godfrey:2015dia}, we find that although most of decay modes have not changed much, the screening effects have demonstrated the power in the $M1$ radiation transitions. Taking the mode $h_b(3P) \to \chi_{bJ}(1P) \gamma$ as an example, there is an order of magnitude difference between the two models.}}
{{Such a large difference mainly comes from the change of bottomonium wave functions rather than the phase space. To illustrate it, we compare the square of the M1 radiation matrix element $\left< \psi_f\left|j_0\left(\frac{\omega r}{2}\right)\right|\psi_i\right>$ in Eq. (\ref{appen:A3}) in two models, where $\omega$ is almost the same for both. For the final states $\chi_{bJ}(1P) \gamma$ with $J$=0, 1, 2, their numerical results calculated by the GI model are $2.56\times10^{-4}$, $9.61\times10^{-4}$ and $1.02\times10^{-3}$  \cite{Godfrey:2015dia}, and those of our modified GI model are $1.38\times10^{-3}$, $1.08\times10^{-4}$ and $1.13\times10^{-4}$, respectively. This comparison once again proves the importance of the screening effects in describing the inner structure of meson states.        }}
In addition, the total widths of the $3P$ states predicted by us are higher on the whole.

In addition to the annihilation decay $ggg$, the decay modes
$\eta_b(1S,2S,3S) \gamma$ and $h_b(1P)\pi\pi$ are also important for $h_b(3P)$ states. Especially, the process of $h_b(3P) \rightarrow\eta_b(3S) \gamma$ has a predicted branching ratio
of 10.4$\%$. If $h_b(3P)$ is confirmed in the future, we suggest experiments to search for the missing $\eta_b(3S)$ state by studying this radiative process of $h_b(3P)$.
Likewise, compared to other radiative transition processes, the decay process to the $S$-wave $\Upsilon$ state is very significant for $\chi_{bJ}(3P)$ states. It should be noted that the $\chi_{b1}(3P)$ state is just found in the chain decay of the radiative transition to $\Upsilon(1S,2S) \gamma$ and to $\Upsilon\rightarrow \mu^+\mu^-$.
It should be also emphasized that our prediction in partial width of hadronic transition $\chi_{b0}(3P)\rightarrow \chi_{b0}(1P)\pi\pi$ gives an opportunity to observe $\chi_{b0}(3P)$ in this channel.

Finally, we discuss the predicted total widths of $P$-wave bottomonia below the open-bottom threshold, where all of $\Gamma_{total}[h_b(nP)]$ with $n$=1,2,3 are about 80-90 keV.
The total widths of $\chi_{b1}(nP)$ and $\chi_{b2}(nP)$ states with $n=1,2,3$ are around 110-145 keV and 196-264 keV, respectively. However, the total widths of
$\chi_{b0}(nP)$ states with $n$=1,2,3 are around 3.13-3.54 MeV, and these differences could possibly distinguish the $P$-wave spin-triplet $\chi_{bJ}$ states from each other in experiments.

\subsection{The $1D$ and $2D$ states}

The ground states and first radial excitations of the $D$-wave bottomonia with orbital angular momentum $L=2$ were estimated as 10150-10170 MeV and 10440-10460 MeV, respectively, which are below the $B\bar{B}$ thresholds. For the $D$-wave states, the annihilation decays to three-gluon or two-gluon are
generally suppressed than the $S$-wave. Therefore, all of the $1D$ and $2D$ states of bottomonium are expected to be narrow states. In 2010, the BaBar Collaboration observed the $D$-wave spin-triplet
 $\Upsilon(1^3D_J)$ states through decays into $\Upsilon(1S)\pi^+\pi^-$ \cite{delAmoSanchez:2010kz} and the member of $J=2$ was confirmed with a significance of 5.8 $\sigma$, although
 other two states $\Upsilon_1(1^3D_1)$ and $\Upsilon_3(1^3D_3)$ have lower significances of standard deviations 1.8 and 1.6, respectively. In general, the experimental studies on the
 $D$-wave bottomonia are presently not enough, just as the total width and the branching ratio of typical decay channels are still unknown.
For the $\Upsilon_2(1^3D_2)$ state, the predicted masses by the GI model \cite{Godfrey:2015dia} and the nonrelativistic constituent
 quark model \cite{Segovia:2016xqb} are 10147 and 10122 MeV, respectively, which are much lower than the measured value $10163.7\pm1.4$ MeV \cite{Olive:2016xmw}. Our theoretical
 mass is 10162 MeV and such a consensus again shows an excellent description of bottomonium mass spectrum
 in this work. Furthermore, we also predict the mass splittings $m[\Upsilon_3(1^3D_3)]-m[\Upsilon_2(1^3D_2)]=8$ MeV and $m[\Upsilon_2(1^3D_2)]-m[\Upsilon_1(1^3D_1)]=9$ MeV.
Ignoring delicate differences of hyperfine spin splittings, the mass of the $2D$ bottomonium is estimated
to be about 10450 MeV, which {{is in good agreement with that calculated in Ref. \cite{Godfrey:2015dia}. Experimental study of $2D$ states is also an interesting issues. }}


Partial widths and branching ratios of
annihilation decay, radiative transition, and hadronic transition and total widths for the $1D$ and $2D$ bottomonia are shown in Table \ref{decay11D} and
Tables \ref{decay12Da}-\ref{decay12Db}, respectively.  {{We want to emphasize that our results for most of decay channels are comparable with those given by the GI model \cite{Godfrey:2015dia}.
However, there is still a significant difference for the $M1$ radiation transition widths calculated from the present work and the GI model, which is mainly due to the difference of the wave functions obtained under the quenched and unquenched quark models.
This situation can be happen when comparing the result from the screened potential model and the GI model for the decays of the $1F$ and $1G$ states, which will be discussed later.}}
For the $1D$ bottomonium states, we notice that the process
$\Upsilon_2(1^3D_2)\rightarrow \Upsilon(1S)\pi^+\pi^-$ can determine the constant $C_2$ of Eq.~(\ref{hadronic amplitude}) and the predicted branching ratios of
the identical hadronic processes for other two members with $J=1,3$ are also consistent with different theoretical models \cite{Godfrey:2015dia,Segovia:2016xqb}. The radiative transition
$\eta_{b2}(1D)\rightarrow h_b(1P) \gamma$ may be an optimal channel to detect the $\eta_{b2}(1D)$ state on account of the estimate of the branching ratio $96 \%$ by us. Accordingly,
the radiative process of $\Upsilon(1^3D_J) \rightarrow \chi_{bJ^\prime}(1^3P_{J^\prime}) \gamma$ with $J=J^\prime+1 $ is the dominant decay mode for spin-triplet states of $1D$ bottomonium,
which has the estimated branching ratios of 44.1 $\%$, 74 $\%$ and 90.5 $\%$ for $J=1,2,3$, respectively. Surely, the non-forbidden radiative decay to
 $P$-wave $\chi_{bJ}$ states
is also significant for $\Upsilon(1^3D_J)$ state. There are similar decay behaviors in $2D$ bottomonium states,
where the radiative transitions to $P$-wave bottomonium states are still dominant except for the $\Upsilon_1(2^3D_1)$ state, whose half of the total width is
contributed by the annihilation mode $ggg$. The analysis of other decay channels of $2D$ bottomonium states from Tables \ref{decay12Da} and \ref{decay12Db} can be summarized as follows:
\begin{enumerate}
\item{The branching ratios of $M1$ radiative transitions of $2D\rightarrow 1D$ and $2D\rightarrow 2D$ with spin-flip are about $10^{-5}\sim10^{-7}$, which  indicate that it is difficult to observe these
decay modes in experiments.}

 \item{The $E1$ spin non-flip radiative transitions of $2D \rightarrow 1F$ with $J_i=J_f-1$ with the total angular momentum of initial (final)
state $J_{i(f)}$ can be used to study first $F$-wave bottomonium, due to the predicted partial width of $0.16 \sim 1.4$ keV and branching ratio of $0.7\% \sim 6\%$.}

\item{Compared with radiative transitions, the contributions of the dipion hadronic decays are much smaller, where the decays to $S$-wave $\eta_b$ or $\Upsilon$ states and $D$-wave $\eta_{b2}$
 or $\Upsilon_J$ states occupy about $10^{-3}\sim10^{-1}$\% and $10^{-10}\sim10^{-2}$\%, respectively.}

\item{It is easy to see that the width of leptonic annihilation decay of $D$-wave $\Upsilon$ state is much smaller than that of $S$-wave $\Upsilon$ states with three orders of magnitude by comparing
Table \ref{decay1upsilon} with Tables \ref{decay11D}-\ref{decay12Da}, which can be used to distinguish two kinds of $\Upsilon$ states with the same $J^{PC}=1^{--}$ in experiments.}
\end{enumerate}

\subsection{The $1F$ and $1G$ states}

As the high spin states, the $F$-wave and $G$-wave bottomonia have no experimental signals at present. If these states can be experimentally observed, it could be a good confirmation for the theoretical calculation of the potential model. The predicted decay properties of $1F$ and $1G$ bottomonia for
partial widths and branching ratios of annihilation decay, electromagnetic transition, and hadronic transition and the total widths are listed in Tables \ref{decay11F} and
\ref{decay11G}, respectively, where one can find a very interesting common decay feature. That is, the dominant decay modes of eight particles are all sorts of the radiative
transition of $1F_{(J)}\rightarrow 1D_{(J-1)}$ or $1G_{(J)} \rightarrow1F_{(J-1)}$ with almost more than $90 \%$ branching ratios, where subindices $J$ / $J-1$ represent the total angular momentum of
a particle. This also indicates that experiments are likely to observe the first $F$-wave bottomonia via these radiative processes of $h_{b3}(1^1F_3)\rightarrow \eta_{b2}(1^1D_2)\gamma$
and  $\chi_{bJ}(1^3F_J) \rightarrow \Upsilon_{J-1}(1^3D_{J-1})\gamma$ once all the ground states of $D$-wave bottomonium are experimentally established. In the same way, search for the
first $G$-wave bottomonia also requires the confirmation of the $F$-wave states in experiment. This forms a chain-like relationship, which means that search for
 these high spin states is probably achieved step by step. Our predicted mass values of $1F$ and $1G$ states are shown in Table \ref{spectrum}, where the average mass values of spin triplet
 states for $1F$ and $1G$ are estimated as 10366 MeV and 10535 MeV, respectively. These are almost the same as those of spin singlet states. We hope that
 these predicted results will be helpful for the future experimental studies on high spin $F$ and $G$-wave states. In addition, it can be seen from Tables \ref{decay11F}-\ref{decay11G}
 that the $1F$ ($1G$) states could also transit to the specific $P$ ($D$)-wave states by emitting two light mesons $\pi\pi$. However, partial widths of these hadronic decay processes are calculated to be
 relatively small. The predicted total widths of the ground states of $F$ and $G$-wave bottomonia are all about 20 keV, which are consistent with the estimates of the GI model \cite{Godfrey:2015dia}.


\section{analysis and predictions of higher bottomonia}\label{sec5}

\begin{table*}[htbp]
\renewcommand\arraystretch{1.4}
\caption{ Partial widths and branching ratios of the OZI-allowed strong decay, annihilation decay, and radiative transition and the total width for $\Upsilon(4S)$.
 Experimental results are taken from the PDG \cite{Olive:2016xmw}. The theoretical results of the GI model \cite{Godfrey:2015dia}
 and the nonrelativistic constituent quark model \cite{Segovia:2016xqb} are summarized in the rightmost columns. The width results are in units of keV.\label{decay2upsilon4s}}
\begin{center}
{\tabcolsep0.03in
\begin{tabular}{llcrcllcllcllc}
\toprule[1pt]\toprule[1pt]
&  & \multicolumn{2}{c}{This work} & & \multicolumn{2}{c}{Expt. \cite{Olive:2016xmw}} & & \multicolumn{2}{c}{GI \cite{Godfrey:2015dia}}  & & \multicolumn{2}{c}{Ref. \cite{Segovia:2016xqb}} \\
\cmidrule[1pt]{3-5}\cmidrule[1pt]{6-7}\cmidrule[1pt]{9-11}\cmidrule[1pt]{12-13}
State   & Channels & Width & $\mathcal{B}$(\%)   & &  Width & $\mathcal{B}$(\%)  & &   Width & $\mathcal{B}$(\%) & &   Width & $\mathcal{B}$(\%) \\
\midrule[1pt]
$\Upsilon(4S)$ & $\ell^+\ell^-$ &  0.431 &  $1.74\times10^{-3}$   & & $0.272\pm0.029$  & $(1.57\pm0.08)\times10^{-3}$    & & 0.39 & $1.8\times10^{-3}$    & & 0.21  &$1.02\times10^{-3}$ & \\
               &  $ggg$          & 16.7 &  $6.76\times10^{-2}$   & & $\cdots$  &  $\cdots$   & & 15.1 & $6.86\times10^{-2}$    & & 15.58 &$7.60\times10^{-2}$ \\
               & $\gamma gg$     & 0.433 &  $1.75\times10^{-3}$   & &  $\cdots$ &$\cdots$     & & 0.40 & $1.8\times10^{-3}$    & & 0.30 & $1.46\times10^{-3}$\\
         & $\gamma\gamma\gamma$ &  $6.36\times10^{-6}$ & $2.57\times10^{-8}$   & & $\cdots$   & $\cdots$    & & $6.0\times10^{-6}$ & $2.7\times10^{-8}$    & & $1.29\times10^{-6}$ & $6.29\times10^{-9}$ \\
               & $\chi_{b0}(1^3P_0)\gamma$ & $5.12\times10^{-4}$ & $2.07\times10^{-6}$ && $\cdots$    & $\cdots$     && $\cdots$       & $\cdots$      && 0.0588   & $2.87\times10^{-4}$    \\
               & $\chi_{b1}(1^3P_1)\gamma$ & 0.0507 & $2.05\times10^{-4}$  &&  $\cdots$    & $\cdots$     &&   $\cdots$     &  $\cdots$     &&  0.0474  & $2.31\times10^{-4}$    \\
               & $\chi_{b2}(1^3P_2)\gamma$ & 0.219 &  $8.87\times10^{-4}$ &&  $\cdots$    & $\cdots$     &&    $\cdots$    & $\cdots$      &&  0.0120  &  $5.85\times10^{-5}$   \\
               & $\chi_{b0}(2^3P_0)\gamma$ & 0.0137 & $5.55\times10^{-5}$  &&$\cdots$      & $\cdots$     &&   $\cdots$     &  $\cdots$     &&  0.17  &   $8.29\times10^{-4}$  \\
               & $\chi_{b1}(2^3P_1)\gamma$ & 0.0138 & $5.59\times10^{-5}$  && $\cdots$     & $\cdots$     &&   $\cdots$     & $\cdots$      && 0.18   &   $8.78\times10^{-4}$  \\
               & $\chi_{b2}(2^3P_2)\gamma$ & 0.226 & $9.15\times10^{-4}$  && $\cdots$     & $\cdots$     &&   $\cdots$     & $\cdots$      &&  0.11  &    $5.37\times10^{-4}$ \\
               & $\chi_{b0}(3^3P_0)\gamma$ & 0.587 & $2.38\times10^{-3}$  && $\cdots$     &  $\cdots$    &&     0.48  & $2.2\times10^{-3}$   && 0.61   &   $2.98\times10^{-3}$  \\
               & $\chi_{b1}(3^3P_1)\gamma$ & 1.14 &  $4.62\times10^{-3}$ && $\cdots$     &   $\cdots$   &&      0.84  &  $3.8\times10^{-3}$  &&  1.17  &  $5.71\times10^{-3}$   \\
               & $\chi_{b2}(3^3P_2)\gamma$ & 1.16 &  $4.70\times10^{-3}$ &&  $\cdots$    &  $\cdots$    &&      0.82  &  $3.7\times10^{-3}$ &&  1.45  &   $7.07\times10^{-3}$  \\
               & $\eta_b(1^1S_0)\gamma$ &  0.0572 &  $2.32\times10^{-4}$ && $\cdots$     & $\cdots$     &&  $\cdots$      &  $\cdots$     && 0.0498   &   $2.43\times10^{-4}$  \\
               & $BB$        & 24.7 MeV & $\sim$100  &&  $\cdots$      & $>$96   &&  22 MeV     & $\sim$100     && 20.59 MeV   & 100    \\
               &Total      & 24.7 MeV & 100   &&  $20.5\pm2.5$ MeV &  $\cdots$   &&  22 MeV     &  100    &&  20.59 MeV  &  100   \\
\bottomrule[1pt]\bottomrule[1pt]
\end{tabular}
}
\end{center}
\end{table*}

\begin{table*}[htbp]
\renewcommand\arraystretch{1.25}
\caption{ Partial widths and branching ratios of the OZI-allowed strong decay, annihilation decay, and radiative transition and the total widths for $\Upsilon(10860)$ and $\Upsilon(11020)$. Experimental results are taken from the PDG \cite{Olive:2016xmw} The theoretical results of the GI model \cite{Godfrey:2015dia}
 and the nonrelativistic constituent quark model \cite{Segovia:2016xqb} are summarized in the rightmost columns. The width results are in units of keV.\label{decay2upsilon5s6s}}
\begin{center}
{\tabcolsep0.018in
\begin{tabular}{llcrcclcllcllc}
\toprule[1pt]\toprule[1pt]
&  & \multicolumn{2}{c}{This work} & & \multicolumn{2}{c}{Expt. \cite{Olive:2016xmw}} & & \multicolumn{2}{c}{GI \cite{Godfrey:2015dia}}  & & \multicolumn{2}{c}{Ref. \cite{Segovia:2016xqb}} \\
\cmidrule[1pt]{3-5}\cmidrule[1pt]{6-7}\cmidrule[1pt]{9-11}\cmidrule[1pt]{12-13}
State   & Channels & Width & $\mathcal{B}$(\%)   & &  Width & $\mathcal{B}$(\%)  & &   Width & $\mathcal{B}$(\%) & &   Width & $\mathcal{B}$(\%) \\
\midrule[1pt]
$\Upsilon(10860)$ & $\ell^+\ell^-$ &  0.348 &  $7.63\times10^{-4}$   & & $0.31\pm0.07$  & $(5.7^{+1.5}_{-1.5})\times10^{-4}$    & & 0.33 & $1.2\times10^{-3}$    & & 0.18  &$3.27\times10^{-4}$ & \\
               &  $ggg$          & 14.2 &  $3.11\times10^{-2}$   & & $\cdots$  &  $\cdots$   & & 13.1 & $4.78\times10^{-2}$    & & 13.33 &$2.42\times10^{-2}$ \\
               & $\gamma gg$     & 0.370 &  $8.11\times10^{-4}$   & & $\cdots$   & $\cdots$  & & $\cdots$ & $\cdots$    & & 0.25 & $4.55\times10^{-4}$\\
         & $\gamma\gamma\gamma$ &  $5.43\times10^{-6}$ & $1.19\times10^{-8}$   & & $\cdots$   & $\cdots$    & & $\cdots$ & $\cdots$    & & $1.10\times10^{-6}$ & $2.00\times10^{-9}$ \\
                  & $\chi_{b0}(2^3P_0)\gamma$ & $2.69\times10^{-4}$ &$5.90\times10^{-7}$   && $\cdots$     &  $\cdots$    && $\cdots$      &$\cdots$      &&  0.22     & $4.00\times10^{-4}$    \\
                  & $\chi_{b1}(2^3P_1)\gamma$ &  0.0532 &$1.17\times10^{-4}$    &&  $\cdots$    &  $\cdots$    &&   $\cdots$    & $\cdots$     && 0.26      & $4.73\times10^{-4}$    \\
\end{tabular}
}
\end{center}
\end{table*}

\begin{table}[htbp]
\renewcommand\arraystretch{1.4}
\caption{ Partial widths and branching ratios of the OZI-allowed strong decay, annihilation decay, and radiative transition and the total widths for $5^1S_0$ and $6^1S_0$
 bottomonium states. The width results are in units of keV.\label{decay2eta5s6s}}
\begin{center}
{\tabcolsep0.16in
\begin{tabular}{cllrc}
\toprule[1pt]\toprule[1pt]
State         &  Channels & Width & $\mathcal{B}$(\%)  \\
\midrule[1pt]
$\eta_b(5^1S_0)$ &  $gg$    & 3.26 MeV & 11.5   \\
        & $\gamma\gamma$  &  0.192    & $6.76\times10^{-4}$  \\
                 & $h_b(1^1P_1)\gamma$ & 0.322 & $1.13\times10^{-3}$                   \\
                 & $h_b(2^1P_1)\gamma$ & 1.00 &  $3.52\times10^{-3}$                    \\
                 & $h_b(3^1P_1)\gamma$ & 1.63 &  $5.74\times10^{-3}$                    \\
                 & $h_b(4^1P_1)\gamma$ & 33.5 &  0.118                    \\
                 & $\Upsilon(1^3S_1)\gamma$ & 0.0642 &  $2.26\times10^{-4}$                   \\
\end{tabular}
}
\end{center}
\end{table}

\begin{table}[htbp]
\renewcommand\arraystretch{1.13}
\caption{ Partial widths and branching ratios of the OZI-allowed strong decay, annihilation decay, and radiative transition and the total widths for the $7S$ bottomonium states.
 The width results are in units of keV.\label{decay27s}}
\begin{center}
{\tabcolsep0.12in
\begin{tabular}{cllrc}
\toprule[1pt]\toprule[1pt]
State         &  Channels & Width & $\mathcal{B}$(\%)  \\
\midrule[1pt]
$\eta_b(7^1S_0)$ &  $gg$    & 2.21 MeV &  2.33  \\
        & $\gamma\gamma$  &  0.130    &  $1.37\times10^{-4}$  \\
                 & $h_b(1^1P_1)\gamma$ & 0.291 &  $3.07\times10^{-4}$    \\
                 & $h_b(2^1P_1)\gamma$ & 0.404 &  $4.26\times10^{-4}$    \\
                 & $h_b(3^1P_1)\gamma$ & 0.444 &  $4.68\times10^{-4}$    \\
                 & $h_b(4^1P_1)\gamma$ & 0.607 &  $6.40\times10^{-4}$    \\
                 & $h_b(5^1P_1)\gamma$ & 0.388 &   $4.09\times10^{-4}$   \\
\end{tabular}
}
\end{center}
\end{table}

\begin{table}[htbp]
\renewcommand\arraystretch{1.13}
\caption{ Partial widths and branching ratios of the OZI-allowed strong decay, annihilation decay, and radiative transition and the total widths for the $8S$ bottomonium states.
 The width results are in units of keV.\label{decay28s}}
\begin{center}
{\tabcolsep0.12in
\begin{tabular}{cllrc}
\toprule[1pt]\toprule[1pt]
State         &  Channels & Width & $\mathcal{B}$(\%)  \\
\midrule[1pt]
$\eta_b(8^1S_0)$ &  $gg$    & 1.81 MeV & 2.81   \\
        & $\gamma\gamma$  &  0.106    &  $1.65\times10^{-4}$  \\
                 & $h_b(1^1P_1)\gamma$ & 0.211 & $3.28\times10^{-4}$      \\
                 & $h_b(2^1P_1)\gamma$ & 0.257 &  $3.99\times10^{-4}$    \\
                 & $h_b(3^1P_1)\gamma$ & 0.336 &  $5.22\times10^{-4}$     \\
                 & $h_b(4^1P_1)\gamma$ & 0.246 &  $3.82\times10^{-4}$     \\
                 & $h_b(5^1P_1)\gamma$ & 0.410 &  $6.37\times10^{-4}$   \\
\end{tabular}
}
\end{center}
\end{table}

\begin{table}[htbp]
\renewcommand\arraystretch{1.13}
\caption{ Partial widths and branching ratios of the OZI-allowed strong decay, annihilation decay, and radiative transition and the total widths for the $4P$ bottomonium states.
The width results are in units of keV.\label{decay24p}}
\begin{center}
{\tabcolsep0.14in
\begin{tabular}{cllrc}
\toprule[1pt]\toprule[1pt]
State         &  Channels & Width & $\mathcal{B}$(\%)  \\
\midrule[1pt]
$h_b(4^1P_1)$ & $ggg$ & 73.2 & 0.152     \\
              & $\eta_b(1^1S_0)\gamma$ & 2.68 &  $5.57\times10^{-3}$    \\
              & $\eta_b(2^1S_0)\gamma$ & 1.83 &   $3.80\times10^{-3}$   \\
              & $\eta_b(3^1S_0)\gamma$ & 3.69 &   $7.67\times10^{-3}$   \\
              & $\eta_b(4^1S_0)\gamma$ & 18.4 &   $3.83\times10^{-2}$   \\
              & $\eta_{b2}(3^1D_2)\gamma$ & 3.17 & $6.59\times10^{-3}$     \\
              & $\chi_{b0}(1^3P_0)\gamma$ & 0.00556 &   $1.16\times10^{-5}$   \\
\end{tabular}
}
\end{center}
\end{table}

\begin{table}[htbp]
\renewcommand\arraystretch{1.4}
\caption{ Partial widths and branching ratios of the OZI-allowed strong decay, annihilation decay, and radiative transition and the total widths for $5^1P_1$ and $5^3P_0$
 bottomonium states. The width results are in units of keV.\label{decay25pa}}
\begin{center}
{\tabcolsep0.13in
\begin{tabular}{cllrc}
\toprule[1pt]\toprule[1pt]
State         &  Channels & Width & $\mathcal{B}$(\%)  \\
\midrule[1pt]
$h_b(5^1P_1)$ & $ggg$ & 76.2 &  0.154    \\
              & $\eta_b(1^1S_0)\gamma$ & 2.28 & $4.62\times10^{-3}$     \\
              & $\eta_b(2^1S_0)\gamma$ & 1.85 & $3.74\times10^{-3}$     \\
              & $\eta_b(3^1S_0)\gamma$ & 1.72 &  $3.48\times10^{-3}$    \\
              & $\eta_b(4^1S_0)\gamma$ & 4.14 &  $8.38\times10^{-3}$    \\
              & $\eta_b(5^1S_0)\gamma$ & 1.34 &  $2.71\times10^{-3}$    \\
              & $\eta_{b2}(4^1D_2)\gamma$ & 3.37 &  $6.82\times10^{-3}$    \\
\end{tabular}
}
\end{center}
\end{table}

\begin{table}[htbp]
\renewcommand\arraystretch{1.4}
\caption{ Partial widths and branching ratios of OZI-allowed strong decay, annihilation decay, and radiative transition and the total widths for $5^3P_1$ and $5^3P_2$
 bottomonium states. The width results are in units of keV.\label{decay25pb}}
\begin{center}
{\tabcolsep0.13in
\begin{tabular}{cllrc}
\toprule[1pt]\toprule[1pt]
State         &  Channels & Width & $\mathcal{B}$(\%)  \\
\midrule[1pt]
$\chi_{b1}(5^3P_1)$& $q\bar{q}g$ & 132 &  0.283    \\
                   & $\Upsilon(1^3S_1)\gamma$ & 0.743 & $1.59\times10^{-3}$     \\
                   & $\Upsilon(2^3S_1)\gamma$ & 0.741 &   $1.59\times10^{-3}$    \\
                   & $\Upsilon(3^3S_1)\gamma$ & 0.824 &   $1.77\times10^{-3}$    \\
                   & $\Upsilon(4^3S_1)\gamma$ & 2.32 &    $4.98\times10^{-3}$   \\
                   & $\Upsilon(5^3S_1)\gamma$ & 0.835 &   $1.79\times10^{-3}$    \\
                   & $\Upsilon_1(4^3D_1)\gamma$ & 0.921 &  $1.98\times10^{-3}$     \\
\end{tabular}
}
\end{center}
\end{table}

\begin{table}[htbp]
\renewcommand\arraystretch{1.25}
\caption{ Partial widths and branching ratios of OZI-allowed strong decay, annihilation decay, and radiative transition and the total widths for $6^1P_1$ and $6^3P_0$
 bottomonium states. The width results are in units of keV.\label{decay26pa}}
\begin{center}
{\tabcolsep0.1in
\begin{tabular}{cllrc}
\toprule[1pt]\toprule[1pt]
State         &  Channels & Width & $\mathcal{B}$(\%)  \\
\midrule[1pt]
$h_b(6^1P_1)$ & $ggg$ & 74.1 &   $6.01\times10^{-2}$    \\
              & $\eta_b(1^1S_0)\gamma$ & 1.98 &  $1.61\times10^{-3}$    \\
              & $\eta_b(2^1S_0)\gamma$ & 1.55 &  $1.26\times10^{-3}$    \\
              & $\eta_b(3^1S_0)\gamma$ & 1.51 &   $1.22\times10^{-3}$   \\
              & $\eta_b(4^1S_0)\gamma$ & 1.17 & $0.949\times10^{-3}$     \\
              & $\eta_b(5^1S_0)\gamma$ & 1.25 &   $1.01\times10^{-3}$   \\
              & $\eta_b(6^1S_0)\gamma$ & 7.50 &     $6.08\times10^{-3}$ \\
\end{tabular}
}
\end{center}
\end{table}

\begin{table}[htbp]
\renewcommand\arraystretch{1.15}
\caption{ Partial widths and branching ratios of OZI-allowed strong decay, annihilation decay, and radiative transition and the total widths for $6^3P_1$ and $6^3P_2$
 bottomonium states. The width results are in units of keV.\label{decay26pb}}
\begin{center}
{\tabcolsep0.11in
\begin{tabular}{cllrc}
\toprule[1pt]\toprule[1pt]
State         &  Channels & Width & $\mathcal{B}$(\%)  \\
\midrule[1pt]
$\chi_{b1}(6^3P_1)$ & $q\bar{q}g$ & 128 &   $9.13\times10^{-2}$   \\
                   & $\Upsilon(1^3S_1)\gamma$ & 0.478 &   $3.41\times10^{-4}$   \\
                   & $\Upsilon(2^3S_1)\gamma$ & 0.394 &   $2.81\times10^{-4}$   \\
                   & $\Upsilon(3^3S_1)\gamma$ & 0.616 &    $4.39\times10^{-4}$  \\
                   & $\Upsilon(4^3S_1)\gamma$ & 0.741 &    $5.29\times10^{-4}$  \\
                   & $\Upsilon(5^3S_1)\gamma$ & 0.706 &    $5.04\times10^{-4}$  \\
                   & $\Upsilon(6^3S_1)\gamma$ & 6.45 &     $4.60\times10^{-3}$ \\
\end{tabular}
}
\end{center}
\end{table}

\subsection{Higher $\eta_b $ radial excitations}

As the pseudoscalar partner of the $S$-wave $\Upsilon$ states, the established members of the $\eta_b$ family are much less than those of the $\Upsilon$ family.
Based on this fact, we need to predict some theoretical results and simultaneously make a systematic study on the higher excited states of the $\eta_b$ family that will
not only provide meaningful clues to experimentally search for them above thresholds but also further reveal their inner nature.

For the $b\bar{b}$ states above thresholds, they usually decay mainly into a pair of positive and negative bottom mesons or bottomed-strange mesons.
{{The OZI-allowed strong decay behaviors could be studied by the QPC model, whose details are given in Appendix \ref{sec3}. In this work, the parameter $\gamma$ for the bottomonium system in the QPC model can be determined by fitting to experimental widths of $\Upsilon(4S)$, $\Upsilon(5S)$, and $\Upsilon(6S)$ states and the fitted results are shown in Table \ref{gammavalue}. For the input masses of bottomonium states, the experimental masses will be considered. If the members of a bottomonium multiplet are partially discovered by experiments, the input masses of the missing states can be obtained by combining the measured mass and predicted splitting. For the predicted bottomonia, we calculate the behaviours of strong decays by utilizing our theoretical masses in Table \ref{spectrum} as an input. Besides the input masses, the spatial wave functions of bottomonium states can be directly derived from our modified GI model. The input masses of bottom and bottomed-strange mesons are summarized in Table \ref{bottommeson}, where two $1^+$ bottom mesons $B(1P_1)$ and $B(1P_1^{\prime})$ are the mixture of $^3P_1$ and $^1P_1$ states, and the mixing angle $\theta_{1P}=-\arcsin(\sqrt{2/3})=-54.7^\circ$ is adopted as in the heavy quark limit \cite{Godfrey:1986wj,Matsuki:2010zy,Barnes:2002mu}. Finally, the wave functions of bottom and bottomed-strange mesons are taken from the calculations of Ref. \cite{Sun:2014wea} and the quark mass parameters are given in Appendix \ref{sec3}.   }}


In this subsection, higher radial excitations from $\eta_b(5S)$ to $\eta_b(8S)$ are systematically studied, whose mass values are predicted in Table \ref{spectrum}. Their corresponding
decay properties including total widths and partial widths and branching ratios of OZI-allowed strong decay, annihilation decay and radiative transition are listed in
Tables \ref{decay2eta5s6s}-\ref{decay28s}. {{Here, we make a comparison of the decay results of $\eta_b(5S)$ and $\eta_b(6S)$ presented in this work and given by the GI model \cite{Godfrey:2015dia}, which generally reflect that these two quark models can achieve the same goal.}}

The mass of $\eta_b(5S)$ state is calculated as 10810 MeV by the modified GI model. We give another estimate of the mass value
of $\eta_b(5S)$ of 10870 MeV utilizing the results of theoretical mass splitting and the measured mass of $\Upsilon(10860)$. At the same time, this mass estimate
is also used as the model parameter of strong decay calculations. In Table \ref{decay2eta5s6s}, one can see that the total width of $\eta_b(5S)$ is predicted to be 28.4 MeV, which
can mainly decay into the $BB^*$, $gg$, $B_s^*B_s^*$, $B_sB_s^*$, and $B^*B^*$, and corresponding branching ratios are 71.5 $\%$, 11.5 $\%$, 10.2 $\%$, 3.91 $\%$, and 2.71 $\%$,
respectively. The dominant radiative decay mode is $h_b(4^1P_1)\gamma$ with a partial width of 33.5 keV. We should notice that the notation $BB^*$ refers to
$B\bar{B^*}$+$B^*\bar{B}$, $B^*B^*$ to $B^*\bar{B^*}$, and so on. The predicted mass of the $\eta_b(6S)$ is 10991 MeV, which should also be improved compared
to the estimates of the GI model, just like $\Upsilon(6S)$. The total width of $\eta_b(6S)$ is estimated to be 20.4 MeV, which is about the same as that of $\eta_b(5S)$. The
dominant decay channel of $\eta_b(6S)$ is still $BB^*$ with a branching ratio of 73 $\%$ and other decay modes $gg$, $BB(1^3P_0)$, $B_sB_s^*$, and $B^*B^*$ also have large
contributions to the total width. In general, our decay results indicate that the $\eta_b(5S)$ and $\eta_b(6S)$ are most likely to be detected in the $BB^*$ mode.

In comparison with the $\Upsilon(nS)$ states with $n$=$7, 8$, which
 are likely to become a new research area for bottomonium physics in the future as their lower radial excited states have been observed experimentally, we also study the $\eta_b(7S)$ and $\eta_b(8S)$ state here. We predict their
mass values as 11149 and 11289 MeV, and hyperfine mass splittings $\Delta m(7S)$$\backslash$$\Delta m(8S)$ as 8$\backslash$7 MeV, respectively.
There are obviously large differences between the predicted mass by the GI model and modified GI model, which can be seen in Table \ref{spectrum}. From Tables \ref{decay27s} and \ref{decay28s},
the total widths of the $\eta_b(7S)$ and $\eta_b(8S)$ are predicted to be 94.8 MeV and 64.4 MeV, respectively, which indicate the $\eta_b(7S)$ is possibly a broad state. Our calculations
show that the dominant decay channels of $\eta_b(7S)$ are $BB^*$, $B^*B(1^3P_2)$, $B^*B(1P_1)$, $B^*B(1P_1^{\prime})$ and $B^*B^*$. The corresponding partial widths are 27.7, 17.9,
15.0, 13.9 and 11.1 MeV, respectively. The decay channels $BB^*$, $B^*B^*$ and $BB(1^3P_2)$ are pivotal for the $\eta_b(8S)$ with partial widths of 29.5, 19.0 and 7.61 MeV, respectively.
From here, we can see the latent importance of $P$-wave bottom mesons in the OZI-allowed open-flavor strong decays of these extremely high bottomonium radial excitations.
In addition, since the contributions of radiative transition and strong decay of bottom-strange mesons are small for the $\eta_b(7S)$ and $\eta_b(8S)$, we suggest future experiments to search for them with the generated final states of bottom mesons.

\subsection{higher $\Upsilon$ radial excitations}

The $S$-wave $\Upsilon$ states with $J^{PC}=1^{--}$ have always been the significant research field of bottomonium physics. Up to now, it is also the unique place to
experimentally study some interesting effects above open-bottom thresholds, which includes
$\Upsilon(4S)$, $\Upsilon(10860)$, and $\Upsilon(11020)$. Therefore, further theoretical exploration for higher $\Upsilon$ bottomonia is necessary. Additionally,
the strong decay into two bottom mesons or bottom-strange mesons will be dominant for the three $\Upsilon$'s mentioned above. Hence, their experimental information
will determine the model parameter $\gamma$ of the strong decay calculation in the QPC model.

There is no controversy in experimental measurements of the $\Upsilon(4S)$ state,
whose total width was first measured as $25\pm2.5$ MeV and $20\pm2\pm4$ MeV by the CUSB \cite{Lovelock:1985nb} and CLEO \cite{Besson:1984bd}
collaborations, respectively. On the other hand, BaBar's measurements in 2005 gave the similar resonance parameters \cite{Aubert:2004pwa} and the current average total width of PDG is
$20.5\pm2.5$ MeV \cite{Olive:2016xmw}.

For the $\Upsilon(5S)$ and $\Upsilon(6S)$, however, the situation is not so optimistic since the early measured data and the recent measurements of
resonance parameters are incompatible with each other.
In short, after 2010, the Belle collaboration released consistent mass values of $\Upsilon(10860)$ by the different production cross sections of
$e^+e^- \rightarrow \Upsilon(nS) \pi^+\pi^-$ with $n$=1,2,3 and $e^+e^- \rightarrow b\bar{b}$, as well as $e^+e^- \rightarrow h_b(nP) \pi^+\pi^-$ with $n$=1,2 \cite{Chen:2008xia,Santel:2015qga,Abdesselam:2015zza}. Their values are larger than the theoretical calculation
and early experimental values. In addition, the measurement of the total width by Belle also disagreed with the previous experimental conclusion of a broad state of $\Upsilon(10860)$
\cite{Lovelock:1985nb,Besson:1984bd}. Although there are a few experimental results of $\Upsilon(11020)$, the center mass values from the BaBar \cite{Aubert:2008ab}
and the recent Belle collaboration \cite{Santel:2015qga,Abdesselam:2015zza} are consistently about 20 MeV smaller than the earlier experimental results \cite{Lovelock:1985nb, Besson:1984bd}.
It can be seen clearly from the PDG \cite{Olive:2016xmw} that the total width of $\Upsilon(11020)$ is measured to be about 30-60 MeV by integrating the present
experimental information. At the same time, because the amount of data is not sufficient, the measurements of these two resonances by the analysis of the $R$ value in the previous experiments are probably unconvincing to a great extent, where
$R=\sigma(e^+e^- \to hadrons)/\sigma(e^+e^- \to \mu^+\mu^-)$. Therefore, we hope to see the more accurate experimental results for the $\Upsilon(10860)$ and $\Upsilon(11020)$ in the forthcoming Belle II experiment.
It is worth mentioning that in the measurements of the Belle \cite{Santel:2015qga},
the resonance parameters from $R_{\Upsilon(nS)\pi\pi}$ and $R_b$ are completely consistent within the error range \cite{Santel:2015qga} although the application of a flat continuum in the $R_b$ fit brings some inconsistencies between the fitted amplitudes for $R_{\Upsilon(nS)\pi\pi}$ and $R_b$. Furthermore, the measured mass values and total widths from $R_{\Upsilon(nS)\pi\pi}$ have too large statistical and systematic errors. So in the calibration of parameter $\gamma $, we select the latest Belle's
measurements of the process $e^+e^- \rightarrow b\bar{b}$, which give the resonance parameters for $\Upsilon(10860)$ and $\Upsilon(11020)$ as
$M[\Upsilon(10860)]=10881.8^{+1.0}_{-1.1}\pm 1.2$ MeV, $\Gamma  [\Upsilon(10860)]=48.5^{+1.9+2.0}_{-1.8-2.8}$ MeV and
$M[\Upsilon(11020)]=11003.0 \pm1.1^{+0.9}_{-1.0}$ MeV, $\Gamma [\Upsilon(11020)]=39.3^{+1.7+1.3}_{-1.6-2.4}$ MeV \cite{Santel:2015qga}, respectively.

Based on the above consideration, in the following, the systematical study will be performed on the $\Upsilon(nS)$ states above the thresholds. Higher bottomonia $\Upsilon(7S)$ and $\Upsilon(8S)$ states are also discussed in this subsection.

\subsubsection{$\Upsilon(4S)$, $\Upsilon(10860)$, and $\Upsilon(11020)$ }

Partial widths and branching ratios of the OZI-allowed strong decay, annihilation decay, and radiative transition and total widths for $\Upsilon(4S)$, $\Upsilon(10860)$ and $\Upsilon(11020)$
are shown in Tables \ref{decay2upsilon4s} and  \ref{decay2upsilon5s6s}, respectively. $\Upsilon(4S)$ is slightly (about 20 MeV) higher than
the $B\bar{B}$ threshold. Hence, its unique dominant decay channel is $B\bar{B}$ and the corresponding experimental branching ratio is larger than $96\%$.
Another measured decay channel of $\Upsilon(4S)$  is leptonic annihilation
decay $\ell^+\ell^-$. Our theoretical branching ratio is $1.74\times10^{-5} $, which is well consistent with $(1.57\pm0.08)\times10^{-5} $ from PDG \cite{Olive:2016xmw}.

The total widths of the $\Upsilon(10860)$ and $\Upsilon(11020)$ are estimated to be 45.6 MeV and 38.3 MeV in our modified GI model, which coincide
well with Belle's results of $48.5^{+1.9+2.0}_{-1.8-2.8}$ and $39.3^{+1.7+1.3}_{-1.6-2.4}$ MeV \cite{Santel:2015qga},  respectively.
Additionally, other theoretical predictions for the total widths of the $\Upsilon(10860)$ and $\Upsilon(11020)$ are not unified \cite{Godfrey:2015dia,Segovia:2016xqb,Ferretti:2013vua}, where there is even a difference of 50 MeV on the prediction of $\Upsilon(11020)$.
Experiments have released the branching ratios of leptonic annihilation decays in the $10^{-6}$ orders of magnitude both for $\Upsilon(5S)$ and $\Upsilon(6S)$ \cite{Olive:2016xmw}. This is in accord with our theoretical predictions. Branching ratios of all six kinematically-allowed open bottom decay modes $BB$, $BB^*$, $B^*B^*$, $B_sB_s$, $B_sB_s^*$, and $B_s^*B_s^*$ of $\Upsilon(5S)$ have
been experimentally measured as $5.5 \pm 1.0 \%$, $13.7 \pm 1.6 \%$, $38.1 \pm 3.4 \%$, $0.5 \pm 0.5 \%$, $1.35 \pm 0.32 \%$, and
$17.6 \pm 2.7 \%$ \cite{Olive:2016xmw}, respectively. One can see that $B^*B^*$ is the leading decay channel and the following $B_s^*B_s^*$ and $BB^*$ are next-to-leading decay
modes. This is not only contradictory with our calculations but most of other theoretical estimates of potential models \cite{Godfrey:2015dia,Segovia:2016xqb,Ferretti:2013vua}. The possible reasons are the following: firstly, as mentioned before, experimental measurement is not enough and one cannot rule out the latent
error in experiment; secondly, the calculated partial width of strong decay is dependent on particle's real mass and a more accurate mass value may reduce the possible
deviation from the phenomenological decay model. Finally, there may be some other physical effects and mechanisms in the strong decay process of $\Upsilon(10860)$. In
Ref. \cite{TorresRincon:2010fu}, authors just discussed the application of the Franck-Condon principle, which is common in molecular physics on the anomalous high branching ratios of
$B_s^*B_s^*$ versus $B_sB_s^*$.

For the $\Upsilon(11020)$ state, there are no experimental data in the open-bottom decay channel at present. We predict that the dominant modes of $\Upsilon(6S)$ are
$BB^*$, $BB(1P_1)$, $BB$, $B^*B^*$ with the corresponding branching ratios as 43 $\%$, 21.6 $\%$ ,20.4 $\%$ and 11.5 $\%$, respectively. These partial widths are quite different from the predictions by the GI model \cite{Godfrey:2015dia}, though both of the predictions of total widths are contiguous, which are expected to be tested by the forthcoming Belle II.

\subsubsection{$\Upsilon(7S)$ and $\Upsilon(8S)$ }

In Ref. \cite{Godfrey:2015dia}, authors did not study the properties of
$\Upsilon(7S)$ and $\Upsilon(8S)$ in the framework of the GI model. Considering the importance of a screening effect for higher bottomonia, we employ the modified GI model with a screening effect to study the nature of the two particles in this work. In Table \ref{spectrum}, we predict
the masses of $\Upsilon(7S)$ and $\Upsilon(8S)$ as 11157 MeV and 11296 MeV, which are raised by 154  and 293 MeV, respectively, compared with the mass value of $\Upsilon(11020)$ from Belle
\cite{Santel:2015qga}. We suggest future experiments to search for these two bottomonia in the vicinity of their corresponding energies mentioned above.
The decay information of partial widths and branching ratios of strong decay, annihilation decay, and radiative transition and the total widths for the $\Upsilon(7S)$
and $\Upsilon(8S)$ are shown in Tables \ref{decay27s} and \ref{decay28s}, respectively. The total width of $\Upsilon(7S)$ is estimated to be 101.4
MeV, which means a broad state. There are thirteen open-bottom modes, among which the important decay channels are $B^*B(1^3P_2)$, $BB^*$,
$B^*B^*$, $B^*B(1P_1^{\prime})$, $B^*B(1P_1)$, and $BB$ with the corresponding partial widths, 28.1, 22.0, 20.4, 9.26, 9.03, and 6.79 MeV, respectively. From Table \ref{decay27s},
we find that the combination of a vector meson $B^*$ and a $P$-wave bottom meson accounts for a large portion of the total width of $\Upsilon(7S)$. Contributions from
the $B+B(1P)$ and bottom-strange meson modes can be almost ignored. Some typical ratios of partial widths are given by
\begin{eqnarray}
\frac{\Gamma[\Upsilon(7^3S_1) \to BB^*]}{\Gamma[\Upsilon(7^3S_1) \to BB]} &=& 3.24,\nonumber\\
\frac{\Gamma[\Upsilon(7^3S_1) \to B^*B^*]}{\Gamma[\Upsilon(7^3S_1) \to BB]} &=& 3.00,\nonumber\\
\frac{\Gamma[\Upsilon(7^3S_1) \to B^*B(1^3P_2)]}{\Gamma[\Upsilon(7^3S_1) \to BB]} &=& 4.14,\nonumber\\
\frac{\Gamma[\Upsilon(7^3S_1) \to B^*B(1P_1^{\prime})]}{\Gamma[\Upsilon(7^3S_1) \to BB]} &=& 1.36,\nonumber\\
\frac{\Gamma[\Upsilon(7^3S_1) \to B^*B(1P_1)]}{\Gamma[\Upsilon(7^3S_1) \to BB]} &=& 1.33,
\end{eqnarray}
which can be tested by future experiments.

In Table \ref{decay28s}, the total width of $\Upsilon(8S)$ is predicted to be 66.6 MeV and the dominant decay channels are $B^*B^*$ and $BB^*$ with branching ratios 40.9 $\%$
and 31.0 $\%$, respectively. The $B^*B^*$ and $BB^*$ modes are excellent decay channels to detect the $\Upsilon(8S)$ bottomonium state in the future experiments. All the decay
modes $BB$, $BB(1P_1^{\prime})$, and $BB(1^3P_2)$ have a consistent partial width of about 4$-$5 MeV, which almost occupies the remaining contributions to the total width. It is difficult to experimentally observe the configuration of
bottom-strange mesons in the $\Upsilon(8S)$ decay.

\subsection{$P$-wave states}

From Fig. \ref{Fig1:spectrum}, the first above-threshold $P$-wave bottomonia are $4P$ states, which include spin-singlet $h_b(4P)$ and spin-triplet $\chi_{bJ}(4P)$ with
$J=0, 1, 2$ and exceed the $B\bar{B}$ threshold by about 200 MeV according to our estimates. These above-threshold particles have not yet been found by experiments. In Table \ref{spectrum}, the mass values of higher radial $P$-wave bottomonia $5P$ and $6P$ states are estimated to be about 10940 and 11100 MeV, respectively,
which are reduced by about 70 and 110 MeV compared to the predictions of the GI model, respectively.
In this subsection, we will give a theoretical analysis of their decay properties including higher radial $P$-wave bottomonia $5P$ and $6P$ states in the
framework of the modified GI model. We hope to provide useful information for the search for these bottomonia in future experiments.

The decay behaviors of $4P$, $5P$, and $6P$ bottomonia are given in Tables \ref{decay24p}-\ref{decay26pb} in succession. For the $4P$ bottomonium states and even higher bottomonia with higher radial quantum number
or higher orbital angular momentum, the screening effect begins to manifestly show its power in the description of corresponding decay properties because the theoretical
calculations of various decay processes are directly dependent on the mass value and the wave function of the related hadrons. Therefore, in the following discussion,
including the subsequent higher $D$, $F$, and $G$-wave states, we will perform a detailed comparison of the predicted decay properties with and without a screening effect.
In Ref. \cite{Godfrey:2015dia}, authors studied the $P$-wave bottomonia only up to $5P$ states by the GI model. Thus, we focus only on the comparison of $4P$ and $5P$ bottomonia,
and our predicted decay properties of $6P$ bottomonia will be illustrated at the end.

According to the numerical results in Tables \ref{decay24p}-\ref{decay25pb} and Ref. \cite{Godfrey:2015dia}, we conclude the followings.
\begin{enumerate}
\item{Compared with the predicted partial widths of the GI model, our values of radiative transitions $4P\to3D$ and $5P\to4D$ are 1.5$\sim$2 times smaller. Our calculated partial widths of the dominant radiative decays $4P\to(4,3)S$ and $5P\to(5,4)S$
are much smaller. Ours of the $5P$ states have an order of magnitude difference. In addition, the predictions of $4P\to2S$ and $5P\to3S$ from the GI model
are almost negligible compared with other radiative transition processes due to corresponding partial widths of $10^{-1} \sim 10^{-4}$ keV on the whole. However, our calculations indicate that there
have consistent numerical results of partial widths between $4P$ or $5P$ states transition to the $S$-wave ground state and the $S$-wave low-excited states.}

\item{For the $4P$ and $5P$ bottomonium states, open-bottom strong decay still is dominated and because the predicted mass of GI model is higher than that of us, so in their calculations some
higher modes of bottom or bottom-strange configuration are included. From the perspective of the total width, the most obvious difference is from the $\chi_{b0}$(4P) state,
whose predicted total widths are estimated to be 112.1 and 34.5 MeV by the modified GI model and GI model, respectively. For the other $4P$ states, the predictions of GI model are
less than about 10 $\sim$ 20 MeV compared with that of modified GI model. After that, the predicted total widths of GI model in the $5P$ bottomonium states are overall  higher than
about 5 $\sim$ 20 MeV except for the $h_b(5P)$ state since the contribution of $P$-wave bottom meson. Although the difference in total width is not conspicuous for most of the
$4P$ and $5P$ states, but the predicted dominant decay modes of $4P$ and $5P$ states from two different models are almost all opposite, which also illustrates that the influence of
the screening effect on the higher bottomonia is quite significant. Thus the present work should be of great value for revealing nature of bottomonium.}

\item{The total width of the $h_b(4P)$ state is predicted to be 48.1 MeV and the dominant decay mode is $B^*B^*$ with branching ratio 90.5 $\%$. The total widths of spin-triplet
$\chi_{bJ}(4P)$ states are estimated to be 112.1, 40.3 and 66.0 MeV corresponding to $J=1,2,3$, respectively. This shows the $\chi_{b0}(4P)$ is a broad state and other two states
are relatively narrow so that the triplet states of $4P$ bottomonium should be easily distinguished in the experiment. In addition, the dominant decay mode of three $\chi_{bJ}(4P)$
states are all $B^*B^*$ with branching ratio 92.8 $\%$, 83.1 $\%$ and 72.0 $\%$, respectively, so we suggest the future experiments to detect $4P$ bottomonium states by the
$B^*B^*$ mode.}

\item{The total width of the $h_b(5P)$ state is predicted to be 49.4 MeV and is identical with that of the $h_b(4P)$ state. The dominant decay channels of the $h_b(5P)$ state are
$BB^*$ and $B^*B^*$, and corresponding branching ratios are 75.7 $\%$ and 21.1 $\%$ respectively. We predict that the total width of $\chi_{bJ}(5P)$ is about 40 $\sim$ 50 MeV,
and decay mode $BB$ and $B^*B^*$ are critical for the $\chi_{b0}(5P)$ state and the dominant decay channels of $\chi_{b1}(5P)$ are $BB^*$ and $B^*B^*$, furthermore, for the
$\chi_{b2}(5P)$, there are three important decay modes $ BB^*$, $B^*B^*$ and $BB$ which all have a great contribution for its total width. The above conclusions could
be examined by experiment in the future.}
\end{enumerate}

The predicted average mass of $6P$ bottomonium states by modified GI model is 11099 MeV, which is about 80 $\sim$ 100 MeV above the experimental mass of the observed $\Upsilon(11020)$
state. Hence, more strong decay channels of $6P$ states are opened. From Table \ref{decay26pa} and \ref{decay26pb}, Our results indicate that the $h_b(6P)$ and $\chi_{bJ}(6P)$ all
are broad bottomonium mesons because of the predicted total width of about 107 $\sim$ 140 MeV. Additionally, $BB^*$, $B^*B^*$, and $B^*B(1^3P_2)$ just simultaneously are the main
decay channels for spin-singlet $h_b(6P)$ and triplet $\chi_{bJ}(6P)$ with $J=1,2$, and the sum of their contributions all are more than 70 $\%$ to the total decay width. The decay
modes $B^*B(1^3P_2)$, $B^*B^*$ and $BB$ are dominant for the $\chi_{b0}(6P)$ state with branching ratios of 28.8 $\%$, 25.7 $\%$ and 24.5 $\%$, respectively. A common decay
feature of $6P$ bottomonium states can clearly be seen, that is, the role of mode $B^*B^*$ and $B^*B(1^3P_2)$ are quite considerable.

\subsection{$D$-wave states}

In this subsection, we will discuss features of $D$-wave bottomonia. Among them, the $D$-wave vector bottomonium with $J^{PC}=1^{--}$ is quite interesting
because unlike charmonium system, there is no clear signal to
show the existence of $D$-wave vector states although many $S$-wave vector bottomonia have been discovered by experiments. Hence it is helpful for understanding this puzzle and further understanding the behaviors of bottomonia that we study
the intrinsic properties of these missing $\Upsilon_1(n^3D_1)$ states. There are only $1D$ and $2D$ bottomonia below the $B\bar{B}$ threshold.
The mass of $3D$ states is located at around 10680 MeV according to our estimates, which is 120 MeV larger than the threshold. We also predict that the masses of $4D$ and $5D$
bottomonia are about 10880 and 11050 MeV, respectively. In Table \ref{spectrum}, we notice that the predicted mass of $\Upsilon_1(4^3D_1)$ is 10871 MeV, which is very close to the measured
value of the $\Upsilon(10860)$ state \cite{Santel:2015qga}. However, the possibility of a such candidate for $\Upsilon(10860)$
can be basically excluded by the following analysis. Next, we discuss the decay properties of $D$-wave bottomonia up to $5D$ states in detail.

\begin{table}[htbp]
\renewcommand\arraystretch{1.4}
\caption{ Partial widths and branching ratios of OZI-allowed strong decay, annihilation decay, and radiative transition and total widths for $3^1D_2$ and $3^3D_1$
 bottomonium states. The width results are in units of keV.\label{decay23da}}
\begin{center}
{\tabcolsep0.15in
\begin{tabular}{cllrc}
\toprule[1pt]\toprule[1pt]
State         &  Channels & Width & $\mathcal{B}$(\%)  \\
\midrule[1pt]
$\eta_{b2}(3^1D_2)$ & $gg$                  & 1.59        & $1.11\times10^{-3}$   \\
                    & $h_{b}(1^1P_1)\gamma$ & 1.03 &    $7.20\times10^{-4}$  \\
                    & $h_{b}(2^1P_1)\gamma$ & 2.96 &    $2.07\times10^{-3}$  \\
                    & $h_{b}(3^1P_1)\gamma$ & 18.2 &     $1.27\times10^{-2}$ \\
                    & $h_{b3}(1^1F_3)\gamma$ & 0.0116 &   $8.11\times10^{-6}$   \\
                    & $h_{b3}(2^1F_3)\gamma$ & 2.01 &     $1.41\times10^{-3}$ \\
                    & $\Upsilon_1(1^3D_1)\gamma$ & $9.97\times10^{-4}$ &  $6.97\times10^{-7}$    \\
\end{tabular}
}
\end{center}
\end{table}

\begin{table}[htbp]
\renewcommand\arraystretch{1.4}
\caption{ Partial widths and branching ratios of OZI-allowed strong decay, annihilation decay, and radiative transition and total widths for $3^3D_2$ and $3^3D_3$
 bottomonium states. The width results are in units of keV.\label{decay23db}}
\begin{center}
{\tabcolsep0.15in
\begin{tabular}{cllrc}
\toprule[1pt]\toprule[1pt]
State         &  Channels & Width & $\mathcal{B}$(\%)  \\
\midrule[1pt]
$\Upsilon_2(3^3D_2)$ & $ggg$                  &  2.27       &  $2.36\times10^{-3}$  \\
                     & $\chi_{b1}(1^3P_1)\gamma$ & 0.903 &    $9.38\times10^{-4}$  \\
                     & $\chi_{b2}(1^3P_2)\gamma$ & 0.122 &    $1.27\times10^{-4}$  \\
                     & $\chi_{b1}(2^3P_1)\gamma$ & 2.44 &     $2.53\times10^{-3}$ \\
                     & $\chi_{b2}(2^3P_2)\gamma$ & 0.467 &    $4.85\times10^{-4}$  \\
                     & $\chi_{b1}(3^3P_1)\gamma$ & 14.0 &     $1.45\times10^{-2}$ \\
                     & $\chi_{b2}(3^3P_2)\gamma$ & 4.23 &     $4.39\times10^{-3}$ \\
\end{tabular}
}
\end{center}
\end{table}

\begin{table}[htbp]
\renewcommand\arraystretch{1.4}
\caption{ Partial widths and branching ratios of OZI-allowed strong decay, annihilation decay, and radiative transition and total widths for $4^1D_2$ and $4^3D_1$
 bottomonium states. The width results are in units of keV.\label{decay24da}}
\begin{center}
{\tabcolsep0.1in
%

\begin{table}[htbp]
\renewcommand\arraystretch{1.122}
\caption{ Partial widths and branching ratios of OZI-allowed strong decay, annihilation decay, and radiative transition and total widths for $5^1D_2$ and $5^3D_1$
 bottomonium states. The width results are in units of keV.\label{decay25da}}
\begin{center}
{\tabcolsep0.13in
\begin{tabular}{cllrc}
\toprule[1pt]\toprule[1pt]
State         &  Channels & Width & $\mathcal{B}$(\%)  \\
\midrule[1pt]
$\eta_{b2}(5^1D_2)$ & $gg$                  &   2.13 &   $1.92\times10^{-3}$ \\
                    & $h_{b}(1^1P_1)\gamma$ & 0.373 &    $3.37\times10^{-4}$  \\
                    & $h_{b}(2^1P_1)\gamma$ & 0.682 &    $6.16\times10^{-4}$  \\
                    & $h_{b}(3^1P_1)\gamma$ & 1.13 &     $1.02\times10^{-3}$ \\
                    & $h_{b}(4^1P_1)\gamma$ & 2.23 &     $2.01\times10^{-3}$ \\
                    & $h_{b}(5^1P_1)\gamma$ & 10.2 &     $9.21\times10^{-3}$ \\
                    & $h_{b3}(2^1F_3)\gamma$ & 0.0202 &  $1.82\times10^{-5}$    \\
\end{tabular}
}
\end{center}
\end{table}

\begin{table}[htbp]
\renewcommand\arraystretch{1.19}
\caption{ Partial widths and branching ratios of the OZI-allowed strong decay, annihilation decay, and radiative transition and total widths for $5^3D_2$ and $5^3D_3$
 bottomonium states. The width results are in units of keV.\label{decay25db}}
\begin{center}
{\tabcolsep0.11in
\begin{tabular}{cllrc}
\toprule[1pt]\toprule[1pt]
State         &  Channels & Width & $\mathcal{B}$(\%)  \\
\midrule[1pt]
$\Upsilon_2(5^3D_2)$ & $ggg$                  &   3.23   &  $3.18\times10^{-3}$  \\
                     & $\chi_{b1}(1^3P_1)\gamma$ & 0.343 &  $3.38\times10^{-4}$    \\
                     & $\chi_{b2}(1^3P_2)\gamma$ & 0.0195 & $1.92\times10^{-5}$     \\
                     & $\chi_{b1}(2^3P_1)\gamma$ & 0.604 &  $5.94\times10^{-4}$    \\
                     & $\chi_{b2}(2^3P_2)\gamma$ & 0.0574 & $5.65\times10^{-5}$     \\
                     & $\chi_{b1}(3^3P_1)\gamma$ & 0.951 &  $9.36\times10^{-4}$    \\
                     & $\chi_{b2}(3^3P_2)\gamma$ & 0.142 &  $1.40\times10^{-4}$    \\
\end{tabular}
}
\end{center}
\end{table}

\begin{table}[htbp]
\renewcommand\arraystretch{1.4}
\caption{ Partial widths and branching ratios of the OZI-allowed strong decay, annihilation decay, and radiative transition and total widths for the $2F$ bottomonium states.
The width results are in units of keV.\label{decay22f}}
\begin{center}
{\tabcolsep0.1in
\begin{tabular}{cllrc}
\toprule[1pt]\toprule[1pt]
State         &  Channels & Width & $\mathcal{B}$(\%)  \\
\midrule[1pt]
$h_{b3}(2^1F_3)$ & $\eta_{b2}(1^1D_2)\gamma$ & 1.99 & 0.482     \\
                 & $\eta_{b2}(2^1D_2)\gamma$ & 17.4 &  4.21    \\
                 & $\eta_{b4}(1^1G_4)\gamma$ & 1.06 &   0.257   \\
                 & $\chi_{b2}(1^3F_2)\gamma$ & $3.70\times10^{-4}$ & $8.96\times10^{-5}$     \\
                 & $\chi_{b3}(1^3F_3)\gamma$ & $1.16\times10^{-4}$ &  $2.81\times10^{-5}$    \\
                 &$BB^*$      & 0.393 MeV &   95.2   \\
               &Total      & 0.413 MeV & 100               \\
\end{tabular}
}
\end{center}
\end{table}

In Tables \ref{decay23da}-\ref{decay24db}, we list the numerical results of $3D$ and $4D$ bottomonium  decays. Comparing ours with the calculation of Ref. \cite{Godfrey:2015dia}, we can conclude
\begin{enumerate}
\item{Compared with the GI model \cite{Godfrey:2015dia}, almost all of our estimated partial widths of radiative transition for $3D$ and $4D$ bottomonia become smaller except for
the electromagnetic processes of $3D\to1P$ and $4D\to2P$, which become larger from 20 $\%$ to 300 $\%$ range.}

\item{Similar to the situation of $\chi_{b0}(4P)$, our prediction of the total width for the $\Upsilon_1(3D)$ is 54.1 MeV, which is largely different from an
estimate of 103.6 MeV by the GI model. Overall, the $3D$ bottomonia are broad resonances, where the predicted total widths of other
three particles $\eta_{b2}(3D)$, $\Upsilon_2(3D)$ and $\Upsilon_3(3D)$ are 143.0, 96.3, and 223.8 MeV, respectively. Our results of $3D$ and $4D$ states on the partial widths and branching ratios of strong decay channels are still quite different from those of the GI model. The
strong decay modes of $\eta_{b2}(3D)$ are only $B^*B^*$ and $BB^*$ with close predicted ratios, which accounts for almost all the contributions to the total width. For the
$\Upsilon_1(3D)$, the dominant decay mode is $B^*B^*$ with a branching ratio 61.8 $\%$. The mode $BB$ with 10.1 $\%$ may not be the best process to
experimentally search for $\Upsilon_1(3D)$. Additionally, its branching ratio of annihilating to the leptonic pair $\ell^+\ell^-$ is three orders of magnitude smaller than $\Upsilon(4S)$ state.
Hence, it is hard to detect this $D$-wave vector bottomonium in the final state events of $\mu^+\mu^-$. This situation is also applied to the $\Upsilon_1(4D)$ and
$\Upsilon_1(5D)$ states. Finally, one can easily find that the modes $B^*B^*$ and $BB^*$ are dominant both for the $\Upsilon_2(3D)$ and $\Upsilon_3(3D)$ at the same time.}

\item{The total widths of $4D$ bottomonia including the spin singlet $\eta_{b2}(4D)$ and three triplets $\Upsilon_J(4D)$ are estimated to be about 70 $\sim$ 90 MeV. Our
results are generally 10 $\sim$ 20 MeV larger than those of the GI model. The $\eta_{b2}(4D)$ mainly decays into $BB^*$ and $B^*B^*$ with the branching ratios
63.2 $\%$ and 33.4 $\%$, respectively. The corresponding partner $\Upsilon_2 (4D)$ has the similar decay behavior. Although the mass of $\Upsilon(10860)$ is compatible with our
prediction of $\Upsilon_1 (4^3D_1)$ state in Table \ref{spectrum}, we can safely rule out this allocation since the two orders of magnitude difference on the branching ratios of the leptonic pair decay $\ell^+\ell^-$ between the theoretical and
measured data. We predict that the main decay channels of $\Upsilon_1 (4D)$ are $B^*B^*$, $BB$ and $BB^*$ with corresponding
partial widths 42.1, 27.4, and 15.1 MeV, respectively. We hope this is validated in future experiments. For the $\Upsilon_3 (4D)$ state, $B^*B^*$ and $BB^*$ are
dominant decay modes. Furthermore, they have a completely consistent  branching ratio 43 $\%$, and the mode $BB$ also has about 10 $\%$ contributions to the total width.}

\item{Compared with the $S$-wave $\Upsilon$'s, the $D$-wave  bottomonia behave more like quite broad states and the leptonic annihilation widths are too small as
mentioned before. Hence, this can also explain why the $S$-wave $\Upsilon$'s are experimentally found in succession but the $D$-wave $\Upsilon_1$'s always have no movement in experiments.}
\end{enumerate}

The detailed decay information of $5D$ bottomonium states are presented in Table \ref{decay25da} and \ref{decay25db}. Our results indicate that the spin-singlet $\eta_{b2}(5D)$ and triplet
$\Upsilon_J(5D)$ with $J=1,2,3$ are all broad states, whose predicted total widths are 110.7, 121.7, 101.6, and 86.0 MeV, respectively. We also notice that
the dominant decay channels of $\eta_{b2}(5^1D_1)$, $\Upsilon_2(5^3D_2)$ and  $\Upsilon_3(5^3D_3)$ are all $B^*B^*$ and $BB^*$ and other relatively important
decay modes are provided by the strong decay channels containing $P$-wave bottom mesons. It is very interesting to study the properties of $\Upsilon_1(5^3D_1)$ because
the mass of $\Upsilon_1(5D)$ is only 40 MeV larger than that of $\Upsilon(6S)$ according to our prediction in Table \ref{spectrum}, which is nearly 100 MeV smaller than that of the GI model.
Our numerical results show that the $\Upsilon_1(5^3D_1)$ is a
broad state, for which there are ten open-bottom decay modes. Furthermore, its main decay channels are $B^*B^*$, $BB$, $BB^*$ , $BB(1P_1^{\prime})$ and $BB(1^3P_2)$, and the
corresponding branching ratios are 38.7 $\%$, 16.4 $\%$, 15.8 $\%$, 14.8 $\%$, and 7.59 $\%$, respectively. The contributions of bottom-strange mesons are too low
to exceed 1 $\%$. Finally, we hope that these results can provide valuable clues for future experiments to search for more $D$-wave bottomonia.


\subsection{$F$-wave states}

In the following, we will focus on the higher $F$-wave bottomonia, i.e., $2F$, $3F$, and $4F$ bottomonia. The theoretical mass values of $F$-wave bottomonia are presented in
Table \ref{spectrum}. Their average mass values are the same as those of the spin-singlet states of 10609, 10812, and 10988 MeV with radial quantum number $n=2, 3, 4$, respectively.
From Fig. \ref{Fig1:spectrum}, it is also easily to find that the mass values of these particles are close to the $S$-wave bottomonium states corresponding to radial quantum number
$n=4, 5, 6$, respectively.

\begin{table}[htbp]
\renewcommand\arraystretch{1.4}
\caption{ Partial widths and branching ratios of the OZI-allowed strong decay, annihilation decay, and radiative transition and the total widths for $3^1F_3$ and $3^3F_2$
 bottomonium states. The width results are in units of keV.\label{decay23fa}}
\begin{center}
{\tabcolsep0.1in
\begin{tabular}{cllrc}
\toprule[1pt]\toprule[1pt]
State         &  Channels & Width & $\mathcal{B}$(\%)  \\
\midrule[1pt]
$h_{b3}(3^1F_3)$ & $\eta_{b2}(1^1D_2)\gamma$ & 0.530 &  $5.14\times10^{-4}$    \\
                 & $\eta_{b2}(2^1D_2)\gamma$ & 2.20 &   $2.13\times10^{-3}$   \\
                 & $\eta_{b2}(3^1D_2)\gamma$ & 14.2 &   $1.38\times10^{-2}$   \\
                 & $\eta_{b4}(2^1G_4)\gamma$ & 1.59 &    $1.54\times10^{-3}$  \\
                 & $\chi_{b2}(2^3F_2)\gamma$ & $3.41\times10^{-4}$ &   $3.30\times10^{-7}$   \\
                 & $\chi_{b3}(2^3F_3)\gamma$ & $1.14\times10^{-4}$ &   $1.10\times10^{-7}$   \\
                 &$BB^*$      & 52.7 MeV & 51.0     \\
                 & $B^*B^*$   & 49.2 MeV & 47.6     \\
                 & $B_sB_s^*$   & 1.35 MeV & 1.31    \\
               &Total      & 103.2 MeV & 100               \\

$\chi_{b2}(3^3F_2)$ & $gg  $                     &  3.17        &   $2.20\times10^{-3}$   \\
                    & $\Upsilon_1(1^3D_1)\gamma$ &  0.540 &     $3.75\times10^{-4}$ \\
                    & $\Upsilon_2(1^3D_2)\gamma$ & 0.0468 &     $3.25\times10^{-5}$ \\
                    & $\Upsilon_3(1^3D_3)\gamma$ & $4.52\times10^{-4}$ &   $3.14\times10^{-7}$   \\
                    & $\Upsilon_1(2^3D_1)\gamma$ &  2.14 &            $1.49\times10^{-3}$   \\
                    & $\Upsilon_2(2^3D_2)\gamma$ & 0.260 &      $1.80\times10^{-4}$ \\
                    & $\Upsilon_3(2^3D_3)\gamma$ & 0.00462 &    $3.21\times10^{-6}$  \\
                    & $\Upsilon_1(3^3D_1)\gamma$ &  12.3 &      $8.54\times10^{-3}$ \\
                    & $\Upsilon_2(3^3D_2)\gamma$ & 2.12 &       $1.47\times10^{-3}$ \\
                    & $\Upsilon_3(3^3D_3)\gamma$ & 0.0569 &     $3.95\times10^{-5}$ \\
                    & $\Upsilon_3(2^3G_3)\gamma$ & 1.53 &       $1.06\times10^{-3}$ \\
                    & $h_{b3}(1^1F_3) \gamma$ &  $3.47\times10^{-5}$ & $2.41\times10^{-8}$     \\
                    & $BB$      & 28.5 MeV &19.8    \\
                   &$BB^*$      & 32.5  MeV &22.5     \\
                  & $B^*B^*$   & 81.1 MeV & 56.3    \\
                  & $B_sB_s$   & 0.898 MeV & 0.623     \\
                  & $B_sB_s^*$   & 1.08 MeV & 0.753    \\
               &Total      & 144.1 MeV & 100        \\

\bottomrule[1pt]\bottomrule[1pt]
\end{tabular}
}
\end{center}
\end{table}

\begin{table}[htbp]
\renewcommand\arraystretch{1.4}
\caption{ Partial widths and branching ratios of OZI-allowed strong decay, annihilation decay, and radiative transition and the total widths for $3^3F_3$ and $3^3F_4$
 bottomonium states. The width results are in units of keV.\label{decay23fb}}
\begin{center}
{\tabcolsep0.1in
\begin{tabular}{cllrc}
\toprule[1pt]\toprule[1pt]
State         &  Channels & Width & $\mathcal{B}$(\%)  \\
\midrule[1pt]
$\chi_{b3}(3^3F_3)$ & $gg  $                     &  0.270   &    $2.32\times10^{-4}$  \\
                    & $\Upsilon_2(1^3D_2)\gamma$ & 0.494 &     $4.24\times10^{-4}$ \\
                    & $\Upsilon_3(1^3D_3)\gamma$ & 0.0289 &    $2.48\times10^{-5}$  \\
                    & $\Upsilon_2(2^3D_2)\gamma$ & 2.01 &     $1.72\times10^{-3}$ \\
                    & $\Upsilon_3(2^3D_3)\gamma$ & 0.169 &    $1.45\times10^{-4}$  \\
                    & $\Upsilon_2(3^3D_2)\gamma$ & 12.5 &     $1.07\times10^{-2}$ \\
                    & $\Upsilon_3(3^3D_3)\gamma$ & 1.48 &     $1.27\times10^{-3}$ \\
                    & $\Upsilon_3(2^3G_3)\gamma$ & 0.105 &    $9.01\times10^{-5}$  \\
                    & $\Upsilon_4(2^3G_4)\gamma$ & 1.49 &     $1.28\times10^{-3}$ \\
                    & $h_{b3}(2^1F_3) \gamma$ &  $7.35\times10^{-5}$ &  $6.30\times10^{-8}$    \\
                   & $BB^*$      & 60.6 MeV &51.9     \\
                 & $B^*B^*$   & 54.3 MeV & 46.5     \\
                 & $B_sB_s^*$   &1.78 MeV & 1.5    \\
               &Total      & 116.6 MeV & 100               \\

$\chi_{b4}(3^3F_4)$ & $gg  $                     &   0.210       & $3.13\times10^{-4}$     \\
                    & $\Upsilon_3(1^3D_3)\gamma$ & 0.487 &     $7.27\times10^{-4}$ \\
                    & $\Upsilon_3(2^3D_3)\gamma$ & 2.05 &      $3.06\times10^{-3}$\\
                    & $\Upsilon_3(3^3D_3)\gamma$ & 13.9 &      $2.07\times10^{-2}$\\
                    & $\Upsilon_3(2^3G_3)\gamma$ & 0.00141 &    $2.10\times10^{-6}$  \\
                    & $\Upsilon_4(2^3G_4)\gamma$ & 0.0848 &     $1.27\times10^{-4}$ \\
                    & $\Upsilon_5(2^3G_5)\gamma$ & 1.64 &      $2.45\times10^{-3}$\\
                    & $h_{b3}(1^1F_3) \gamma$ & $3.17\times10^{-4}$ &  $4.73\times10^{-7}$    \\
                   & $BB$        & 0.426 MeV &0.636    \\
                  & $BB^*$        & 14.3 MeV &21.3    \\
                  & $B^*B^*$   & 51.9 MeV & 77.5     \\
                  & $B_sB_s$   & 0.365 MeV & 0.545   \\
                  & $B_sB_s^*$   & 0.0605 MeV & $9.03\times10^{-2}$   \\
               &Total      & 67.0 MeV & 100        \\

\bottomrule[1pt]\bottomrule[1pt]
\end{tabular}
}
\end{center}
\end{table}

\begin{table}[htbp]
\renewcommand\arraystretch{1.40}
\caption{ Partial widths and branching ratios of OZI-allowed strong decay, annihilation decay, and radiative transition and the total widths for $4^1F_3$ and $4^3F_2$
 bottomonium states. The width results are in units of keV.\label{decay24fa}}
\begin{center}
{\tabcolsep0.1in
\begin{tabular}{cllrc}
\toprule[1pt]\toprule[1pt]
State         &  Channels & Width & $\mathcal{B}$(\%)  \\
\midrule[1pt]
$h_{b3}(4^1F_3)$ & $\eta_{b2}(1^1D_2)\gamma$ & 0.238 &    $3.51\times10^{-4}$  \\
                 & $\eta_{b2}(2^1D_2)\gamma$ & 0.701 &    $1.03\times10^{-3}$  \\
                 & $\eta_{b2}(3^1D_2)\gamma$ & 2.09 &     $3.08\times10^{-3}$ \\
                 & $\eta_{b2}(4^1D_2)\gamma$ & 12.2 &    $1.80\times10^{-2}$  \\
                 & $\eta_{b4}(3^1G_4)\gamma$ & 2.00 &     $2.95\times10^{-3}$ \\
                 & $\chi_{b2}(2^3F_2)\gamma$ & $3.67\times10^{-4}$ & $5.41\times10^{-7}$     \\
                 & $\chi_{b3}(2^3F_3)\gamma$ & $9.41\times10^{-5}$ &  $1.39\times10^{-7}$    \\
                 & $\chi_{b3}(2^3F_4)\gamma$ & $5.96\times10^{-6}$ &   $8.79\times10^{-9}$   \\
                 &$BB^*$      & 29.3 MeV &43.2     \\
                 & $BB(1^3P_0)$  & 0.0617 MeV & $9.10\times10^{-2}$     \\
                 & $B^*B^*$   & 37.6 MeV &55.5     \\
                 & $B_sB_s^*$   & 0.235 MeV & 0.346    \\
                 & $B_s^*B_s^*$   & 0.548 MeV & 0.808    \\
               &Total      & 67.8 MeV & 100               \\
$\chi_{b2}(4^3F_2)$ & $\Upsilon_1(1^3D_1)\gamma$ &  0.283 &  $3.62\times10^{-4}$    \\
                    & $\Upsilon_2(1^3D_2)\gamma$ & 0.0183 &    $2.34\times10^{-5}$  \\
                    & $\Upsilon_3(1^3D_3)\gamma$ & $7.59\times10^{-5}$ &  $9.71\times10^{-8}$    \\
                    & $\Upsilon_1(2^3D_1)\gamma$ &  0.708 &            $9.05\times10^{-4}$   \\
                    & $\Upsilon_2(2^3D_2)\gamma$ & 0.0675 &           $8.63\times10^{-5}$\\
                    & $\Upsilon_3(2^3D_3)\gamma$ & $7.79\times10^{-4}$ &   $9.96\times10^{-7}$   \\
                    & $\Upsilon_1(3^3D_1)\gamma$ & 1.99 &                $2.54\times10^{-3}$  \\
                    & $\Upsilon_2(3^3D_2)\gamma$ & 0.253 &                $3.24\times10^{-4}$\\
                    & $\Upsilon_3(3^3D_3)\gamma$ & 0.00471 &             $6.02\times10^{-6}$\\
                    & $\Upsilon_1(4^3D_1)\gamma$ &  10.4 &             $1.33\times10^{-2}$ \\
                    & $\Upsilon_2(4^3D_2)\gamma$ & 1.81 &             $2.31\times10^{-3}$\\
                    & $\Upsilon_3(4^3D_3)\gamma$ & 0.0490 &           $6.27\times10^{-5}$\\
                    & $\Upsilon_3(3^3G_3)\gamma$ & 1.82 &             $2.33\times10^{-3}$\\
                    & $h_{b3}(1^1F_3) \gamma$ &  $4.94\times10^{-5}$ &  $6.32\times10^{-8}$    \\
                   & $BB$      & 13.7 MeV &17.5    \\
                   &$BB^*$      & 24.7  MeV &31.6     \\
                  & $B^*B^*$   & 39.3 MeV & 50.2     \\
                  & $B_sB_s$   & 0.0849 MeV & 0.109    \\
                  & $B_sB_s^*$   & 0.0947 MeV & 0.121    \\
                  & $B_s^*B_s^*$   & 0.373 MeV & 0.477    \\
               &Total      & 78.2 MeV & 100        \\

\bottomrule[1pt]\bottomrule[1pt]
\end{tabular}
}
\end{center}
\end{table}

\begin{table}[htbp]
\renewcommand\arraystretch{1.4}
\caption{ Partial widths and branching ratios of OZI-allowed strong decay, annihilation decay, and radiative transition and the total widths for $4^3F_3$ and $4^3F_4$
 bottomonium states. The width results are in units of keV.\label{decay24fb}}
\begin{center}
{\tabcolsep0.1in
\begin{tabular}{cllrc}
\toprule[1pt]\toprule[1pt]
State         &  Channels & Width & $\mathcal{B}$(\%)  \\
\midrule[1pt]
$\chi_{b3}(4^3F_3)$ & $\Upsilon_2(1^3D_2)\gamma$ & 0.232 &    $3.28\times10^{-4}$  \\
                    & $\Upsilon_3(1^3D_3)\gamma$ & 0.00934 &   $1.32\times10^{-5}$    \\
                    & $\Upsilon_2(2^3D_2)\gamma$ & 0.649 &     $9.17\times10^{-4}$ \\
                    & $\Upsilon_3(2^3D_3)\gamma$ & 0.0417 &   $5.89\times10^{-5}$   \\
                    & $\Upsilon_2(3^3D_2)\gamma$ & 1.90 &    $2.68\times10^{-3}$  \\
                    & $\Upsilon_3(3^3D_3)\gamma$ & 0.166 &    $2.34\times10^{-4}$  \\
                    & $\Upsilon_2(4^3D_2)\gamma$ & 10.8 &     $1.53\times10^{-2}$ \\
                    & $\Upsilon_3(4^3D_3)\gamma$ & 1.29 &     $1.82\times10^{-3}$ \\
                    & $\Upsilon_3(3^3G_3)\gamma$ & 0.127 &     $1.79\times10^{-4}$ \\
                    & $\Upsilon_4(3^3G_4)\gamma$ & 1.88 &     $2.66\times10^{-3}$ \\
                    & $h_{b3}(3^1F_3) \gamma$ &  $6.49\times10^{-5}$ &   $9.17\times10^{-8}$   \\
                  &$BB^*$      & 36.9 MeV &52.1     \\
                 & $BB(1^3P_0)$  & $1.11\times10^{-4}$ MeV & $1.57\times10^{-4}$    \\
                 & $B^*B^*$   & 33.3 MeV &47.0     \\
                 & $B_sB_s^*$   & 0.214 MeV & 0.303    \\
                 & $B_s^*B_s^*$   & 0.400 MeV & 0.566    \\
               &Total      & 70.8 MeV & 100               \\
$\chi_{b4}(4^3F_4)$ & $\Upsilon_3(1^3D_3)\gamma$ &  0.202 &   $3.34\times10^{-4}$    \\
                    & $\Upsilon_3(2^3D_3)\gamma$ & 0.637 &    $1.05\times10^{-3}$   \\
                    & $\Upsilon_3(3^3D_3)\gamma$ & 1.94 &     $3.21\times10^{-3}$  \\
                    & $\Upsilon_3(4^3D_3)\gamma$ & 11.8 &     $1.95\times10^{-2}$  \\
                    & $\Upsilon_3(3^3G_3)\gamma$ & 0.00166 &  $2.74\times10^{-6}$     \\
                    & $\Upsilon_4(3^3G_4)\gamma$ & 0.103 &    $1.70\times10^{-4}$   \\
                    & $\Upsilon_5(3^3G_5)\gamma$ & 1.89 &     $3.12\times10^{-3}$  \\
                    & $h_{b3}(2^1F_3) \gamma$ &  $3.65\times10^{-4}$ &    $6.03\times10^{-7}$   \\
                   & $BB$        & 0.00769 MeV &$1.27\times10^{-2}$    \\
                  & $BB^*$        & 3.33 MeV &5.5    \\
                  & $B^*B^*$   & 55.7 MeV & 92.1     \\
                  & $B_sB_s$   & 0.272 MeV & 0.449   \\
                  & $B_sB_s^*$   & 0.192 MeV & 0.317   \\
                  & $B_s^*B_s^*$   &0.960 MeV & 1.59    \\
               &Total      & 60.5 MeV & 100        \\

\bottomrule[1pt]\bottomrule[1pt]
\end{tabular}
}
\end{center}
\end{table}

In Tables \ref{decay22f}-\ref{decay23fb}, we present the numerical decay results of $2F$ and $3F$ bottomonia, which are also studied by Ref. \cite{Godfrey:2015dia}
in the framework of the GI model. By analyzing and comparing our calculation with the GI model \cite{Godfrey:2015dia}, we conclude
\begin{enumerate}
\item{The radiative transitions of $F$-wave states also have a rule very similar to $P$-wave and $D$-wave bottomonia, which can be seen by comparing the predictions of the
 modified GI model and GI model. That is, our partial widths of radiative decays are smaller than those of the GI model on the whole except for the radiative processes $2F\to1D$ and $3F\to2D$.}

\item{$2F$ bottomonia are states near the threshold, and their corresponding open-bottom channels are either $BB$ or $BB^*$, whose partial widths from the GI model are much larger  than our predictions. We predict that the total widths of $h_{b3}(2F)$ and $\chi_{b3}(2F)$ states are 0.413 and 0.543 MeV, respectively, which are about one-30th times the results of the GI model. The total widths of the other two particles $\chi_{b2}(2F)$ and $\chi_{b4}(2F)$ are calculated as 41.4 and 1.03 MeV, respectively. This result is about 2 to 3 times smaller than those of the GI model. From the above results, it can be easily noticed that the $\chi_{b2}(2F)$ is a relatively broad state but the other $2F$ states are all narrow states, which means that it is not difficult to distinguish the spin-triplet with $J=2$ of $2F$ bottomonia from the experiment. Some important radiative decay modes of these narrow states may also have large contributions to the total width. Hence, the processes $ h_{b3}(2F)\rightarrow \eta_{b2}(2D)\gamma$, $\chi_{b3}(2F) \rightarrow \Upsilon_2(2D)\gamma$, and $\chi_{b4}(2F) \rightarrow \Upsilon_3(2D)\gamma$ are possibly one of promising ways to detect these narrow $2F$ bottomonia.}

\item{Different from the decay behaviors of $2F$ bottomonia, the total widths of $h_{b3}(3F)$ and $\chi_{bJ}(3F)$ with $J=2,3,4$ states are estimated to be 103.2, 144.1, 116.6, and 67.0 MeV, respectively, from the modified GI model. It is apparent that $3F$ bottomonia are broad on the whole, which is consistent with the GI model. The dominant decay modes of four $3F$ bottomonia are $B^*B^*$ and $BB^*$ and their contributions to the total width reach 80 $ \%$ to 98 $\%$. The decay mode $BB$ should also be important for the $\chi_{b2}(3^3F_2)$ state, where our branching ratio of $BB$ is 19.8 $\%$ while the GI model gives only 7.85 $\%$. Finally, for the $\chi_{b4}(3^3F_4)$ state we need to emphasize that the mode $BB^*$ is important and $B_sB_s^*$ is negligible because of the predicted branching ratios of 21.3 $\%$ and $9.03\times 10^{-4}$, respectively. However, the prediction of GI model is completely inconsistent with our result, which is to be identified in future experiments.}
\end{enumerate}

The numerical results of $4F$ bottomonia decays are shown in Tables \ref{decay24fa} and \ref{decay24fa} and the predicted total widths of $4F$ states are located at near 70 MeV. In their open-bottom decay channels, the contributions from bottom-strange mesons are all small, among which the largest is not more than 3 $\%$. The dominant decay modes of $h_{b3}(4F)$ and $\chi_{bJ}(4F)$ with $J=2,3$ are $B^*B^*$ and $BB^*$, whose the sum of branching ratios is more than 80 $\%$. Only the mode $B^*B^*$ with branching ratio 92.1 $\%$ governs the $\chi_{b4}(4^3F_4)$ state. The $BB$ channel is also important for the $\chi_{b2}(4^3F_2)$ state but is negligible for the $\chi_{b4}(4^3F_4)$ state on account of the branching ratios 17.5 $\%$ and $1.27\times10^{-4}$, respectively.


\subsection{$G$-wave states}

\begin{table}[htbp]
\renewcommand\arraystretch{1.4}
\caption{ Partial widths and branching ratios of the OZI-allowed strong decay, and radiative transition and the total widths for the $2G$ bottomonium states.
The width results are in units of keV.\label{decay22g}}
\begin{center}
{\tabcolsep0.1in
\begin{tabular}{cllrc}
\toprule[1pt]\toprule[1pt]
State         &  Channels & Width & $\mathcal{B}$(\%)  \\
\midrule[1pt]
$\eta_{b4}(2^1G_4)$ & $h_{b3}(1^1F_3) \gamma$ &  1.44 &    $1.01\times10^{-3}$  \\
                    & $h_{b3}(2^1F_3) \gamma$ &  17.1 &    $1.19\times10^{-2}$  \\
                    & $\Upsilon_3(1^3G_3)\gamma$ & $1.70\times10^{-4}$ &   $1.19\times10^{-7}$   \\
                    & $\Upsilon_4(1^3G_4)\gamma$ & $6.57\times10^{-5}$ &   $4.59\times10^{-8}$   \\
                    & $\Upsilon_5(1^3G_5)\gamma$ & $5.19\times10^{-6}$ &   $3.63\times10^{-9}$   \\
                   &$BB^*$      & 77.3 MeV &54.0     \\
                 & $B^*B^*$   & 65.8 MeV    &46.0     \\
               &Total      & 143.1 MeV & 100               \\
$\Upsilon_3(2^3G_3)$ & $\chi_{b2}(1^3F_2)\gamma$ & 1.43 &   $1.29\times10^{-3}$   \\
                     & $\chi_{b3}(1^3F_3)\gamma$ & 0.0874 &  $7.89\times10^{-5}$    \\
                     & $\chi_{b4}(1^3F_4)\gamma$ & $9.33\times10^{-4}$ &  $8.42\times10^{-7}$    \\
                     & $\chi_{b2}(2^3F_2)\gamma$ & 16.0 &     $1.44\times10^{-2}$ \\
                     & $\chi_{b3}(2^3F_3)\gamma$ & 1.33 &      $1.20\times10^{-3}$\\
                     & $\chi_{b4}(2^3F_4)\gamma$ & 0.0203 &    $1.83\times10^{-5}$  \\
                     & $\eta_{b4}(1^1G_4)\gamma$ & $7.15\times10^{-7}$ &  $6.45\times10^{-10}$    \\
                     & $BB$      & 19.4 MeV &17.5    \\
                   &$BB^*$      & 56.4  MeV &50.9     \\
                  & $B^*B^*$   & 35.0 MeV & 31.6     \\
                  & $B_sB_s$   & 0.00886 MeV & $8.00\times10^{-3}$    \\
               &Total      & 110.8 MeV & 100        \\
$\Upsilon_4(2^3G_4)$ & $\chi_{b3}(1^3F_3)\gamma$ & 1.37 &  $9.82\times10^{-4}$    \\
                     & $\chi_{b4}(1^3F_4)\gamma$ & 0.0651 &  $4.67\times10^{-5}$    \\
                     & $\chi_{b3}(2^3F_3)\gamma$ & 16.0 &    $1.15\times10^{-2}$  \\
                     & $\chi_{b4}(2^3F_4)\gamma$ & 1.03 &    $7.38\times10^{-4}$  \\
                     & $\eta_{b4}(1^1G_4)\gamma$ & $5.62\times10^{-5}$ &   $4.03\times10^{-8}$   \\
                    &$BB^*$      & 87.3 MeV &62.5     \\
                 & $B^*B^*$   & 52.3 MeV &37.5     \\
               &Total      & 139.5 MeV & 100               \\
$\Upsilon_5(2^3G_5)$ & $\chi_{b4}(1^3F_4)\gamma$ & 1.37 &   $8.59\times10^{-4}$   \\
                     & $\chi_{b4}(2^3F_4)\gamma$ & 16.6 &   $1.04\times10^{-2}$   \\
                     & $\eta_{b4}(1^1G_4)\gamma$ & $1.87\times10^{-4}$ &  $1.17\times10^{-7}$    \\
                    & $BB$        & 15.7 MeV &9.87    \\
                  & $BB^*$        & 21.3 MeV &13.4    \\
                  & $B^*B^*$   & 122 MeV & 76.8     \\
                  & $B_sB_s$   & $1.50\times10^{-5}$ MeV & $9.38\times10^{-6}$   \\
               &Total      & 159.4 MeV & 100        \\

\bottomrule[1pt]\bottomrule[1pt]
\end{tabular}
}
\end{center}
\end{table}

\begin{table}[htbp]
\renewcommand\arraystretch{1.4}
\caption{ Partial widths and branching ratios of the OZI-allowed strong decay, and radiative transition and the total widths for $3^1G_4$ and $3^3G_3$
 bottomonium states. The width results are in units of keV.\label{decay23ga}}
\begin{center}
{\tabcolsep0.1in
\begin{tabular}{cllrc}
\toprule[1pt]\toprule[1pt]
State         &  Channels & Width & $\mathcal{B}$(\%)  \\
\midrule[1pt]
$\eta_{b4}(3^1G_4)$ & $h_{b3}(1^1F_3) \gamma$ &  0.317 &   $5.95\times10^{-4}$   \\
                    & $h_{b3}(2^1F_3) \gamma$ &  1.74 &    $3.26\times10^{-3}$  \\
                    & $h_{b3}(3^1F_3) \gamma$ &  14.2 &    $2.66\times10^{-2}$  \\
                    & $\Upsilon_3(2^3G_3)\gamma$ & $1.74\times10^{-4}$ &  $3.26\times10^{-7}$    \\
                    & $\Upsilon_4(2^3G_4)\gamma$ & $7.11\times10^{-5}$ &   $1.33\times10^{-7}$   \\
                    & $\Upsilon_5(2^3G_5)\gamma$ & $8.31\times10^{-6}$ &    $1.56\times10^{-8}$  \\
                   &$BB^*$      & 21.7 MeV &40.6     \\
                 & $B^*B^*$   & 29.5 MeV    &55.3     \\
                 & $B_sB_s^*$   & 0.973 MeV & 1.82    \\
                 & $B_s^*B_s^*$   &1.18 MeV & 2.21    \\
               &Total      & 53.3 MeV & 100               \\
$\Upsilon_3(3^3G_3)$ & $\chi_{b2}(1^3F_2)\gamma$ & 0.328 &    $8.24\times10^{-4}$  \\
                     & $\chi_{b3}(1^3F_3)\gamma$ & 0.0152 &    $3.82\times10^{-5}$  \\
                     & $\chi_{b4}(1^3F_4)\gamma$ & $1.02\times10^{-4}$ &   $2.56\times10^{-7}$   \\
                     & $\chi_{b2}(2^3F_2)\gamma$ & 1.75 &      $4.40\times10^{-3}$ \\
                     & $\chi_{b3}(2^3F_3)\gamma$ & 0.111 &     $2.79\times10^{-4}$ \\
                     & $\chi_{b4}(2^3F_4)\gamma$ & 0.00126 &   $3.17\times10^{-6}$   \\
                     & $\chi_{b2}(3^3F_2)\gamma$ & 13.5 &     $3.39\times10^{-2}$ \\
                     & $\chi_{b3}(3^3F_3)\gamma$ & 1.14 &     $2.86\times10^{-3}$ \\
                     & $\chi_{b4}(3^3F_4)\gamma$ & 0.0173 &   $4.35\times10^{-5}$   \\
                     & $\eta_{b4}(1^1G_4)\gamma$ & $2.18\times10^{-6}$ &   $5.48\times10^{-9}$   \\
                     & $BB$      & 4.88 MeV &12.2    \\
                   &$BB^*$      & 20.1  MeV &50.4     \\
                  & $B^*B^*$   & 13.2 MeV & 33.3     \\
                  & $B_sB_s$   & 0.336 MeV & 0.844    \\
                  & $B_sB_s^*$   & 0.00500 MeV & $1.26\times10^{-2}$    \\
                  & $B_s^*B_s^*$   & 1.27 MeV & 3.20   \\
               &Total      & 39.8 MeV & 100        \\

\bottomrule[1pt]\bottomrule[1pt]
\end{tabular}
}
\end{center}
\end{table}

\begin{table}[htbp]
\renewcommand\arraystretch{1.4}
\caption{ Partial widths and branching ratios of OZI-allowed strong decay and, radiative transition and the total widths for $3^3G_4$ and $3^3G_5$
 bottomonium states. The width results are in units of keV.\label{decay23gb}}
\begin{center}
{\tabcolsep0.1in
\begin{tabular}{cllrc}
\toprule[1pt]\toprule[1pt]
State         &  Channels & Width & $\mathcal{B}$(\%)  \\
\midrule[1pt]
$\Upsilon_4(3^3G_4)$ & $\chi_{b3}(1^3F_3)\gamma$ & 0.304 &    $6.03\times10^{-4}$  \\
                     & $\chi_{b4}(1^3F_4)\gamma$ & 0.0108 &   $2.14\times10^{-5}$   \\
                     & $\chi_{b3}(2^3F_3)\gamma$ & 1.66 &     $3.29\times10^{-3}$ \\
                     & $\chi_{b4}(2^3F_4)\gamma$ & 0.0821 &   $1.63\times10^{-4}$   \\
                     & $\chi_{b3}(3^3F_3)\gamma$ & 13.3 &     $2.64\times10^{-2}$ \\
                     & $\chi_{b4}(3^3F_4)\gamma$ & 0.852 &    $1.69\times10^{-3}$  \\
                     & $\eta_{b4}(2^1G_4)\gamma$ & $5.91\times10^{-5}$ &  $1.17\times10^{-7}$    \\
                    &$BB^*$      & 26.1 MeV &51.7     \\
                 & $B^*B^*$   & 22.4 MeV &44.5     \\
                 & $B_sB_s^*$   & 0.780 MeV & 1.55    \\
                 & $B_s^*B_s^*$   & 1.11 MeV & 2.21    \\
               &Total      & 50.4 MeV & 100               \\
$\Upsilon_5(3^3G_5)$ & $\chi_{b4}(1^3F_4)\gamma$ & 0.299 &   $4.43\times10^{-4}$   \\
                     & $\chi_{b4}(2^3F_4)\gamma$ & 1.67 &    $2.47\times10^{-3}$  \\
                     & $\chi_{b4}(3^3F_4)\gamma$ & 14.1 &    $2.09\times10^{-2}$  \\
                     & $\eta_{b4}(2^1G_4)\gamma$ & $1.90\times10^{-4}$ &   $2.81\times10^{-7}$   \\
                   & $BB$        & 5.16 MeV &7.65    \\
                  & $BB^*$        & 3.02 MeV &4.47    \\
                  & $B^*B^*$   & 56.5 MeV & 83.7     \\
                  & $B_sB_s$   & 0.306 MeV & 0.453   \\
                  & $B_sB_s^*$   & 0.950 MeV & 1.41   \\
                  & $B_s^*B_s^*$   &1.56 MeV & 2.31    \\
               &Total      & 67.5 MeV & 100        \\

\bottomrule[1pt]\bottomrule[1pt]
\end{tabular}
}
\end{center}
\end{table}

$G$-wave bottomonia are typical high spin
 states, which are difficult to be experimentally found in recent years. However, we still give our predictions of decay behaviors for the higher $G$-wave bottomonia with radial quantum number up to 3, which have an extra calculation of $3G$ bottomonium states compared to Ref. \cite{Godfrey:2015dia}. The hyperfine mass splittings among the spin-singlet and spin-triplet states of $2G$ and $3G$ are quite small and
 the average mass values of $2G$ and $3G$ bottomonia in Table \ref{spectrum} are 10747 and 10929 MeV, respectively, which are close to those of $4P$ and $5P$ bottomonia, respectively. The decay information on partial widths of several decay types and total widths of $2G$ and $3G$ bottomonia are shown in Tables \ref{decay22g}-\ref{decay23gb}. For the radiative transition of $2G$ states,
 we obtain the behaviors similar to $P, D, F$-wave states. Hence, we no longer explain it here. Furthermore, the total widths of $2G$ bottomonia are estimated to be about 110 $\sim$ 160 MeV, which means that they are all broad states although these values are significantly small relative to those of the GI model. Since the mass values of $2G$ bottomonia do not reach the threshold line of including a $P$-wave bottom meson, the main decay modes of $2G$ states are $BB$, $BB^*$ and $B^*B^*$. The mode $BB^*$ is dominant for the $\Upsilon_3(2^3G_3)$ and $\Upsilon_4(2^3G_4)$ with branching ratios 50.9 $\%$ and 62.5 $\%$, respectively. The mode $B^*B^*$ with a branching ratio 76.8 $\%$ is critical for the $\Upsilon_5(2^3G_5)$. Finally, the modes $BB^*$ and $B^*B^*$ have the same importance for the $\eta_{b4}(2^1G_4)$ state.

Similarly, the $3G$ bottomonium states cannot decay into $P$-wave bottom mesons on account of the mass values. In Tables \ref{decay23ga} and \ref{decay23gb}, the predicted total widths of $\eta_{b4}(3G)$ and $\Upsilon_J (3G)$ with $J=3,4,5$ states are 53.3, 39.8, 50.4, and 67.5 MeV, respectively. All the contributions of bottom-strange mesons to the total widths of $3G$ states occupy about 4 $\%$. In addition, the strong decay behaviors of channels $BB$, $BB^*$, and $B^*B^*$ of $3G$ bottomonia are almost exactly the same as the case of $2G$ bottomonia, which also indicates that the dominant decay modes of each $2G$ bottomonium state and corresponding radially excited $3G$ state are similar to each other.

\section{Comparison between the screening potential and coupled-channel quark models }\label{sec:ccm}

{{The screening potential model adopted in this work and the coupled-channel quark model are typical unquenched quark models.
In Ref. \cite{Li:2009ad}, the authors compared the results from the screening potential model and the coupled-channel model by taking a charmonium spectrum as an example, which, to some extent, reflects the equivalence between the screening potential model and the coupled-channel quark model.

We notice that Ferretti and Santopinto \cite{Ferretti:2013vua} studied the bottomonium spectrum below 10.6 GeV by the coupled-channel quark model, which makes us have a chance to test the equivalence of two unquenched quark models further.
In Fig. \ref{Fig4:comparison}, we compare the results, which are from the present work obtained by the screening potential model and from a coupled-channel quark model  \cite{Ferretti:2013vua}. In general, these two unquenched quark models present the comparable results, which again supports the conclusions of Ref. \cite{Li:2009ad}. Surely, we also find the differences in the results from two unquenched quark models for the $3P$ and $1D$ states. This fact shows some differences between the screening potential and the coupled-channel quark models. After all, they are two approaches to phenomenologically describe the unquenched effects.}} {{Besides, the effects of nearby thresholds are important for some higher bottomonia, which cannot be normally reflected in the screening potential model. To explore the strength of this effect in the bottomonium system, we will take $\Upsilon(4S)$ and $\Upsilon(5S)$ states as an example to calculate their mass shifts in the coupled-channel model \cite{Zhou:2011sp}. In Ref. \cite{Zhou:2011sp}, the inverse meson propagator can be presented as
\begin{equation}
\mathbb{P}^{-1}(s)=m_0^2-s+\sum_n\left(\mathrm{Re}\Pi_n(s)+i\mathrm{Im}\Pi_n(s)\right),
\end{equation}
where $m_0$ is the mass of a bare state, and $\mathrm{Re}\Pi_n(s)$ and $\mathrm{Im}\Pi_n(s)$ are real and imaginary parts of the self-energy function for the $n$-th decay channel, respectively. The bare mass $m_0$ can be obtained from the GI model \cite{Godfrey:1985xj}. To obtain the mass of a physical state, the $\mathbb{P}^{-1}(s)$ can be expressed in the Breit-Wigner representation
\begin{equation}
\mathbb{P}^{-1}(s)=m(s)^2-s+im_{BW}\Gamma_{tot}(s),
\end{equation}
where $m(s)^2=m_0^2+\sum_n\mathrm{Re}\Pi_n(s)$ and $\Gamma_{tot}(s)=\frac{\sum_n\mathrm{Im}\Pi_n(s)}{m_{BW}}$. The physical mass $m_{BW}$ can be determined by solving an equation $m(s)^2-s=0$. Furthermore, the imaginary of the self-energy function $\mathrm{Im}\Pi_n(s)$ can be related to the amplitude of the $^3P_0$ model by applying the Cutkosky rule and the corresponding $\mathrm{Re}\Pi_n(s)$ can also be obtained by the dispersion relation. The parameter $\gamma_0$ in the $^3P_0$ model can be fixed by fitting the experimental width of $\Upsilon(4S)$ and $\Upsilon(5S)$, which gives us $\gamma_0=0.337$ for the bottomonium system. By calculation of the coupled-channel model, we obtain the physical mass $m_{BW}(4S)=10.592$ GeV and $m_{BW}(5S)=10.838$ GeV. Comparing the above results and those from our screening potential model, we find that mass differences between two models are 20 MeV and 16 MeV for $\Upsilon(4S)$ and $\Upsilon(5S)$ states, respectively.
These results show that the effects of nearby thresholds are essential for the bottomonium system, but on the other hand, they are also within our expectations.
Generally, the contributions of effects of nearby thresholds to higher bottomonia should be systematically calculated in the coupled-channel model, which can be considered for further research in the future.          }}

{In addition, when comparing our results of average mass of these $3P$ states with those from other models \cite{Godfrey:2015dia,Segovia:2016xqb,Ferretti:2013vua}, we notice that
our results are in agreement with those given in Refs. \cite{Godfrey:2015dia,Segovia:2016xqb,Ferretti:2013vua}. In Ref. \cite{Ferretti:2013vua}, the authors specified the coupled-channel effect on the mass splittings for the $\chi_{bJ}(3P)$ states. If further focusing on the mass splittings of $\chi_{bJ}(3P)$ states, there exist differences among different model calculations \cite{Godfrey:2015dia,Segovia:2016xqb,Ferretti:2013vua}. It is obvious that the theoretical and experimental study on this mass splittings of $\chi_{bJ}(3P)$ states will be an intriguing research topic in future. }

\begin{figure}[tbp]
\begin{center}
\scalebox{0.36}{\includegraphics{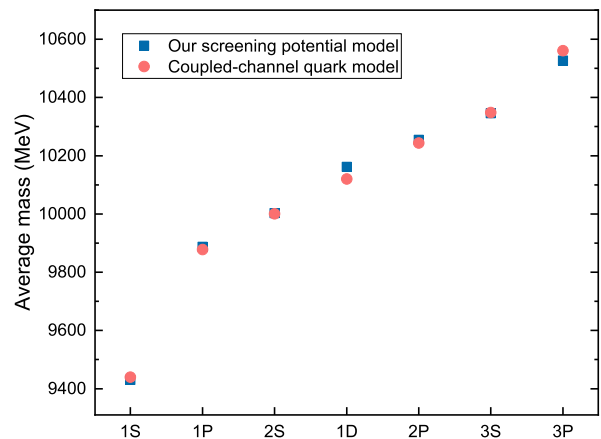}}
\caption{ A comparison of the results of our screening potential model and those obtained by a coupled-channel quark model \cite{Ferretti:2013vua}. \label{Fig4:comparison}}
 \end{center}
\end{figure}


\section{summary}\label{sec6}

As we all know, the screening effect usually plays a very important role for highly excited mesons. It affects the mass values, wave functions of mesons and hence, estimates of corresponding decay behaviors. Motivated by the recent studies of bottomonium properties in the framework of the Godfrey-Isgur relativized quark model \cite{Godfrey:2015dia}, we have performed the most comprehensive study on the properties of bottomonium family by using the modified Godfrey-Isgur model with a screening effect. We have studied radiative transition, annihilation decay, hadronic transition, and OZI-allowed two-body strong decay of $b\bar{b}$ states. Our calculated numerical results indicate that our predictions for the properties of higher bottomonia are quite different from the conclusion of GI model. Hence, we also expect that this work could provide some valuable results to the future research of bottomonium in experiment.

Our work in this paper can be divided into two parts, study of mass spectrum and study of decay behaviors of bottomonium states. Furthermore, we have focused our main attention on the prediction and analysis of higher bottomonia due to significant reflection of the screening effect. Firstly, we have taken advantage of the measured mass of 18 observed bottomonium states in Table \ref{fitting} to fit eight undetermined parameters of the modified GI model in Table \ref{MGIpara}.  It can be found that our theoretical mass values have been greatly improved compared to those of the GI model. At the same time, our results have been well matched with experimental results. Based on the above preparation, the predicted mass spectrum of bottomonium states has been given in Table \ref{spectrum}. It is interesting to note that the mass values of $\Upsilon(7S)$ and $\Upsilon(8S)$ are predicted as 11157 and 11296 MeV, respectively, which are higher than the measured mass of $\Upsilon(11020)$ only by 154 and 293 MeV, respectively.

Classifed in $L$, the decay properties of bottomonium states have been separately discussed in accordance with mass values above and below the open-bottom threshold. We have found that a screening effect is weak for the decay behaviors of the most bottomonium states below the threshold, whose estimates are similar to those of the GI model. For the higher bottomonium states above the threshold, the screening effect has become important. We have obtained fairly inconsistent conclusion on characteristic decay behaviors of bottomonium mesons between the GI model with and without screening effects.
Here, we have provided results to check the validity of our model in future experiments.

In the following years, exploration for higher bottomonium states will become a major topic in the future LHCb and forthcoming Belle II experiments. Until then, the highly excited states that are still missing are likely to be found. Moreover, some hidden experimental information of observed $b\bar{b}$ states can be further perfected. In this work, we have provided abundant theoretical information for higher bottomonia, which is helpful for piloting  experiments to search for these missing bottomonium states.

\section*{Acknowledgments}
This paper appears in arXiv just on 2018 Chinese New Year's Eve. With it, we want to celebrate Chinese New Year. Good luck and great success in the coming new year! {{We would like to thank Ming-Xiao Duan for his help in the calculations of the coupled-channel model.}} This project is partly supported by the National Natural Science Foundation of China under Grant Nos. 11222547, 11175073, and 11647301. Xiang Liu is also supported in part by the National Program for Support of Top-notch Young Professionals and the Fundamental Research Funds for the Central Universities.

\appendix
\section{Theoretical models of decay behaviors}\label{sec3}
In this Appendix, we give all the formulas necessary for calculating radiative transitions, annihilation decays, hadronic transitions, and two-body OZI-allowed strong decays.

\subsection{Radiative transitions}
The radiative transitions of a heavy quarkonium are important in the sense that radiative decays not only are main decay channels of some particles below the open-flavor threshold, but help us better understand the inner structure of a quarkonium, i.e., wave functions and interactions of $Q\bar{Q}$. For the E1 transition process, $n^{2S+1}L_J\rightarrow {n'}^{2S+1}{L'}_{J'}+\gamma $ of charmonium, the partial width is given by \cite{Kwong:1988ae},
\begin{equation}
\Gamma_{E1}=\frac{4}{3}\alpha e_b^2 \omega^3\delta_{L,L'\pm1}C_{if}\left|\langle \psi_f|r|\psi_i\rangle\right|^2
\end{equation}
with
\begin{equation}
C_{if}=max(L,L')(2J'+1) {\begin{Bmatrix} L'&J'&S\\ J&L&1 \end{Bmatrix}}^2,
\end{equation}
where the $e_b$ is a bottom quark charge in units of $|e|$ and $\alpha$ is a fine-structure constant, $\omega$ is an emitted photon energy and $\langle \psi_f|r|\psi_i\rangle$ is the transition matrix element which has the integral form $\int_0^{\infty}R_{n'L'}(r)rR_{nL}(r)r^2dr$. Here the radial wave function $R_{nL}(r)$ is obtained from the modified GI model using  parameters listed in Table \ref{MGIpara}, and they are the same in the calculation of M1 radiative transitions.

The partial width of the M1 radiative transion with the spin flip from the initial state $n^{2S+1}L_J$ to the final state ${n'}^{2S'+1}{L}_{J'}$ can be written as \cite{Novikov:1977dq}
\begin{equation}
\Gamma_{M1}=\frac{4\alpha e_b^2\omega^3}{3m_b^2}\delta_{S,S'\pm1}\frac{2J'+1}{2L+1}\left|\left< \psi_f\left|j_0\left(\frac{\omega r}{2}\right)\right|\psi_i\right>\right|^2 \label{appen:A3}
\end{equation}
with
\begin{equation}
\left< \psi_f\left|j_0\left(\frac{\omega r}{2}\right)\right|\psi_i\right>=\int_0^{\infty}R_{n'L'}(r)j_0\left(\frac{\omega r}{2}\right)R_{nL}(r)r^2dr,
\end{equation}
where $m_b$ is the mass of a bottom quark, $j_0(\frac{\omega r}{2})$ is the spherical Bessel function and other parameters have been defined above.
The results of radiative transition will be discussed in Secs. \ref{sec4} and \ref{sec5}.

\subsection{Annihilation decays}

The annihilation decay of a heavy quarkonium is important especially for those low excited states below the open-flavor threshold, where the annihilation decay to gluons is dominant. Furthermore, the decay mode is generally available in experiments, in which a vector meson $^3S_1$ or $^3D_1$ generates lepton pairs, i.e., $e^+e^-$, $\mu^+\mu^-$, and $\tau^+\tau^-$. The measured branching ratios of lepton-pair decays can also be used to judge whether experimental $XYZ$ exotic states are treated as conventional mesons or not because this ratio is usually very small for the multiquark states \cite{Badalian:1985es}. It is important to study the annihilation decays of bottomonium states into gluons, light quarks, leptons, and photons in this paper, and the general formulas for annihilation decays of a heavy quarkonium have been extensively studied by using perturbative QCD methods \cite{Appelquist:1974zd,DeRujula:1974rkb,Chanowitz:1975ee,Barbieri:1975am,Barbieri:1976fp,Barbieri:1979be,Kwong:1987ak,Ackleh:1991dy,Ackleh:1991ws,Belanger:1987cg,
Bergstrom:1991dp,Robinett:1992px,Bradley:1980eh}.
The most important feature of annihilation decays is that the probability of annihilation is related to the zero point of the meson wave function or its $n$-th order derivative where $n$ corresponds to the orbital angular momentum of a meson, i.e., $n=L$. For the lepton pair annihilation decay of $^3S_1$ or $^3D_1$ bottomonium state, this process occurs via a virtual photon in the tree level, and their width expression with the first-order QCD radiative corrections can be found in Ref. \cite{Kwong:1987ak,Bradley:1980eh}. The formulas for annihilation processes $n^3S_1 \rightarrow ggg$, $n^3S_1 \rightarrow \gamma\gamma\gamma$ and $n^3S_1 \rightarrow gg\gamma$ and the annihilation modes of $P$-wave states including QCD radiative corrections are given in Ref. \cite{Kwong:1987ak}. The general expressions for the decays into two gluons and two photon of spin singlet state with an arbitrary total angular momentum are given in Ref. \cite{Ackleh:1991dy}. Finally, for the bottomonium states with the higher angular momentum, the authors of Ref. \cite{Belanger:1987cg,Bergstrom:1991dp} have given complete expressions for the annihilation decay to three gluons for the $D$-wave spin-triplet states. In Refs. \cite{Ackleh:1991ws,Robinett:1992px}, the annihilation decay of the $F$-wave spin-triplet states into two gluons was also studied. Fortunately, the authors of Ref. \cite{Godfrey:2015dia} have summarized all these lowest-order annihilation decay formulas with the first order QCD corrections of a heavy quarkonium from $S$-wave to $G$-wave states. Hence, we do not specifically list these formulas here. It is worth noticing that according to the formula of Ref. \cite{Kwong:1987ak} and our model input, we recalculate each coefficient of first order radiative corrections for the processes $\chi_{b0}(nP)\rightarrow gg$ and $\chi_{b2}(nP)\rightarrow gg$. The radiative correction terms for $\chi_{b0}(5P,6P)$ and $\chi_{b2}(5P,6P)$ states are also added in this work. We find that the coefficient $C(nP)$ in the radiative correction term ${C(nP)\alpha_s}/{\pi}$ of the above processes are modified only for the radial excited states, $\chi_{b0}(3P)$ and $\chi_{b2}(2,3,4P)$. The corresponding constants $C(nP)$'s are 10.4, 0.89, 1.59 and 2.10, respectively. The correction constants for $\chi_{b0}(5,6P)$ and $\chi_{b2}(5,6P)$ are also estimated to be 10.6, 10.7, 2.50 and 2.85, respectively. Finally, we need to emphasize that some of parameters in the annihilation decay calculations are given by $m_b$=5.027 GeV, $\alpha_s(b\bar{b})$=0.18 and $\alpha=1/137$.

\subsection{hadronic transitions}

The hadronic transition of a heavy quarkonium usually refers to the release of a light hadron when the $Q\bar{Q}$ state moves to a lower energy level, which is very
critical in the search for some bottomonium particles below open-bottom thresholds. In this work, the hadronic transitions of bottomonia will be studied in the framework of the QCD multipole
expansion method, in which the hadronic transition is described as first emitting one gluon from a heavy quark to form the intermediate hybrid state with a
color octet $Q\bar{Q}$ pair and then recombine themselves into a light hadron with another emitted gluon via the hadronization process. Since the mass difference of a heavy quarkonium between
before and after the transition is usually small, the wavelengths of emitted gluons are generally far larger than the radius of the heavy quarkonium. Similar to electromagnetic
radiation, gluon field can be treated in the multipole expansion form in this situation ,which was first proposed in Ref. \cite{Gottfried:1977gp}.
Next, we briefly introduce the QCD multipole expansion method and more details can be found in the review article Ref. \cite{Kuang:2006me}.

The quark and gluon fields are assumed to be $\psi(x)$ and $A_\mu^a(x)$, respectively and are transformed as
\begin{align}
\Psi(\mathbf{x},t)=U&^{-1}(\mathbf{x},t)\psi(x), \nonumber \\
\frac{\lambda_a}{2}A_\mu^{a\prime}(\mathbf{x},t)&=U^{-1}(\mathbf{x},t)\frac{\lambda_a}{2}A_\mu^a(x) U(\mathbf{x},t) \nonumber \\
&-\frac{i}{g_s}U^{-1}(\mathbf{x},t)\partial_\mu U(\mathbf{x},t)
\end{align}
by the introduction of an operator $U(\mathbf{x},t)$ defined as
\begin{align}
U(\mathbf{x},t)=P \exp \left[ ig_s \int_{\mathbf{X}}^{\mathbf{x}}\frac{\lambda_a}{2}\mathbf{A}^a(\mathbf{x}{^\prime},t) d\mathbf{x}^\prime \right],
\end{align}
where $P$ is the path-ordering operator and $\mathbf{x}$ is the mass center of a quarkonium. In fact, this transformation indicates that $\Psi(\mathbf{x},t)$ dressed by
gluons plays a role of a constituent quark in the effective Lagrangian of system which can be obtained in Ref. \cite{Yan:1980uh}. As mentioned above, the process
of launching gluons of heavy quarkonium can be treated by the multipole expansion method where in the zero position, the emitted transformed gluon field $A_\mu^{\prime}(\mathbf{x},t)$
can be expanded as
\begin{align}
&A_0^{\prime}(\mathbf{x},t)=A_0^{\prime}(0,t)-\mathbf{x}\cdot \mathbf{E}(0,t)+\cdots, \nonumber \\
&\mathbf{A}^{\prime}(\mathbf{x},t)=-\frac{1}{2}\mathbf{x}\times \mathbf{B}(0,t)+\cdots
\end{align}
on the basis of effective Lagrangian, the corresponding Hamiltonian can be derived as the follow form \cite{Yan:1980uh}
\begin{align}
H_{QCD}^{eff}=H_{QCD}^{(0)}+H_{QCD}^{(1)}+H_{QCD}^{(2)},
\end{align}
where the $H_{QCD}^{(0)}$ is part of kinetic and potential energy of heavy quarkonium field which is not the simple Hamiltonian of free field but already contains relatively
strong interaction, and the $H_{QCD}^{(1)}$ and $H_{QCD}^{(2)}$ are usually seen as a perturbation which consist of the interactions of color charge, color-electric dipole moment
and color-magnetic dipole moment as well as higher order multipole momentum of quarkonium field, respectively. The general formula of S-matrix element in the QCD multipole expansion
 is given in Ref.\cite{Kuang:1990kd}. Here, we only focus on the spin-nonflip $\pi\pi$ transitions which were dominated by double electric-dipole (E1-E1) transitions since other
 transitions including spin-flip $\pi\pi$ processes where E1-M1 transition is main and spin-nonflip $\eta$ decays which are contributed by E1-M2 and M1-M1 transitions are usually
 suppressed compared with the E1-E1 transitions. Starting from the general formula of S-matrix, the amplitude of spin-nonflip $\pi\pi$ transitions can be written as
 \cite{Yan:1980uh,Kuang:1990kd,Kuang:1981se}
\begin{align}
{\cal M}_{E1E1}=i\frac{g_{E}^{2}}{6} \langle\! \Phi_{F}h \,
| \mathbf{\bar{x}}\cdot \mathbf{E} \, \frac{1}{E_{I}-H^{(0)}_{QCD}-iD_{0}} \,
\mathbf{\bar{x}}\cdot \mathbf{E} | \, \Phi_{I} \! \rangle,
\label{eq:E1E1}
\end{align}
where $|\Phi_{I}\rangle$ and $\langle \Phi_{F}h | $ are initial quarkonium and final quarkonium and light hadron, respectivrly. $\mathbf{\bar{x}}$
is the separation of heavy quark and anti-quark and $(D_0)_{bc}=\delta_{bc}\partial_0-g_sf_{abc}A_0^a$. After inserting a complete set of intermediate states, this transition amplitude
can be divided into two parts which are a heavy quark MGE (multipole gluon emission) factor and an H (hadronization) factor, respectively and the concrete form is given by \cite{Kuang:1981se}
\begin{equation}
{\cal M}_{E1E1}=i\frac{g_{E}^{2}}{6} \sum_{KL}
\frac{\left\langle\right.\!\! \Phi_{F}|\bar{x}_k|KL \!\!\left.\right\rangle
\left\langle\right.\!\! KL|\bar{x}_l|\Phi_I \!\!\left.\right\rangle}{E_I-E_{KL}}
\left\langle\right.\!\! \pi\pi|E^{a}_{k} E^{a}_{l}|0 \!\!\left.\right\rangle.
\end{equation}
As for the MGE factor which has two electric dipole factors, the initial state $|\Phi_{I}\rangle$ first transforms to the intermediate vibrational state $\langle KL|$ formed by the color-octet quarkonium and gulon called as a hybrid state. Since this three-body bound state cannot be solved by the QCD, we use the quark confining string (QCS) model \cite{Tye:1975fz,Giles:1977mp,Buchmuller:1979gy} as a viable approach to calculate the intermediate hybrid, which will be mentioned later. The part of an MGE factor can be calculated by appllying the eigenvalue and the wave function of an intermediate hybrid, initial and final quarkonium states.
The H factor $\langle \pi\pi|E^{a}_{k} E^{a}_{l}|0\rangle$ clearly reflects the process of two emitted gluons transforming to the light hadrons after hadronization. This H factor is highly nonperturbative due to the low scale of energy. Hence, this matrix element can not be also directly obtained by the QCD, however, a phenomenological approximation can be given by using
the partially conserved axial-vector current and soft pion theorem \cite{Brown:1975dz,Kuang:1981se}. Based on the above treatments,
 the final transition rate is given by \cite{Kuang:1981se}
 \begin{align}
\Gamma(\phi_I \rightarrow \phi_F +\pi\pi)=&\delta_{l_{I}l_{F}}\delta_{J_IJ_F}
\left(G|C_{1}|^{2}-\frac{2}{3}H|C_{2}|^{2}\right)\left|\mathcal{A}_1\right|^2 \nonumber \\
&+(2l_{I}+1)(2l_{F}+1)(2J_{F}+1) \nonumber\\
&\times \sum_{k}(2k+1)\left[1+(-1)^{k}\right] \left\lbrace\begin{matrix} s & l_{F} & J_{F} \\ k & J_{I} & l_{I} \end{matrix}\right\rbrace^{2} \nonumber\\
& \times  H |C_{2}|^{2} |\mathcal{A}_2|^2
\label{hadronic amplitude}
\end{align}
with
\begin{align}
\mathcal{A}_1=\sum_{L}(2L+1) \left(\begin{matrix} l_{I} & 1 & L \\ 0 & 0 & 0
\end{matrix}\right) \left(\begin{matrix} L & 1 & l_{I} \\ 0 & 0 & 0
\end{matrix}\right) f_{IF}^{L11},
\end{align} \begin{align}
\mathcal{A}_2=\sum_{L} (2L+1) \left(\begin{matrix} l_{F} & 1 & L \\ 0 & 0 & 0
\end{matrix}\right) \left(\begin{matrix} L & 1 & l_{I} \\ 0 & 0 & 0
\end{matrix}\right) \left\lbrace\begin{matrix} l_{I} & L & 1 \\ 1 & k & l_{F}
\end{matrix}\right\rbrace f_{IF}^{L11},
\end{align}
where $C_1$ and $C_2$ are parameters which are determined by the processes of $\Upsilon(2S)\rightarrow \Upsilon(1S)\pi\pi$ and
$\Upsilon(1D)\rightarrow \Upsilon(1S)\pi\pi$, respectively. The symbols $l_I(l_F)$, $J_I(J_F)$ are the orbital and total angular momentum of initial (final) state, respectively and the spin $s$ does not
change after the reaction. The $f_{IF}^{LP_{I}P_{F}}$ has the following structure     \begin{align}
f_{IF}^{LP_{I}P_{F}} =& \sum_{K} \frac{1}{M_{I}-M_{KL}} \int dr\,
r^{2+P_{F}} R_{F}(r)R_{KL}(r) \nonumber\\
& \times \int dr' r'^{2+P_{I}} R_{KL}(r') R_{I}(r'),
\end{align}
where $M_{KL}$ and $R_{KL}(r)$ are the mass and radial wave function of the intermediate state, respectively. The phase-space factors $G$ and $H$ are written as  \begin{align}
G=&\frac{3}{4}\frac{M_{F}}{M_{I}}\pi^{3}\int
dM_{\pi\pi}^{2}\,K\,\left(1-\frac{4m_{\pi}^{2}}{M_{\pi\pi}^{2}}\right)^{1/2}(M_{
\pi\pi}^{2}-2m_{\pi}^{2})^{2},
\end{align} \begin{align}
H=&\frac{1}{20}\frac{M_{F}}{M_{I}}\pi^{3}\int
dM_{\pi\pi}^{2}K\left(1-\frac{4m_{\pi}^{2}}{M_{\pi\pi}^{2}}\right)^{1/2} \Biggl[(M_{\pi\pi}^{2}-4m_{\pi}^{2})^{2}   \nonumber \\
&\left.\quad\,\, \times \left(1+\frac{2}{3}\frac{K^{2}}{M_{
\pi\pi}^{2}}\right)+\frac{8K^{4}}{15M_{\pi\pi}^{4}}(M_{\pi\pi}^{4}+2m_{\pi}^{2} M_{\pi\pi}^{2}+6m_{\pi}^{4})\right],
\end{align}
with $K$ given by
\begin{equation}
K = \frac{\sqrt{\left[(M_{I}+M_{F})^{2}-M_{\pi\pi}^{2}\right]
\left[(M_{I}-M_{F})^{2}-M_{ \pi\pi}^{2}\right]}}{2M_{I}}.
\end{equation}

The intermediate hybrid states can be described by the quark confining string (QCS) model \cite{Tye:1975fz,Giles:1977mp,Buchmuller:1979gy}, in which we consider that the quark and anti-quark are connected by an appropriate color electric flux tube or string. If the string is in the ground state, the system of a quark-antiquark pair is a meson where the string corresponds to the strong confinement interaction. The vibration of the string means a new state with gluon excitation effects, which is composed of the excited gluon field and quark-antiquark pair, i.e., the so-called hybrid state. For this vibrational mode, assuming both ends of a string are fixed because of too heavy quark
masses, then the effective vibrational potential can be given by \cite{Giles:1977mp}
 \begin{align}
 V_n(r)=\sigma(r)r\left(1-\frac{2n\pi}{2n\pi+\sigma(r)[(r-2d)^2+4d^2]}\right)^{-\frac{1}{2}}
 \end{align}
 with
 \begin{align}
 d=\frac{1}{4}\sigma(r)r^2\;\frac{\alpha_n}{2m_b+\sigma(r)r\alpha_n},
 \end{align}
where $d$ is the correction of finite heavy quark mass and $n$ indicates the excitation level. The $\alpha_n$ related to the shape of the vibrational string \cite{Giles:1977mp} is taken as
$\sqrt{1.5}$, which is consistent with Ref. \cite{Segovia:2016xqb} and is insensitive to our mass spectrum of hybrid states.

The potential of a hybrid meson can be expressed as \cite{Buchmuller:1979gy}

\begin{align}
V_{hyb}(r)=V_{G}(r)+V_{S}(r)+\left[V_{n}(r)-\sigma(r) r\right],
\end{align}
where $V_{G}(r)$ is one-gluon exchange potential and $V_{S}(r)$ is a color confining potential. It is easy to see that the above potential becomes a general $Q\bar{Q}$ interaction when $n=0$ for the vibrational potential $V_n$. For theoretical self-consistency, forms of $V_{G}(r)$ and $V_{S}(r)$ are taken from our modified GI model and due to a screening effect, the effective string tension $\sigma(r)$ is not a constant but rather a function of a distance $r$ of $Q$ and $\bar{Q}$. The specific expressions of potentials $V_{G}(r)$ and $V_{S}(r)$ can be written as
\begin{align}
V_{G}(r)=&- \frac{4\alpha_s(r)}{3r}, \nonumber \\
V_{S}(r)=&\;\sigma(r)r+c
\end{align}
with
\begin{align}
\alpha_s(r)=&\;\sum_k \alpha_k \frac{2}{\sqrt{\pi}}\int_0^{\gamma_kr}e^{-x^2}dx,  \nonumber \\
\sigma(r)=&\;b(1-e^{-\mu r})/(\mu r).
\end{align}

Solving the Schr\"odinger equation for a hybrid meson, one obtains the mass spectrum and corresponding wave function of a hybrid state, which are used to calculate the width of hadronic transition by Eq. (\ref{hadronic amplitude}). Nevertheless, we have to emphasize that the QCD multipole expansion is dependent on the inputs and has its own error due to theoretical uncertainties. Hence, we should regard the calculated width of hadronic transition as rough estimates rather than precise results. In addition, considering that there may be a more complex mechanism of hadronic transition for highly excited states, we focus on the hadronic transition of the bottomonium states only below open-bottom threshold in this work. The numerical results of hadronic decay will be discussed in Secs. \ref{sec4} and it should be noted that the GI's results of hadronic transition Ref. \cite{Godfrey:2015dia} are derived from the reduced matrix elements, which are obtained by measured transition rates rather than direct calculations. Here, we adopt the direct QCD multipole-expansion method to calculate a partial width of hadronic transition of bottomonium states and it is also useful to make a comparison with the results of Ref. \cite{Godfrey:2015dia}.

\subsection{Two-body OZI-allowed strong decays}
Quark-Pair Creation (QPC) model is applicable to the calculation of OZI allowed hadron strong decays. This model is proposed by the Micu \cite{Micu:1968mk} at the earliest in 1968 and it has been further developed by the Orsay Group \cite{LeYaouanc:1972vsx,LeYaouanc:1973ldf,LeYaouanc:1974cvx,LeYaouanc:1977fsz} which is one of the most popular phenomenological method to deal with the OZI allowed strong decays and has been greatly used in the calculation of strong decay. The model assumes that a created quark-antiquark pair $q\bar{q}$ from the vacuum is a $ ^3P_0$ state which has spin-parity $J^{PC}=0^{++}$, so the model also known as $^3P_0$ model. In the following, we will briefly introduce this model. For the OZI allowed strong decay process $A\rightarrow B+C$, the transition operators $\mathcal{T}$ can be expressed as
\begin{eqnarray}
\mathcal{T}& = &-3\gamma \sum_{m}\langle 1m;1-m|00\rangle\int d \mathbf{p}_3d\mathbf{p}_4 \delta ^3 (\mathbf{p}_3+\mathbf{p}_4) \nonumber \\
&& \times \mathcal{Y}_{1m}\left( \frac{\mathbf{p}_3-\mathbf{p}_4}{2} \right) \chi _{1,-m}^{34} \phi _{0}^{34}
\omega_{0}^{34} b_{3i}^{\dag} (\mathbf{p}_3) d_{4j}^{\dag}(\mathbf{p}_4).\label{QPC1}
\end{eqnarray}
where $\mathcal{Y}_{lm}\left( \mathbf{p}\right)=p^lY_{lm}(\theta_p,\phi_p)$ is a solid spherical harmonic function and $b_3^{\dag}(d_{4}^{\dag})$ is quark (antiquark) creation operator, $\phi _{0}^{34}=(u\bar{u}+d\bar{d}+s\bar{s})/\sqrt{3}$ and $\omega_{0}^{34} $ are SU(3) flavor and color wave function of vacuum quark pair respectively, and the dimensionless parameter $\gamma$ describes the strength of creating a quark-antiquark pair from the vacuum.  $\gamma$ value for $s\bar{s}$ pair creation is generally more than a factor of $1/\sqrt{3}$ compared to that of $u\bar{u}/d\bar{d}$ pair creation. The reason of the existence of factor $1/\sqrt{3}$ is in order to show the SU(3) symmetry breaking \cite{LeYaouanc:1972vsx,LeYaouanc:1973ldf,LeYaouanc:1974cvx,LeYaouanc:1977fsz,LeYaouanc:1977gm,Song:2015fha}.
The transition matrix of decay process can be written as
\begin{eqnarray}
\langle BC | \mathcal{T}| A\rangle = \delta^3(\mathbf{P}_B+\mathbf{P}_C) \mathcal{M}^{M_{J_A}M_{J_B}M_{J_c}}\label{QPC2},
\end{eqnarray}
where $\mathbf{P}_B$ and $\mathbf{P}_C$ are the momenta of final meson $B$ and $C$ in the center frame of initial meson $A$, respectively, and $\mathcal{M}^{M_{J_A}M_{J_B}M_{J_c}}$ is the decay amplitude. The mock state $|D\rangle$ where $D$ is an arbitrary state including initial state $A$ and final state $B,C$ \cite{Hayne:1981zy} has the form
\begin{eqnarray}
&&|D(n^{2S+1}L_{JM_J})(\mathbf{P}_D)\rangle\nonumber\\
&&=\sqrt{2E}\sum\limits_{ {M_{S}},{M_{L}}}\langle LM_L;SM_{S}|JM_{J}\rangle\chi^{D}_{S,M_s} \nonumber\quad\\
&&\times \phi^D \omega^D\int d\mathbf{p}_1 d\mathbf{p}_2 \delta^3(
\mathbf{P}_D-\mathbf{p}_1-\mathbf{p}_2) \nonumber\\ &&\quad\times\Psi^D_{nLM_L}(\mathbf{p}_1,\mathbf{p}_2)|q_1(\mathbf{p}_1) \bar{q}_2(\mathbf{p}_2)\rangle,\label{QPC3}
\end{eqnarray}
where the front factor $E$ is particle energy and $\Psi^D_{nLM_L}(\mathbf{p}_1,\mathbf{p}_2)$, $\chi^D$, $\phi^D$ and $\omega^D$ donate spatial, spin, flavor and color wave function of meson $D$, respectively, and $\langle LM_L ;SM_{S}|JM_{J}\rangle$ is Clebsch-Gordan coefficients. For the spatial wave function of initial and final states, we use the exact eigenfunctions by solving Schr$\ddot{\text{o}}$dinger equation in potential models rather than simple harmonic oscillator (SHO) wave function.
Combined Eqs. (\ref{QPC1})-(\ref{QPC2}) and Eq. (\ref{QPC3}), the decay amplitude $\mathcal{M}^{M_{J_A}M_{J_B}M_{J_c}}$ can be derived.

For the convenience of experimental measurement the decay amplitudes could be related to the helicity partial wave amplitudes by Jacob-Wick formula \cite{Jacob:1959at}
\begin{eqnarray}
\mathcal{M}^{JL}(\mathbf{P})&=&\frac{\sqrt{2L+1}}{2J_A+1}\sum_{M_{J_B}M_{J_C}}\langle L0;JM_{J_A}|J_AM_{J_A}\rangle \nonumber \\
&&\times \langle J_BM_{J_B};J_CM_{J_C}|{J_A}M_{J_A}\rangle \mathcal{M}^{M_{J_{A}}M_{J_B}M_{J_C}},
\end{eqnarray}
 where $J$ and $L$ are the total and orbital angular momenta between final state $B$ and $C$ respectively and $\mathbf{P}=\mathbf{P}_B$. Finally, the partial width of the $A\to BC$ can be written as
\begin{eqnarray}
\Gamma_{A\to BC} &=& \pi ^2\frac{|\mathbf{P}_B|}{m_A^2}\sum_{J,L}|\mathcal{M}^{JL}(\mathbf{P})|^2,
\end{eqnarray}
where $m_{A}$ is the mass of the initial state $A$. In addition, in the calculations of strong decay, the constituent quark mass of bottom, up/down and strange quark are taken as 5.027, 0.22 and 0.419 GeV, respectively.

\section{Interaction potentials in the modified GI model}\label{appendix:B}

{{ In this Appendix, we will list the semi-relativistic effective potentials in the modified GI model with a screening effect for the bottomonium system \cite{Song:2015nia}
\begin{equation}
\tilde{V}_{eff}(\mathbf{p},\mathbf{r})=\tilde{V}^{conf}+\tilde{V}^{hyp}+\tilde{V}^{so}.
\label{Eq:Htot1}
\end{equation}
Here, the first term, spin-independent potential $\tilde{V}^{conf}$, can be written as
\begin{equation}
\tilde{V}^{conf}=\left(1+\frac{p^2}{E_bE_{\bar{b}}}\right)^{1/2}\tilde{G}(r)\left(1 +\frac{p^2}{E_bE_{\bar{b}}}\right)^{1/2}+\tilde{S}(r),
\end{equation}
where $\tilde{G}(r)$ and $\tilde{S}(r)$ are the smearing one-gluon exchange and screening confinement interactions, respectively.
$E_b$=$\sqrt{m_b^2+p^2}$ and $E_{\bar{b}}$=$\sqrt{m_{\bar{b}}^2+p^2}$ are the energy of bottom and anti-bottom quark in bottomonium, respectively.
The second term $\tilde{V}^{hyp}$ is the color-hyperfine interaction, which reads
\begin{equation}
\tilde{V}^{hyp}=\tilde{V}^{tensor}+\tilde{V}^{c}
\end{equation}
with
\begin{eqnarray}
\tilde{V}^{tensor}&=&-\left(\frac{\textbf{S}_b\cdot \textbf{r}~ \textbf{S}_{\bar{b}}\cdot \textbf{r}/r^2-\frac{1}{3}\textbf{S}_b\cdot \textbf{S}_{\bar{b}}}{m_{b}m_{\bar{b}}}\right)\left(\frac{\partial^2}{\partial r^2}-\frac{1}{r}\frac{\partial}{\partial r}\right)\tilde{G}^t_{12},\nonumber \\ \\
%
\tilde{V}^{c}&=&\frac{2\textbf{S}_b\cdot \textbf{S}_{\bar{b}}}{3m_{b}m_{\bar{b}}}\nabla^2\tilde{G}^c_{12},
\end{eqnarray}
where $\tilde{V}^{tensor}$ and $\tilde{V}^{c}$ are the tensor and contact hyperfine potentials, respectively. The $\tilde{G}^t_{12}$ and $\tilde{G}^c_{12}$ reflect a relativistic momentum-independent correction to the one-gluon exchange potential, which can be obtained from Eq. (\ref{eq:momf}) and their specific expressions are
\begin{equation}
\tilde{G}^{t}_{12}=\left(\frac{m_bm_{\bar{b}}}{E_bE_{\bar{b}}}\right)^{1/2+\epsilon_{t}} \tilde{G}(r)\left(\frac{m_bm_{\bar{b}}}{E_bE_{\bar{b}}}\right)^{1/2+\epsilon_{t}},
\end{equation}
\begin{equation}
\tilde{G}^{c}_{12}=\left(\frac{m_bm_{\bar{b}}}{E_bE_{\bar{b}}}\right)^{1/2+\epsilon_{c}} \tilde{G}(r)\left(\frac{m_bm_{\bar{b}}}{E_bE_{\bar{b}}}\right)^{1/2+\epsilon_{c}},
\end{equation}
respectively, with constant parameters $\epsilon_{t}$ and $\epsilon_{c}$. The last term $\tilde{V}^{so}$ is the spin-orbit coupling and it includes vector and scalar spin-orbit potentials, namely,
\begin{equation}
\tilde{V}^{so}=\tilde{V}^{so(v)}+\tilde{V}^{so(s)},
\end{equation}
where
\begin{eqnarray}
\tilde{V}^{so(v)}&=&\frac{\textbf{S}_b \cdot\textbf{L} }{2m_{b}^2r}\frac{\partial\tilde{G}_{11}^{so(v)}}{\partial r}+\frac{\textbf{S}_{\bar{b}} \cdot\textbf{L}}{2m_{\bar{b}}^2r}\frac{\partial\tilde{G}_{22}^{so(v)}}{\partial r} \nonumber \\
&&+\frac{(\textbf{S}_b+\textbf{S}_{\bar{b}}) \cdot\textbf{L} }{m_{b}m_{\bar{b}}r}\frac{\partial \tilde{G}_{12}^{so(v)}}{\partial r},
\end{eqnarray}
\begin{eqnarray}
\tilde{V}^{so(s)}&=&-\frac{\textbf{S}_b \cdot\textbf{L}}{2m_{b}^2r}\frac{\partial\tilde{S}_{11}^{so(s)}}{\partial r}-\frac{\textbf{S}_{\bar{b}} \cdot\textbf{L}}{2m_{\bar{b}}^2r}\frac{\partial\tilde{S}_{22}^{so(s)}}{\partial r}.
\end{eqnarray}
Here, the $\tilde{G}^{so(v)}_{ij}$ and $\tilde{S}^{so(s)}_{ii}$ are also the momentum-dependent corrections for the vector and scalar spin-orbit interactions, respectively, which are given by
\begin{eqnarray}
\tilde{G}^{so(v)}_{ij}&=&\left(\frac{m_im_j}{E_iE_j}\right)^{1/2+\epsilon_{so(v)}} \tilde{G}(r)\left(\frac{m_im_j}{E_iE_j}\right)^{1/2+\epsilon_{so(v)}}, \\
%
\tilde{S}^{so(s)}_{ii}&=&\left(\frac{m_i^2}{E_i^2}\right)^{1/2+\epsilon_{so(s)}} \tilde{S}(r)\left(\frac{m_i^2}{E_i^2}\right)^{1/2+\epsilon_{so(s)}},
\end{eqnarray}
where the subscripts $i(j)$=1, 2 denote bottom and anti-bottom quark, respectively.

}}


\begin{thebibliography}{99}



\bibitem{Aubert:1974js}
  J.~J.~Aubert {\it et al.} [E598 Collaboration],
  Experimental Observation of a Heavy Particle J,
  \href{http://dx.doi.org/10.1103/PhysRevLett.33.1404}{Phys.\ Rev.\ Lett.\  {\bf 33}, 1404 (1974)}.

\bibitem{Augustin:1974xw}
  J.~E.~Augustin {\it et al.} [SLAC-SP-017 Collaboration],
  Discovery of a Narrow Resonance in $e^+$ $e^-$ Annihilation,
  \href{http://dx.doi.org/10.1103/PhysRevLett.33.1406}{Phys.\ Rev.\ Lett.\  {\bf 33}, 1406 (1974)}
  [Adv.\ Exp.\ Phys.\  {\bf 5}, 141 (1976)].

\bibitem{Herb:1977ek}
  S.~W.~Herb {\it et al.},
  Observation of a Dimuon Resonance at 9.5-GeV in 400-GeV Proton-Nucleus Collisions,
  \href{http://dx.doi.org/10.1103/PhysRevLett.39.252}{Phys.\ Rev.\ Lett.\  {\bf 39}, 252 (1977)}.

\bibitem{Innes:1977ae}
  W.~R.~Innes {\it et al.},
  Observation of structure in the $\Upsilon$ region,
  \href{http://dx.doi.org/10.1103/PhysRevLett.39.1240}{Phys.\ Rev.\ Lett.\  {\bf 39}, 1240 (1977)}
  \href{http://dx.doi.org/10.1103/PhysRevLett.39.1640.2}{Erratum: [Phys.\ Rev.\ Lett.\  {\bf 39}, 1640 (1977)]}.

\bibitem{Brambilla:2010cs}
  N.~Brambilla {\it et al.},
  Heavy quarkonium: progress, puzzles, and opportunities,
  \href{http://dx.doi.org/10.1140/epjc/s10052-010-1534-9}{Eur.\ Phys.\ J.\ C {\bf 71}, 1534 (2011)}.

\bibitem{Besson:1984bd}
  D.~Besson {\it et al.} [CLEO Collaboration],
  Observation of New Structure in the $e^+$ $e^-$ Annihilation Cross-Section Above B anti-B Threshold,
  \href{http://dx.doi.org/10.1103/PhysRevLett.54.381}{Phys.\ Rev.\ Lett.\  {\bf 54}, 381 (1985)}.

\bibitem{Lovelock:1985nb}
  D.~M.~J.~Lovelock {\it et al.},
  Masses, Widths, And Leptonic Widths Of The Higher Upsilon Resonances,
  \href{http://dx.doi.org/10.1103/PhysRevLett.54.377}{Phys.\ Rev.\ Lett.\  {\bf 54}, 377 (1985)}.

\bibitem{Klopfenstein:1983nx}
  C.~Klopfenstein {\it et al.},
  Observation of the Lowest $P$-Wave $b \bar{b}$ Bound States,
  \href{http://dx.doi.org/10.1103/PhysRevLett.51.160}{Phys.\ Rev.\ Lett.\  {\bf 51}, 160 (1983)}.

\bibitem{Pauss:1983pa}
  F.~Pauss {\it et al.},
  Observation of $\chi_b$ Production in the Exclusive Reaction $\Upsilon^\prime \to \gamma \chi_b \to \gamma \gamma \Upsilon \to \gamma \gamma (e^+ e^-$ or $\mu^+ \mu^-$),
  \href{http://dx.doi.org/10.1016/0370-2693(83)91539-3}{Phys.\ Lett.\  {\bf 130B}, 439 (1983)}.

\bibitem{Han:1982zk}
  K.~Han {\it et al.},
  Observation Of P Wave B Anti-b Bound States,
  \href{http://dx.doi.org/10.1103/PhysRevLett.49.1612}{Phys.\ Rev.\ Lett.\  {\bf 49}, 1612 (1982)}.

\bibitem{Eigen:1982zm}
  G.~Eigen {\it et al.},
  Evidence For $\chi_b^\prime$ Production In The Exclusive Reaction $\Upsilon^{\prime\prime} \to \gamma \chi_b^\prime \to (\gamma \gamma \Upsilon^\prime$ or $\gamma \gamma \Upsilon)$,
  \href{http://dx.doi.org/10.1103/PhysRevLett.49.1616}{Phys.\ Rev.\ Lett.\  {\bf 49}, 1616 (1982)}.

\bibitem{Aubert:2008ba}
  B.~Aubert {\it et al.} [BaBar Collaboration],
  Observation of the bottomonium ground state in the decay $\Upsilon(3S) \to \gamma \eta_b$,
  \href{http://dx.doi.org/10.1103/PhysRevLett.101.071801}{Phys.\ Rev.\ Lett.\  {\bf 101}, 071801 (2008)}
  \href{http://dx.doi.org/10.1103/PhysRevLett.102.029901}{Erratum: [Phys.\ Rev.\ Lett.\  {\bf 102}, 029901 (2009)]}.

\bibitem{Lees:2011mx}
  J.~P.~Lees {\it et al.} [BaBar Collaboration],
  Study of radiative bottomonium transitions using converted photons,
  \href{http://dx.doi.org/10.1103/PhysRevD.84.099901}{Phys.\ Rev.\ D {\bf 84}, 072002 (2011)}.

\bibitem{Bonvicini:2009hs}
  G.~Bonvicini {\it et al.} [CLEO Collaboration],
  Measurement of the $\eta_b(1S)$ mass and the branching fraction for $\Upsilon(3S) \to \gamma \eta_b(1S)$,
  \href{http://dx.doi.org/10.1103/PhysRevD.81.031104}{Phys.\ Rev.\ D {\bf 81}, 031104 (2010)}.

  \bibitem{Dobbs:2012zn}
  S.~Dobbs, Z.~Metreveli, K.~K.~Seth, A.~Tomaradze and T.~Xiao,
  Observation of $\eta_b(2S)$ in $\Upsilon(2S) \to \gamma \eta_b(2S)$, $\eta_b(2S) \to $ hadrons, and Confirmation of $\eta_b(1S)$,
  \href{http://dx.doi.org/10.1103/PhysRevLett.109.082001}{Phys.\ Rev.\ Lett.\  {\bf 109}, 082001 (2012)}.

\bibitem{Mizuk:2012pb}
  R.~Mizuk {\it et al.} [Belle Collaboration],
  Evidence for the $\eta_b(2S)$ and observation of $h_b(1P) \to \eta_b(1S) \gamma$ and $h_b(2P) \to \eta_b(1S) \gamma$,
  \href{http://dx.doi.org/10.1103/PhysRevLett.109.232002}{Phys.\ Rev.\ Lett.\  {\bf 109}, 232002 (2012)}.


\bibitem{Aad:2011ih}
  G.~Aad {\it et al.} [ATLAS Collaboration],
  Observation of a new $\chi_b$ state in radiative transitions to $\Upsilon(1S)$ and $\Upsilon(2S)$ at ATLAS,
  \href{http://dx.doi.org/10.1103/PhysRevLett.108.152001}{Phys.\ Rev.\ Lett.\  {\bf 108}, 152001 (2012)}.

\bibitem{Abazov:2012gh}
  V.~M.~Abazov {\it et al.} [D0 Collaboration],
  Observation of a narrow mass state decaying into $\Upsilon(1S) + \gamma$ in $p\bar{p}$ collisions at $\sqrt{s} = 1.96$ TeV,
  \href{http://dx.doi.org/10.1103/PhysRevD.86.031103}{Phys.\ Rev.\ D {\bf 86}, 031103 (2012)}.

\bibitem{Lees:2011zp}
  J.~P.~Lees {\it et al.} [BaBar Collaboration],
  Evidence for the $h_b(1P)$ meson in the decay $\Upsilon(3S) \to \pi^0 h_b(1P)$,
  \href{http://dx.doi.org/10.1103/PhysRevD.84.091101}{Phys.\ Rev.\ D {\bf 84}, 091101 (2011)}.

\bibitem{Adachi:2011ji}
  I.~Adachi {\it et al.} [Belle Collaboration],
  First observation of the $P$-wave spin-singlet bottomonium states $h_b(1P)$ and $h_b(2P)$,
  \href{http://dx.doi.org/10.1103/PhysRevLett.108.032001}{Phys.\ Rev.\ Lett.\  {\bf 108}, 032001 (2012)}.

\bibitem{Bonvicini:2004yj}
  G.~Bonvicini {\it et al.} [CLEO Collaboration],
  First observation of a $\Upsilon(1D)$ state,
  \href{http://dx.doi.org/10.1103/PhysRevD.70.032001}{Phys.\ Rev.\ D {\bf 70}, 032001 (2004)}.

\bibitem{delAmoSanchez:2010kz}
  P.~del Amo Sanchez {\it et al.} [BaBar Collaboration],
  Observation of the $\Upsilon(1^3D_J)$ Bottomonium State through Decays to $\pi^+\pi^-\Upsilon(1S)$,
  \href{http://dx.doi.org/10.1103/PhysRevD.82.111102}{Phys.\ Rev.\ D {\bf 82}, 111102 (2010)}.

\bibitem{Brambilla:2004wf}
  N.~Brambilla {\it et al.} [Quarkonium Working Group],
  Heavy quarkonium physics,
  hep-ph/0412158.

\bibitem{Lewis:2012ir}
  R.~Lewis and R.~M.~Woloshyn,
  Higher angular momentum states of bottomonium in lattice NRQCD,
  \href{http://dx.doi.org/10.1103/PhysRevD.85.114509}{Phys.\ Rev.\ D {\bf 85}, 114509 (2012)}.

\bibitem{Dowdall:2013jqa}
  R.~J.~Dowdall {\it et al.} [HPQCD Collaboration],
  Bottomonium hyperfine splittings from lattice nonrelativistic QCD including radiative and relativistic corrections,
  \href{http://dx.doi.org/10.1103/PhysRevD.89.031502}{Phys.\ Rev.\ D {\bf 89}, no. 3, 031502 (2014)}
  \href{http://dx.doi.org/10.1103/PhysRevD.92.039904}{Erratum: [Phys.\ Rev.\ D {\bf 92}, 039904 (2015)]}.

\bibitem{Godfrey:2015dia}
  S.~Godfrey and K.~Moats,
  Bottomonium Mesons and Strategies for their Observation,
  \href{http://dx.doi.org/10.1103/PhysRevD.92.054034}{Phys.\ Rev.\ D {\bf 92}, no. 5, 054034 (2015)}.

\bibitem{Segovia:2016xqb}
  J.~Segovia, P.~G.~Ortega, D.~R.~Entem and F.~Fern��ndez,
  Bottomonium spectrum revisited,
  \href{http://dx.doi.org/10.1103/PhysRevD.93.074027}{Phys.\ Rev.\ D {\bf 93}, no. 7, 074027 (2016)}.

\bibitem{Li:2009nr}
  B.~Q.~Li and K.~T.~Chao,
  Bottomonium Spectrum with Screened Potential,
  \href{http://dx.doi.org/10.1088/0253-6102/52/4/20}{Commun.\ Theor.\ Phys.\  {\bf 52}, 653 (2009)}.

\bibitem{Deng:2016ktl}
  W.~J.~Deng, H.~Liu, L.~C.~Gui and X.~H.~Zhong,
  Spectrum and electromagnetic transitions of bottomonium,
  \href{http://dx.doi.org/10.1103/PhysRevD.95.074002}{Phys.\ Rev.\ D {\bf 95}, no. 7, 074002 (2017)}.

\bibitem{Ebert:2002pp}
  D.~Ebert, R.~N.~Faustov and V.~O.~Galkin,
  Properties of heavy quarkonia and $B_c$ mesons in the relativistic quark model,
  \href{http://dx.doi.org/10.1103/PhysRevD.67.014027}{Phys.\ Rev.\ D {\bf 67}, 014027 (2003)}.

\bibitem{Ferretti:2013vua}
  J.~Ferretti and E.~Santopinto,
  Higher mass bottomonia,
  \href{http://dx.doi.org/10.1103/PhysRevD.90.094022}{Phys.\ Rev.\ D {\bf 90}, no. 9, 094022 (2014)}.

\bibitem{Li:2015zda}
  Y.~Li, P.~Maris, X.~Zhao and J.~P.~Vary,
  Heavy Quarkonium in a Holographic Basis,
  \href{http://dx.doi.org/10.1016/j.physletb.2016.04.065}{Phys.\ Lett.\ B {\bf 758}, 118 (2016)}.

\bibitem{Vijande:2004he}
  J.~Vijande, F.~Fernandez and A.~Valcarce,
  Constituent quark model study of the meson spectra,
  \href{http://dx.doi.org/10.1088/0954-3899/31/5/017}{J.\ Phys.\ G {\bf 31}, 481 (2005)}.

\bibitem{Gupta:1986xt}
  S.~N.~Gupta, S.~F.~Radford and W.~W.~Repko,
  Semirelativistic Potential Model for Heavy Quarkonia,
  \href{http://dx.doi.org/10.1103/PhysRevD.34.201}{Phys.\ Rev.\ D {\bf 34}, 201 (1986)}.

\bibitem{Richardson:1978bt}
  J.~L.~Richardson,
  The Heavy Quark Potential and the $\Upsilon$, $J/\psi$ Systems,
  \href{http://dx.doi.org/10.1016/0370-2693(79)90753-6}{Phys.\ Lett.\  {\bf 82B}, 272 (1979)}.

\bibitem{Buchmuller:1980su}
  W.~Buchmuller and S.~H.~H.~Tye,
  Quarkonia and Quantum Chromodynamics,
  \href{http://dx.doi.org/10.1103/PhysRevD.24.132}{Phys.\ Rev.\ D {\bf 24}, 132 (1981)}.

\bibitem{Martin:1980jx}
  A.~Martin,
  A FIT of Upsilon and Charmonium Spectra,
  \href{http://dx.doi.org/10.1016/0370-2693(80)90527-4}{Phys.\ Lett.\  {\bf 93B}, 338 (1980)}.

\bibitem{Radford:2007vd}
  S.~F.~Radford and W.~W.~Repko,
  Potential model calculations and predictions for heavy quarkonium,
  \href{http://dx.doi.org/10.1103/PhysRevD.75.074031}{Phys.\ Rev.\ D {\bf 75}, 074031 (2007)}.

\bibitem{Motyka:1997di}
  L.~Motyka and K.~Zalewski,
  Mass spectra and leptonic decay widths of heavy quarkonia,
  \href{http://dx.doi.org/10.1007/s100529800743}{Eur.\ Phys.\ J.\ C {\bf 4}, 107 (1998)}.

\bibitem{Gonzalez:2003gx}
  P.~Gonzalez, A.~Valcarce, H.~Garcilazo and J.~Vijande,
  Heavy meson description with a screened potential,
  \href{http://dx.doi.org/10.1103/PhysRevD.68.034007}{Phys.\ Rev.\ D {\bf 68}, 034007 (2003)}.

\bibitem{Ding:1995he}
  Y.~B.~Ding, K.~T.~Chao and D.~H.~Qin,
  Possible effects of color screening and large string tension in heavy quarkonium spectra,
  \href{http://dx.doi.org/10.1103/PhysRevD.51.5064}{Phys.\ Rev.\ D {\bf 51}, 5064 (1995)}.

\bibitem{Beyer:1992nd}
  M.~Beyer, U.~Bohn, M.~G.~Huber, B.~C.~Metsch and J.~Resag,
  Relativistic effects and the constituent quark model of heavy quarkonia,
  \href{http://dx.doi.org/10.1007/BF01482594}{Z.\ Phys.\ C {\bf 55}, 307 (1992)}.

\bibitem{Ding:1993uy}
  Y.~B.~Ding, K.~T.~Chao and D.~H.~Qin,
  Screened Q anti-Q potential and spectrum of heavy quarkonium,
  \href{http://dx.doi.org/10.1088/0256-307X/10/8/004}{Chin.\ Phys.\ Lett.\  {\bf 10}, 460 (1993)}.

\bibitem{Wei-Zhao:2013sta}
  T.~Wei-Zhao, C.~Lu, Y.~You-Chang and C.~Hong,
  Bottomonium states versus recent experimental observations in the QCD-inspired potential model,
  \href{http://dx.doi.org/10.1088/1674-1137/37/8/083101}{Chin.\ Phys.\ C {\bf 37}, 083101 (2013)}.

\bibitem{Gonzalez:2014nka}
  P.~Gonzalez,
  Generalized screened potential model,
  \href{http://dx.doi.org/10.1088/0954-3899/41/9/095001}{J.\ Phys.\ G {\bf 41}, 095001 (2014)}.

\bibitem{Godfrey:1985xj}
  S.~Godfrey and N.~Isgur,
  Mesons in a Relativized Quark Model with Chromodynamics,
  \href{http://dx.doi.org/10.1103/PhysRevD.32.189}{Phys.\ Rev.\ D {\bf 32}, 189 (1985)}.

\bibitem{Song:2015nia}
  Q.~T.~Song, D.~Y.~Chen, X.~Liu and T.~Matsuki,
  Charmed-strange mesons revisited: mass spectra and strong decays,
  \href{http://dx.doi.org/10.1103/PhysRevD.91.054031}{Phys.\ Rev.\ D {\bf 91}, 054031 (2015)}.


\bibitem{Born:1989iv}
  K.~D.~Born, E.~Laermann, N.~Pirch, T.~F.~Walsh and P.~M.~Zerwas,
  Hadron Properties in Lattice {QCD} With Dynamical Fermions,
  \href{http://dx.doi.org/10.1103/PhysRevD.40.1653}{Phys.\ Rev.\ D {\bf 40}, 1653 (1989)}.



\bibitem{Li:2009zu}
  B.~Q.~Li and K.~T.~Chao,
  Higher Charmonia and X,Y,Z states with Screened Potential,
  \href{http://dx.doi.org/10.1103/PhysRevD.79.094004}{Phys.\ Rev.\ D {\bf 79}, 094004 (2009)}.



\bibitem{Li:2009ad}
  B.~Q.~Li, C.~Meng and K.~T.~Chao,
  Coupled-Channel and Screening Effects in Charmonium Spectrum,
  \href{http://dx.doi.org/10.1103/PhysRevD.80.014012}{Phys.\ Rev.\ D {\bf 80}, 014012 (2009)}.



\bibitem{Song:2015fha}
  Q.~T.~Song, D.~Y.~Chen, X.~Liu and T.~Matsuki,
  Higher radial and orbital excitations in the charmed meson family,
  \href{http://dx.doi.org/10.1103/PhysRevD.92.074011}{Phys.\ Rev.\ D {\bf 92}, no. 7, 074011 (2015)}.



\bibitem{Olive:2016xmw}
  C.~Patrignani {\it et al.} [Particle Data Group],
  Review of Particle Physics,
  \href{http://dx.doi.org/10.1088/1674-1137/40/10/100001}{Chin.\ Phys.\ C {\bf 40}, no. 10, 100001 (2016)}.










\bibitem{Meinel:2010pv}
  S.~Meinel,
  Bottomonium spectrum at order $v^6$ from domain-wall lattice QCD: Precise results for hyperfine splittings,
  \href{http://dx.doi.org/10.1103/PhysRevD.82.114502}{Phys.\ Rev.\ D {\bf 82}, 114502 (2010)}.

\bibitem{Lees:2011bv}
  J.~P.~Lees {\it et al.} [BaBar Collaboration],
  Study of di-pion bottomonium transitions and search for the $h_b(1P)$ state,
  \href{http://dx.doi.org/10.1103/PhysRevD.84.011104}{Phys.\ Rev.\ D {\bf 84}, 011104 (2011)}.


\bibitem{Kornicer:2010cb}
  M.~Kornicer {\it et al.} [CLEO Collaboration],
  Measurements of branching fractions for electromagnetic transitions involving the $\chi_{bJ}(1P)$ states,
  \href{http://dx.doi.org/10.1103/PhysRevD.83.054003}{Phys.\ Rev.\ D {\bf 83}, 054003 (2011)}.

\bibitem{Tamponi:2015xzb}
  U.~Tamponi {\it et al.} [Belle Collaboration],
  First observation of the hadronic transition $ \Upsilon(4S) \to \eta h_{b}(1P)$ and new measurement of the $h_b(1P)$ and $\eta_b(1S)$ parameters,
  \href{http://dx.doi.org/10.1103/PhysRevLett.115.142001}{Phys.\ Rev.\ Lett.\  {\bf 115}, no. 14, 142001 (2015)}.





\bibitem{Godfrey:1986wj}
  S.~Godfrey and R.~Kokoski,
  The Properties of $P$-Wave Mesons with One Heavy Quark,
  \href{http://dx.doi.org/10.1103/PhysRevD.43.1679}{Phys.\ Rev.\ D {\bf 43}, 1679 (1991)}.

\bibitem{Matsuki:2010zy}
  T.~Matsuki, T.~Morii and K.~Seo,
  Mixing angle between $^3P_1$ and $^1P_1$ in HQET,
  \href{http://dx.doi.org/10.1143/PTP.124.285}{Prog.\ Theor.\ Phys.\  {\bf 124}, 285 (2010)}.

\bibitem{Barnes:2002mu}
  T.~Barnes, N.~Black and P.~R.~Page,
  Strong decays of strange quarkonia,
  \href{http://dx.doi.org/10.1103/PhysRevD.68.054014}{Phys.\ Rev.\ D {\bf 68}, 054014 (2003)}.


\bibitem{Sun:2014wea}
  Y.~Sun, Q.~T.~Song, D.~Y.~Chen, X.~Liu and S.~L.~Zhu,
  Higher bottom and bottom-strange mesons,
  \href{http://dx.doi.org/10.1103/PhysRevD.89.054026}{Phys.\ Rev.\ D {\bf 89}, no. 5, 054026 (2014)}.


\bibitem{Aubert:2004pwa}
  B.~Aubert {\it et al.} [BaBar Collaboration],
  A measurement of the total width, the electronic width, and the mass of the $\Upsilon(10580)$ resonance,
  \href{http://dx.doi.org/10.1103/PhysRevD.72.032005}{Phys.\ Rev.\ D {\bf 72}, 032005 (2005)}.


\bibitem{Chen:2008xia}
  K.-F.~Chen {\it et al.} [Belle Collaboration],
  Observation of an enhancement in $e^+e^- \to \Upsilon(1S)\pi^+ \pi^-$, $\Upsilon(2S)\pi^+ \pi^-$, and $\Upsilon(3S)\pi^+ \pi^-$ production around $\sqrt{s}=10.89$ GeV at Belle,
  \href{http://dx.doi.org/10.1103/PhysRevD.82.091106}{Phys.\ Rev.\ D {\bf 82}, 091106 (2010)}.


\bibitem{Santel:2015qga}
  D.~Santel {\it et al.} [Belle Collaboration],
  Measurements of the $\Upsilon$(10860) and $\Upsilon$(11020) resonances via $\sigma(e^+e^-\to \Upsilon(nS)\pi^+ \pi^-)$,
  \href{http://dx.doi.org/10.1103/PhysRevD.93.011101}{Phys.\ Rev.\ D {\bf 93}, no. 1, 011101 (2016)}.


\bibitem{Abdesselam:2015zza}
  A.~Abdesselam {\it et al.} [Belle Collaboration],
  Energy scan of the $e^+e^- \to h_b(nP)\pi^+\pi^-$ $(n=1,2)$ cross sections and evidence for $\Upsilon(11020)$ decays into charged bottomonium-like states,
  \href{http://dx.doi.org/10.1103/PhysRevLett.117.142001}{Phys.\ Rev.\ Lett.\  {\bf 117}, no. 14, 142001 (2016)}.


\bibitem{Aubert:2008ab}
  B.~Aubert {\it et al.} [BaBar Collaboration],
  Measurement of the $e^{+} e^{-} \to b \bar{b}$ cross section between $\sqrt{s}$ = 10.54-GeV and 11.20-GeV,
  \href{http://dx.doi.org/10.1103/PhysRevLett.102.012001}{Phys.\ Rev.\ Lett.\  {\bf 102}, 012001 (2009)}.


\bibitem{TorresRincon:2010fu}
  J.~M.~Torres-Rincon and F.~J.~Llanes-Estrada,
  Heavy Quark Fluorescence,
  \href{http://dx.doi.org/10.1103/PhysRevLett.105.022003}{Phys.\ Rev.\ Lett.\  {\bf 105}, 022003 (2010)}.



\bibitem{Zhou:2011sp}
  Z.~Y.~Zhou and Z.~Xiao,
  Hadron loops effect on mass shifts of the charmed and charmed-strange spectra,
  \href{http://dx.doi.org/10.1103/PhysRevD.84.034023}{Phys.\ Rev.\ D {\bf 84}, 034023 (2011)}.






\bibitem{Kwong:1988ae}
  W.~Kwong and J.~L.~Rosner,
  $D$-Wave Quarkonium Levels of the $\Upsilon$ Family,
  \href{http://dx.doi.org/10.1103/PhysRevD.38.279}{Phys.\ Rev.\ D {\bf 38}, 279 (1988)}.

\bibitem{Novikov:1977dq}
  V.~A.~Novikov, L.~B.~Okun, M.~A.~Shifman, A.~I.~Vainshtein, M.~B.~Voloshin and V.~I.~Zakharov,
  Charmonium and Gluons: Basic Experimental Facts and Theoretical Introduction,
  \href{http://dx.doi.org/10.1016/0370-1573(78)90120-5}{Phys.\ Rept.\  {\bf 41}, 1 (1978)}.



\bibitem{Badalian:1985es}
  A.~M.~Badalian, B.~L.~Ioffe and A.~V.~Smilga,
  Four Quark States In The Heavy Quark System,
  \href{http://dx.doi.org/10.1016/0550-3213(87)90248-3}{Nucl.\ Phys.\ B {\bf 281}, 85 (1987)}.


\bibitem{Appelquist:1974zd}
  T.~Appelquist and H.~D.~Politzer,
  Orthocharmonium and $e^+ e^-$ Annihilation,
  \href{http://dx.doi.org/10.1103/PhysRevLett.34.43}{Phys.\ Rev.\ Lett.\  {\bf 34}, 43 (1975)}.

\bibitem{DeRujula:1974rkb}
  A.~De Rujula and S.~L.~Glashow,
  Is Bound Charm Found?,
  \href{http://dx.doi.org/10.1103/PhysRevLett.34.46}{Phys.\ Rev.\ Lett.\  {\bf 34}, 46 (1975)}.

\bibitem{Chanowitz:1975ee}
  M.~S.~Chanowitz,
  Comment on the Decay of $\psi$ (3.1) Into Even G - Parity States,
  \href{http://dx.doi.org/10.1103/PhysRevD.12.918}{Phys.\ Rev.\ D {\bf 12}, 918 (1975)}.

\bibitem{Barbieri:1975am}
  R.~Barbieri, R.~Gatto and R.~Kogerler,
  Calculation of the Annihilation Rate of P Wave Quark - anti-Quark Bound States,
  \href{http://dx.doi.org/10.1016/0370-2693(76)90419-6}{Phys.\ Lett.\  {\bf 60B}, 183 (1976)}.

\bibitem{Barbieri:1976fp}
  R.~Barbieri, R.~Gatto and E.~Remiddi,
  Singular Binding Dependence in the Hadronic Widths of $1^{++}$ and $1^{+-}$ Heavy Quark anti-Quark Bound States,
  \href{http://dx.doi.org/10.1016/0370-2693(76)90729-2}{Phys.\ Lett.\  {\bf 61B}, 465 (1976)}.

\bibitem{Barbieri:1979be}
  R.~Barbieri, E.~d'Emilio, G.~Curci and E.~Remiddi,
  Strong Radiative Corrections to Annihilations of Quarkonia in QCD,
  \href{http://dx.doi.org/10.1016/0550-3213(79)90047-6}{Nucl.\ Phys.\ B {\bf 154}, 535 (1979)}.

\bibitem{Kwong:1987ak}
  W.~Kwong, P.~B.~Mackenzie, R.~Rosenfeld and J.~L.~Rosner,
  Quarkonium Annihilation Rates,
  \href{http://dx.doi.org/10.1103/PhysRevD.37.3210}{Phys.\ Rev.\ D {\bf 37}, 3210 (1988)}.

\bibitem{Ackleh:1991dy}
  E.~S.~Ackleh and T.~Barnes,
  Two photon widths of singlet positronium and quarkonium with arbitrary total angular momentum,
  \href{http://dx.doi.org/10.1103/PhysRevD.45.232}{Phys.\ Rev.\ D {\bf 45}, 232 (1992)}.

\bibitem{Belanger:1987cg}
  G.~Belanger and P.~Moxhay,
  Three-Gluon Annihilation of $D$-Wave Quarkonium,
  \href{http://dx.doi.org/10.1016/0370-2693(87)91630-3}{Phys.\ Lett.\ B {\bf 199}, 575 (1987)}.

\bibitem{Bergstrom:1991dp}
  L.~Bergstrom and P.~Ernstrom,
  Decays of D-wave quarkonium states into ggg and $\gamma$gg,
  \href{http://dx.doi.org/10.1016/0370-2693(91)90533-V}{Phys.\ Lett.\ B {\bf 267}, 111 (1991)}.

\bibitem{Ackleh:1991ws}
  E.~S.~Ackleh, T.~Barnes and F.~E.~Close,
  Two photon helicity selection rules and widths for positronium and quarkonium states with arbitrary angular momenta,
  \href{http://dx.doi.org/10.1103/PhysRevD.46.2257}{Phys.\ Rev.\ D {\bf 46}, 2257 (1992)}.

\bibitem{Robinett:1992px}
  R.~W.~Robinett and L.~Weinkauf,
  Covariant formalism for $F$-wave quarkonium production and annihilation: Application to $^3F_J \to gg$ decays,
  \href{http://dx.doi.org/10.1103/PhysRevD.46.3832}{Phys.\ Rev.\ D {\bf 46}, 3832 (1992)}.


\bibitem{Bradley:1980eh}
  A.~Bradley and A.~Khare,
  {QCD} Correction to the Leptonic Decay Rate of $D$-Wave Vector Mesons,
  \href{http://dx.doi.org/10.1007/BF01547876}{Z.\ Phys.\ C {\bf 8}, 131 (1981)}.






\bibitem{Gottfried:1977gp}
  K.~Gottfried,
  Hadronic Transitions Between Quark - anti-Quark Bound States,
  \href{http://dx.doi.org/10.1103/PhysRevLett.40.598}{Phys.\ Rev.\ Lett.\  {\bf 40}, 598 (1978)}.

\bibitem{Kuang:2006me}
  Y.~P.~Kuang,
  QCD multipole expansion and hadronic transitions in heavy quarkonium systems,
  \href{http://dx.doi.org/10.1007/s11467-005-0012-6}{Front.\ Phys.\ China {\bf 1}, 19 (2006)}.

\bibitem{Yan:1980uh}
  T.~M.~Yan,
  Hadronic Transitions Between Heavy Quark States in Quantum Chromodynamics,
  \href{http://dx.doi.org/10.1103/PhysRevD.22.1652}{Phys.\ Rev.\ D {\bf 22}, 1652 (1980)}.

\bibitem{Kuang:1990kd}
  Y.~P.~Kuang, Y.~P.~Yi and B.~Fu,
  Multipole Expansion in Quantum Chromodynamics and the Radiative Decays $J/\psi \to \gamma + \eta$ and $J/\psi \to \gamma + \pi^0$,
  \href{http://dx.doi.org/10.1103/PhysRevD.42.2300}{Phys.\ Rev.\ D {\bf 42}, 2300 (1990)}.

\bibitem{Kuang:1981se}
  Y.~P.~Kuang and T.~M.~Yan,
  Predictions for Hadronic Transitions in the $b\bar{b}$ System,
  \href{http://dx.doi.org/10.1103/PhysRevD.24.2874}{Phys.\ Rev.\ D {\bf 24}, 2874 (1981)}.

  \bibitem{Tye:1975fz}
    S.~H.~H.~Tye,
    A Quark-Binding String,
    \href{http://dx.doi.org/10.1103/PhysRevD.13.3416}{Phys.\ Rev.\ D {\bf 13}, 3416 (1976)}.

    \bibitem{Giles:1977mp}
      R.~Giles and S.~H.~H.~Tye,
      The Application of the Quark-Confining String to the $\psi$ Spectroscopy,
      \href{http://dx.doi.org/10.1103/PhysRevD.16.1079}{Phys.\ Rev.\ D {\bf 16}, 1079 (1977)}.

     \bibitem{Buchmuller:1979gy}
       W.~Buchmuller and S.~H.~H.~Tye,
       Vibrational States in the $\Upsilon$ Spectroscopy,
       \href{http://dx.doi.org/10.1103/PhysRevLett.44.850}{Phys.\ Rev.\ Lett.\  {\bf 44}, 850 (1980)}.

    \bibitem{Brown:1975dz}
      L.~S.~Brown and R.~N.~Cahn,
      Chiral Symmetry and $\psi^\prime \to \psi\pi\pi$ Decay,
      \href{http://dx.doi.org/10.1103/PhysRevLett.35.1}{Phys.\ Rev.\ Lett.\  {\bf 35}, 1 (1975)}.





\bibitem{Micu:1968mk}
  L.~Micu,
  Decay rates of meson resonances in a quark model,
  \href{http://dx.doi.org/10.1016/0550-3213(69)90039-X}{Nucl.\ Phys.\ B {\bf 10}, 521 (1969)}.

\bibitem{LeYaouanc:1972vsx}
  A.~Le Yaouanc, L.~Oliver, O.~Pene and J.~C.~Raynal,
  Naive quark pair creation model of strong interaction vertices,
  \href{http://dx.doi.org/10.1103/PhysRevD.8.2223}{Phys.\ Rev.\ D {\bf 8}, 2223 (1973)}.

\bibitem{LeYaouanc:1973ldf}
  A.~Le Yaouanc, L.~Oliver, O.~Pene and J.-C.~Raynal,
  Naive quark pair creation model and baryon decays,
  \href{http://dx.doi.org/10.1103/PhysRevD.9.1415}{Phys.\ Rev.\ D {\bf 9}, 1415 (1974)}.

\bibitem{LeYaouanc:1974cvx}
  A.~Le Yaouanc, L.~Oliver, O.~Pene and J.~C.~Raynal,
  Resonant Partial Wave Amplitudes in $\pi N \to \pi \pi N$ According to the Naive Quark Pair Creation Model,
  \href{http://dx.doi.org/10.1103/PhysRevD.11.1272}{Phys.\ Rev.\ D {\bf 11}, 1272 (1975)}.

\bibitem{LeYaouanc:1977fsz}
  A.~Le Yaouanc, L.~Oliver, O.~Pene and J.-C.~Raynal,
  Strong Decays of $\psi^{\prime\prime}$ (4.028) as a Radial Excitation of Charmonium,
  \href{http://dx.doi.org/10.1016/0370-2693(77)90250-7}{Phys.\ Lett.\  {\bf 71B}, 397 (1977)}.

\bibitem{LeYaouanc:1977gm}
  A.~Le Yaouanc, L.~Oliver, O.~Pene and J.~C.~Raynal,
  Why Is $\psi^{\prime\prime\prime}$ (4.414) SO Narrow?,
  \href{http://dx.doi.org/10.1016/0370-2693(77)90062-4}{Phys.\ Lett.\  {\bf 72B}, 57 (1977)}.

\bibitem{Hayne:1981zy}
  C.~Hayne and N.~Isgur,
  Beyond the Wave Function at the Origin: Some Momentum Dependent Effects in the Nonrelativistic Quark Model,
  \href{http://dx.doi.org/10.1103/PhysRevD.25.1944}{Phys.\ Rev.\ D {\bf 25}, 1944 (1982)}.


\bibitem{Jacob:1959at}
  M.~Jacob and G.~C.~Wick,
  On the general theory of collisions for particles with spin,
  \href{http://dx.doi.org/10.1016/0003-4916(59)90051-X}{Annals Phys.\  {\bf 7}, 404 (1959)}
  [Annals Phys.\  {\bf 281}, 774 (2000)].




\end{thebibliography}
\end{document}